\documentclass[pra,aps,twocolumn,superscriptaddress,floatfix,tightenlines]{revtex4-1}
\usepackage{amsmath}

\usepackage{graphicx}

\usepackage{amsfonts}
\usepackage{hyperref}
\usepackage[caption=false]{subfig}

\usepackage{float}
\usepackage{comment} 
\usepackage{braket}
\usepackage{empheq}

\usepackage{lipsum}

\usepackage{stackengine}
\newcommand\IncG[2][]{\addstackgap{%
		\raisebox{-.5\height}{\includegraphics[#1]{#2}}}}

\usepackage{changepage}

\usepackage{color,soul}
\definecolor{nblue}{rgb}{0.06,0.3,0.73}
\definecolor{nblack}{rgb}{0,0,0}

\usepackage{dirtytalk}
\begin{document}
	\title{Phase estimation of definite photon number states by using quantum circuits}
	\author{Peyman Najafi}
	\affiliation{Dept. of Physics, Institute for Advanced Studies in Basic Sciences (IASBS), Zanjan, Iran}
	
	\author{Ghasem Naeimi}
	\affiliation{Physics Groups, Qazvin Branch, Islamic Azad University, Qazvin, Iran}
	
	\author{Shahpoor Saeidian}
	\affiliation{Dept. of Physics, Institute for Advanced Studies in Basic Sciences (IASBS), Zanjan, Iran}

	\begin{abstract}
	We propose a method to map the conventional optical interferometry setup into quantum circuits. The unknown phase shift inside a Mach-Zehnder interferometer in the presence of photon loss is estimated by simulating the quantum circuits. For this aim, we use the Bayesian approach in which the likelihood functions are needed, and they are obtained by simulating the appropriate quantum circuits. The precision of four different definite photon-number states of light, which all possess six photons, is compared. In addition, the fisher information for the four definite photon-number states in the setup is also estimated to check the optimality of the chosen measurement scheme.
	\end{abstract}
	
	\maketitle
	
	\section{Introduction}
	\label{introduction}
	Precisely estimating an unknown parameter is important both in science and engineering. Depending on the parameter that one wants to estimate, one can use different approaches. In some cases, they may be able to measure the quantity directly, however, sometimes quantities are not directly accessible either in principle or due to experimental impediments, and they need to be estimated indirectly.
	
	Different estimation approaches have different precisions in estimating the unknown quantity. Generally speaking, these approaches use the resources of atoms and/or photons to estimate the desired quantity, and by increasing the resources, they can reveal the unknown quantity with better precision. Nevertheless, increasing the resources as much as one wants is not possible in some cases (for example biological samples~\cite{taylor_2016_quantum_metrology_and}), so it is essential to design an apparatus that gives the best possible estimation of the quantity by using a specified number of resources.
	
	The optical phase is the most significant parameter in the context of quantum metrology, especially in the optical interferometry~\cite{grote_2013_first, taylor_2013_biological, joo_2012_quantum}.
	Optical interferometers use photons as resources and are one of the most precise tools for measuring quantities such as length, index of refraction, velocity, etc~\cite{polino2020photonic, taylor_2016_quantum_metrology_and}, by mapping them into a phase shift. 
	By using classical states of light, the best precision that one can estimate the phase shift scales like $1/\sqrt{N}$~\cite{demkowicz_2009_quantum-phase-estimation-with-lossy-interferometer}, in which $N$ is the number of photons used. This limit is called the standard quantum limit (SQL). However, this is not the ultimate precision limit; we can enhance the precision by a factor of $1/\sqrt{N}$ and reach the Heisenberg limit ($1/N$) by employing quantum mechanical strategies like using entangled photonic states.   
	
	In a typical phase estimation setup, the input state together with the strategy that we use to extract the information and estimate the phase shift, determine the precision of the estimation~\cite{giovannetti_2006_quantum-metrology, giovannetti_2011_advances-in-quantum-metrology, mitchell_2004_super, leibfried_2004_toward, jin_2011_optimal, woolley_2008_nonlinear, choi_2008_bose, boixo_2008_quantum, lucke_2011_twin, anisimov_2010_quantum, campos_2003_optical, piera_2021_experimental}. 
	Choosing the best input state depends on the decoherence existing in the interferometer~\cite{dorner_2009_optimal, demkowicz_2009_quantum-phase-estimation-with-lossy-interferometer} and also on the number of resources (for example number of photons) that we can use. Apart from the above factors, our prior knowledge (before making any measurement) about the unknown quantity can also affect the optimal estimation strategy~\cite{demkowicz_2011_optimal}. 
	
	Consider a two-mode interferometer like Mach-Zehnder for instance. If prior to measurement we know that the unknown phase is around a specific value with slight deviation and the number of photons we can use is $N$, and there is no photon loss and decoherence in the interferometer,  $NOON$ states give the best possible precision and achieve the Heisenberg limit~\cite{kolodynski_2010_phase, lee_2002_quantum}. However, when there is photon loss in the arms of the interferometer, $NOON$ states become extremely fragile~\cite{escher_2011_general, chin_2012_quantum, genoni_2011_optical, ma_2011_quantum, kacprowicz_2010_experimental, kolodynski_2010_phase} and there exist other states which outperform $NOON$ states in different photon loss conditions. 
	
	A class of path-entangled photon Fock states, called $mm^{\prime}$ states, were proposed in~\cite{Huver_2008} to outperform $NOON$ states in the presence of photon loss. Huver \emph{et al.} have introduced a theoretically optimal detection operator and computed the precision of their phase estimation scheme by calculating the expectation value of the detection operator and using the error propagation formula. However, experimental realization of their optimal measurement scheme has not been reported yet. Performance of $NOON$ and $mm^{\prime}$ states by using parity detection, has also been investigated in the presence of photon loss~\cite{jiang2012strategies} and phase fluctuation~\cite{dowling_2013}. 
	Another class of definite photon number states that act better than the $NOON$ states in the presence of loss are Holland-Burnett ($HB$) states~\cite{holland_1993_interferometric}. Unlike $NOON$ states, $HB$ states can be prepared easily by using a beam splitter and photon pairs from spontaneous parametric down-conversion (SPDC)~\cite{HB}. $HB$ states are almost optimally robust not only to photon loss but also to inefficiencies in state preparation and photon detection and can beat SQL in realistic phase estimation scenarios~\cite{HB, datta_2011_quantum, sun_2008_experimental, thekkadath_2020_quantum, xiang_2011_entanglement, xiang_2013_optimal}.  
	
	In this work, we focus on four different definite photon number input states of light, which all possess six photons and belong to three different classes of non-classical states: $NOON$ states, $mm^{\prime}$ states, and $HB$ states. We will compare their phase estimation precision in different photon loss conditions in a Mach-Zehnder interferometer by mapping the corresponding quantum optical setups into quantum circuits and simulating them to obtain the information needed. This is a new approach towards investigating phase estimation precision of definite photon number input states in the presence of loss. We will choose a 50-50 beam splitter with two photon number-resolving detectors (PNRD) as our measurement scheme and use Bayesian estimation to estimate the unknown phase shift. Likelihood functions that we need in the Bayesian estimation for different input states will also be approximated by simulating appropriate quantum circuits rather than calculating them theoretically, which can be arduous in the presence of photon loss.
	
	In detail, we proceed as follows: Section \ref{model} explains our model and shows how to map the corresponding quantum optical setups into quantum circuits. In section \ref{results} our results are presented and analyzed. Finally, a summary and conclusion are given in section \ref{summary}. 
	
	\section{Describing Our Model and designing corresponding quantum circuits}
	\label{model}
\begin{figure}[h!]
		\includegraphics[width=\linewidth]{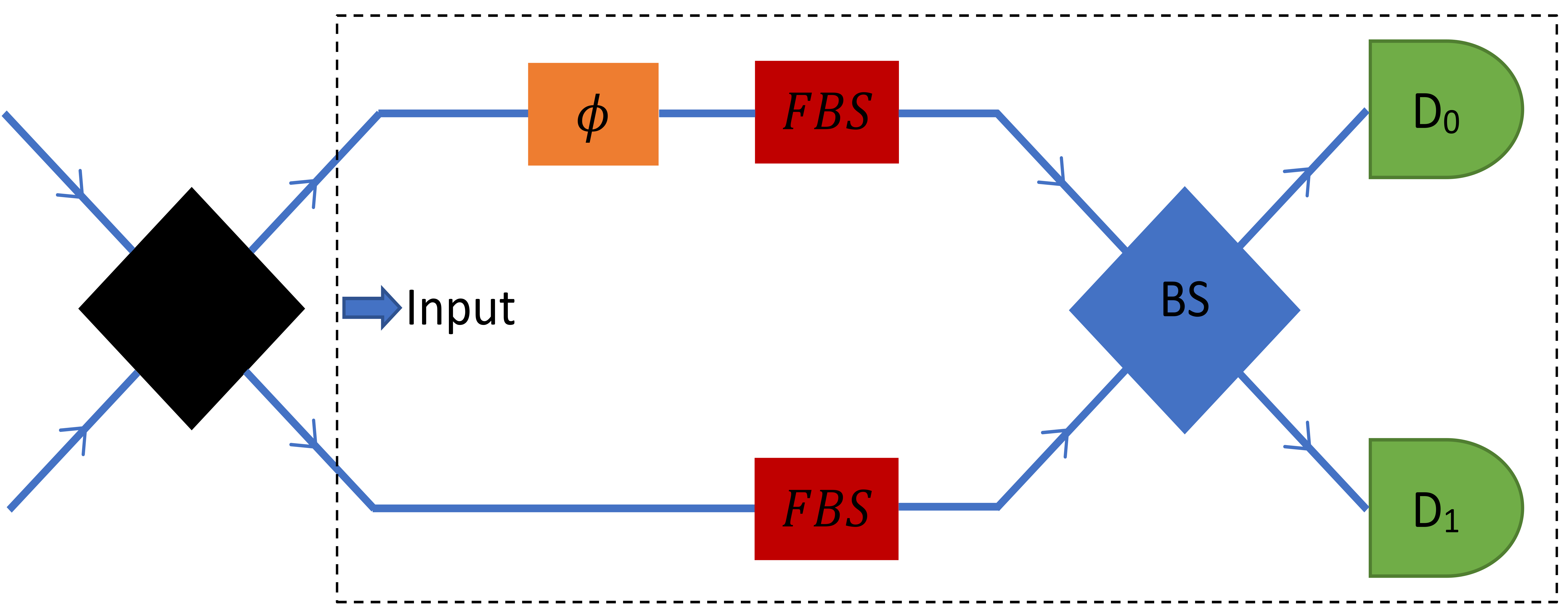}
		\caption{The optical interferometry setup. Black box prepares the desired input state. $\phi$ indicates the phase shift between two arms and photon loss is modeled by adding a fictitious beam splitter (red box) in each arm. The blue box represents a 50-50 beam splitter and $D_0$ and $D_1$ are photon-number-resolving detectors.}
		\label{Fig1}
		
	\end{figure}

	The phase estimation setup considered here is similar to the ordinary Mach-Zehnder interferometer in which the first beam splitter is replaced by a black box that can generate any desired input photonic state.
	To model the photon loss, two extra fictitious beam splitters are used in both arms of the interferometer~\cite{knott2014attaining, jiang2012strategies, Huver_2008, cooper_2012_robust, jarzyna_2012_quantum}. Different rates of photon loss can simply be created in the interferometer by adjusting the transmission coefficients of the beam splitters. Figure. \ref{Fig1} shows the corresponding optical interferometry setup in which the detectors ($D_0$ and $D_1$) are photon-number-resolving detectors (PNRD).
	
	To simulate the optical interferometry setup, an appropriate quantum circuit is designed. At first, we map the input photonic state into a corresponding state in qubits. There are two modes for input photons: The lower and the upper paths are the first and the second modes, respectively. So an arbitrary pure input state with a definite photon number ($N$) would be like
	\begin{equation}
		\ket{\psi_{input}} = \sum_{m=0}^{N} c_m \ket{m}_l \ket{N-m}_u,
		\label{mode-rep}
	\end{equation}
where $\sum_{m=0}^{N} |c_m|^2 = 1$. Eq.~(\ref{mode-rep}) is in mode representation and $\ket{m}_l \ket{N-m}_u$ means that $m$ out of $N$ photons are in the lower arm and the remaining $N-m$ photons are in the upper arm. The input state is a superposition of states like $\ket{m}_l \ket{N-m}_u$, so if we can map these states into qubits of our circuit appropriately, we can represent the input state by the superposition of these corresponding qubit states.

By writing $\ket{m}_l \ket{N-m}_u$  in the particle representation (and applying bosonic symmetry) we arrive at~\cite{quantum_limits}:
\begin{equation}
	\begin{aligned}
		&\ket{m}_l \ket{N-m}_u \xrightarrow{\text{map to particle representation}}\\
		&\frac{1}{\sqrt{\binom{N}{m}}} \sum_{\prod} \prod(\ket{l}_1\ket{l}_2 ... \ket{l}_m \ket{u}_{m+1}\ket{u}_{m+2} ... \ket{u}_N).				
	\end{aligned}
	\label{particle-rep}
\end{equation} 
Here, $\ket{l}_j (\ket{u}_j)$ means that the $jth$ photon is in the lower (upper) arm and $\prod$ is an arbitrary permutation of `$l$'s and `$u$'s, and summation is over nontrivial permutations. Eq.~(\ref{particle-rep}) clarifies how to show the input state in qubits. We choose $N$ qubits and label them from $1$ to $N$ by assuming that each qubit corresponds to a photon in Eq.~(\ref{particle-rep}). 
Then we create the bosonic symmetric state of these qubits. In each term of the total superposition state (Eq.~(\ref{particle-rep})), if the photon is in the lower arm, we put the analogous qubit in $\ket{0}$, and if the photon is in the upper arm, we put the qubit in $\ket{1}$. The interferometer's two arms are distinguishable, which means that considering the two orthogonal states (computational bases) in this mapping is legitimate. By this map, we can only devote two distinguishable modes for photons at each stage of the circuit, which can only show the case where there is no loss. When we introduce photon loss in both arms, we need to have two other distinguishable modes to each photon so that they can be in four distinguishable modes: lower arm, upper arm, lost from the lower arm, and lost from the upper arm.

Therefore, we change the mapping, and two qubits for each photon are allocated. Our map in the presence of the photon loss is as follows:
	\begin{equation*}
		\begin{cases}
			\text{photon in the lower arm} &\rightarrow \ket{0}\ket{0} \\
			\text{photon in the upper arm} & \rightarrow \ket{1}\ket{1} \\
			\text{photon lost from the lower arm} & \rightarrow \ket{0}\ket{1}\\
			\text{photon lost from the upper arm} & \rightarrow \ket{1}\ket{0}\\
		\end{cases} 
	\end{equation*}
	As a result, if the input photonic state has $N$ photons, we allocate $2N$ qubits to them in our quantum circuit.

	\begin{figure}[h!]
		\includegraphics[width=\linewidth, height=5.5cm]{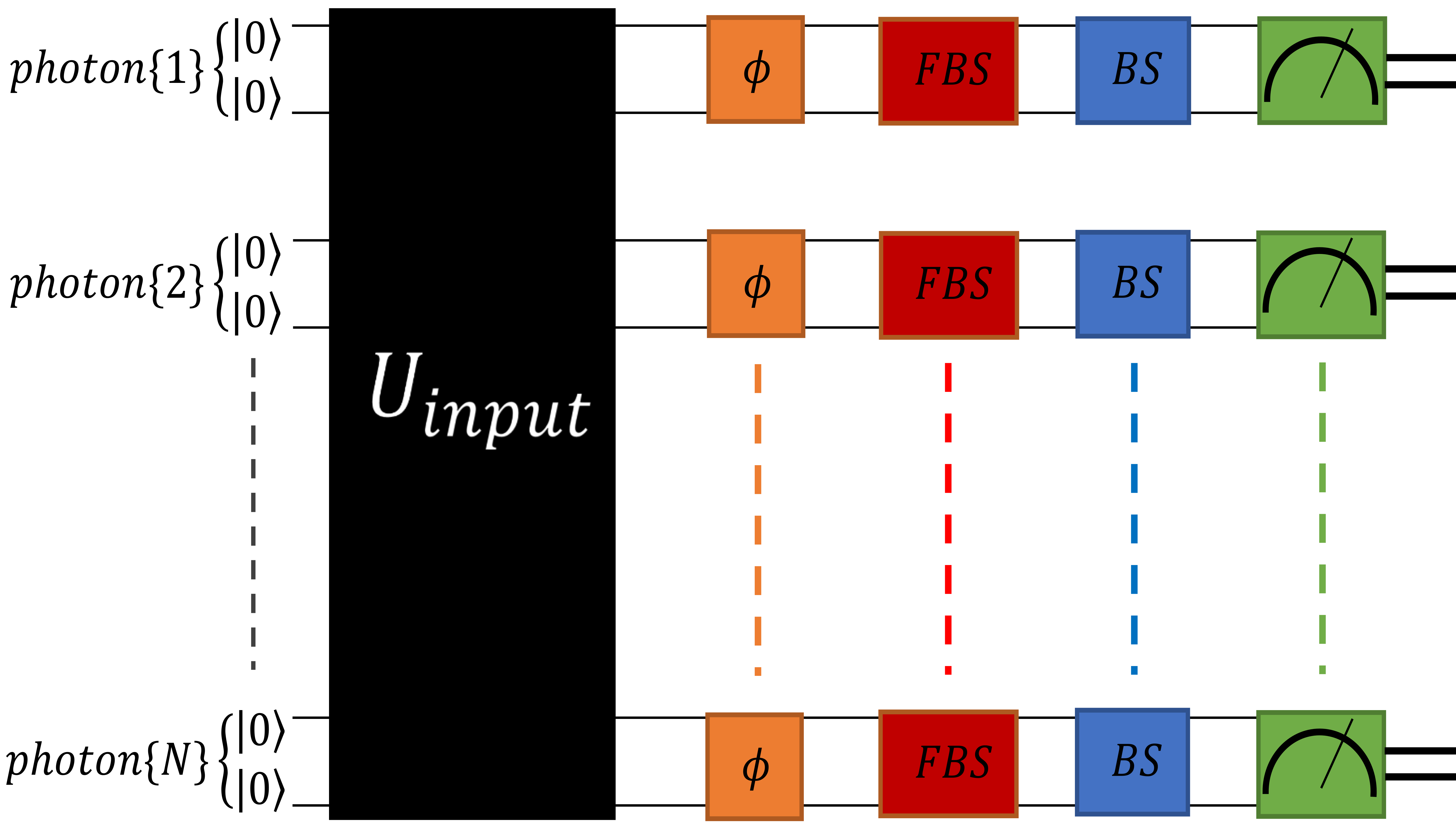}
		\caption{Our quantum circuit to simulate setup of Fig.~\ref{Fig1}. $2N$ qubits are chosen to represent an input state corresponding to $N$ photons. $U_{input}$ (black box) implements the unitary evolution to prepare the desired input state. Orange, red and blue boxes implement unitary transformations of phase shift, photon loss and the final 50-50 beam splitter respectively. Finally qubits are measured in computational bases.}
		\label{general-q-circuit}
	\end{figure}
	
	Figure.~\ref{general-q-circuit} shows the design of our quantum circuit. We assume that all qubits are initially at $\ket{0}$ state. $U_{input}$ is a unitary evolution that transforms $\ket{0}_1 \ket{0}_2 ... \ket{0}_{2N}$ to our desired input state. Considering that our input state is like Eq.~(\ref{mode-rep}) in photon representation, the analogous state of the qubits would be
	\begin{equation}
		\begin{aligned}
			\sum_{m=0}^{N} \frac{c_m}{\sqrt{\binom{N}{m}}}&\sum_{\prod^{\prime}} \sideset{}{'}\prod (\overset{photon \{1\}}{\overbrace{\ket{0}_1\ket{0}_2}} \; 
			\overset{photon \{2\}}{\overbrace{\ket{0}_3\ket{0}_4}} ... \overset{photon \{m\}}{\overbrace{\ket{0}_{2m-1}\ket{0}_{2m}}} \\
			&\overset{photon \{m+1\}}{\overbrace{\ket{1}_{2m+1}\ket{1}_{2m+2}}} ...
			\overset{photon \{N\}}{\overbrace{\ket{1}_{2N-1}\ket{1}_{2N}}}),
		\end{aligned}
		\label{general-input-qubit}
	\end{equation}
	
	where $\prod^{\prime}$ arbitrarily permutes the corresponding qubit pairs' states of different photons and as in Eq.~(\ref{particle-rep}), $\sum_{\prod^{\prime}}$ means summation over nontrivial permutations. As mentioned earlier, $U_{input}$ transforms initial state $\ket{0}_1...\ket{0}_{2N}$ to the state of Eq.~(\ref{general-input-qubit}). To implement the $U_{input}$, we have to decompose it into products of gates that are available (we call them basis gates here) to us. Appendix (\ref{basis-gates}) illustrates the list of the gates that we have used in our quantum circuits, and appendix (\ref{U-input}) demonstrates how to do this decomposition. 
	After preparing the input state, the phase shift, fictitious beam splitters, and the final beam splitter of the interferometer are implemented in order. 
	As it is clear from Fig.~\ref{general-q-circuit}, each of the above three parts (phase shift, fictitious beam splitters, and the final beam splitter) consist of $N$ similar subroutines. Each particular subroutine acts only on the two qubits corresponding to a specific photon. Subroutines of the three parts transform the state of a qubit pair corresponding to a photon as below~\cite{simon_2017_quantum_metrology}
	\begin{equation}
		\label{eq_phase_shift_transformation } 
		\begin{aligned}
			\phi 
			&\begin{cases}
				&\ket{0}\ket{0}\rightarrow \ket{0}\ket{0}\\
				&\ket{1}\ket{1}\rightarrow \exp(i\phi)\ket{1}\ket{1}
			\end{cases} \\
			FBS
			&\begin{cases}
				&\ket{0}\ket{0}\rightarrow \sqrt{t_0}\ket{0}\ket{0} + i\sqrt{1 - t_0}\ket{0}\ket{1}\\
				&\ket{1}\ket{1}\rightarrow \sqrt{t_1}\ket{1}\ket{1} + i\sqrt{1 - t_1}\ket{1}\ket{0}
			\end{cases} \\
			BS 
			&\begin{cases}
				&\ket{0}\ket{0}\rightarrow \frac{1}{\sqrt{2}}\ket{0}\ket{0} + \frac{i}{\sqrt{2}}\ket{1}\ket{1}\\
				&\ket{1}\ket{1}\rightarrow \frac{i}{\sqrt{2}}\ket{0}\ket{0} + \frac{1}{\sqrt{2}}\ket{1}\ket{1}
			\end{cases} 
		\end{aligned}
	\end{equation}
	
	In Eq.~(\ref{eq_phase_shift_transformation }), $\phi$ is the phase shift in the interferometer, and $t_0$ and $t_1$ are the transmissivity of the lower and upper fictitious beam splitter, respectively. $L_0 = 1-t_0$ and $L_1 = 1-t_1$ determine the photon loss rate in the lower and upper arms respectively. We interpret the state after the final beam splitter in a way that $\ket{0}\ket{0}$ and $\ket{1}\ket{1}$ mean photon is in the path going to $D_0$ and $D_1$ respectively. Appendix (\ref{ap-phase}) shows the sequence of the basis gates to implement these transformations.
	
	At the end of the circuit, we measure the qubits in the computational basis and end up with a string of `$0$'s and `$1$'s. Finally, we convert this string to a result that shows us the number of photons detected by $D_0$ and $D_1$. 
	
	Now our quantum circuit simulates the analogous quantum optical interferometer, and we can compare different input states' precision in estimating phase shift in different photon loss rates. First, we implement the appropriate $U_{input}$ to create our desired input state, then we put $\phi = \phi^*$ (which is the value of the quantity that the optical interferometer is going to estimate). Then we adjust $t_0$ and $t_1$ to create the photon loss rates that we want. Finally, we simulate the circuit for $N_r$ times, which means that the input state is sent through the interferometer for $N_r$ times to find a probability distribution $P(\phi)$ for phase. The variance of this probability distribution shows the error in estimating the phase shift; the lower the variance, the better the precision. 
	
	To estimate the properties of an object in a real experiment, it is placed in one of the arms of the interferometer and then the phase shift (induced by the object) is estimated by sending the input states (photons) through the interferometer for a specified number of times. 
	The phase shift's actual value ($\phi^*$) is unknown to the experimenter, but the number of photons that get detected at different detectors each time is known, and the phase shift can be estimated by this information. In this work, it is assumed that before sending the input states through the interferometer, the experimenter initially knows that the unknown phase shift is limited to a finite range,
	\begin{equation}
		P_{pri}(\phi) = \begin{cases}
			\frac{1}{\phi_f - \phi_i} \quad &\phi_i \leq \phi \leq \phi_f \\
			0 \quad & otherwise
		\end{cases}.		
	\end{equation}
	We show the measurement result of each run of the experiment by $D(n_0,n_1)$, which means that $n_0$ and $n_1$ photons are detected at $D_0$ and $D_1$ respectively; note that $n_0 + n_1 \leq N$. If we know the conditional likelihood functions $P_{n_0,n_1}(\phi) \equiv \frac{P(D(n_0,n_1)|\phi)}{\int_{0}^{2\pi} d\phi^{\prime} \; P(D(n_0,n_1)|\phi^{\prime})} $ for different possible values of $n_0$ and $n_1$, the probability distribution could be estimated by the following algorithm~\cite{knott2015robust, ly_2017_tutorial-on-fisher}
	\begin{enumerate}
		\item $P(\phi) = P_{pri}(\phi)$
		\item  run the experiment and multiply $P(\phi)$ by $P_{n_0,n_1}(\phi)$ 
		\item $P(\phi) = normalized(P(\phi)*P_{n_0,n_1}(\phi))$
		\item repeat steps 2 and 3 for $N_r-1$ times
	\end{enumerate} 
	The output $P(\phi)$ of this algorithm is the probability distribution of the unknown phase shift ($\phi^*$) after sending the input photonic state for $N_r$ times, and is a function of the $N_r$ measurement results ($\vec{D}$). We refer to it by $P_{final}(\phi|\vec{D})$. We can estimate the $\phi^*$ by the following formula:
	
	\begin{equation}
		\hat{\phi}_{est}(\vec{D}) = \int_{\phi_i}^{\phi_f}d\phi \; P_{final}(\phi|\vec{D})\, \phi.
	\end{equation}  
	The standard deviation is given by
	\begin{equation}
		\Delta \hat{\phi}_{est}(\vec{D}) = \sqrt{\int_{\phi_i}^{\phi_f}d\phi \; P_{final}(\phi|\vec{D})\, (\phi - \hat{\phi}_{est}(\vec{D}))^2}.
	\end{equation}  
	$\Delta \hat{\phi}_{est}$ is a function of the specific measurement results, $\vec{D}$, that has been obtained by doing the experiment for $N_r$ times~\cite{li_2018_frequentist}. Repeating the same experiment may yield different measurement results $\vec{D^{\prime}}$, which could result in a slightly different $\Delta \hat{\phi}_{est}$. Therefore, in order to remove the dependency on specific $\vec{D}$, we will calculate the average standard deviation
	\begin{equation}
		(\Delta \hat{\phi}_{est})_{avg} = \sum_{\vec{D}} P(\vec{D}|\phi^*)\Delta \hat{\phi}_{est}(\vec{D}).
	\end{equation}  
	Now we focus on how to find $P(D(n_0,n_1)|\phi)$ for each input state in specified condition of photon loss. We can find it by calculating it analytically or using the quantum circuit that we have designed to approximate it. Calculating $P(D(n_0,n_1)|\phi)$ analytically can sometimes be difficult, especially when there is photon loss, but our method avoids this difficulty and approximate the probability distribution up to the precision that we want. To do that, we prepare the input state, fix the values of $t_0$ and $t_1$, which creates the desired photon loss condition, and choose $n$ equally spaced points from $\phi_i$ to $\phi_f$. Now by running the circuit for $W$ times ($W \gg 1$) at each of these n points, we can approximate $P(D(n_0,n_1)|\phi)$ by the following formula:
	\begin{equation}
		P(D(n_0,n_1)|\phi) \approx \frac{R(n_0,n_1)}{W},
	\end{equation}   
	where $R(n_0,n_1)$ is the number of times that $n_0$ and $n_1$ photons are detected at lower and upper detectors respectively. By increasing the number of points ($n$) and the number of times that we simulate the circuit at each point ($W$), we can approach the exact $P(D(n_0,n_1)|\phi)$.

As mentioned in section \ref{introduction}, we consider definite photon number input states of three different classes, which all possess the same number of photons. The states of $NOON$ and $mm^{\prime}$ classes for $N$ photons are written as~\cite{jiang2012strategies}: 
\begin{equation}
	\ket{N::0} = \frac{1}{\sqrt{2}} (\ket{N}_l \ket{0}_u + \ket{0}_l \ket{N}_u),	
\end{equation}   

\begin{equation}
	\ket{m::m^{\prime}} = \frac{1}{\sqrt{2}} (\ket{m}_l \ket{m^\prime}_u + \ket{m^\prime}_l \ket{m}_u).	
\end{equation}   
Here $m$ and $m^{\prime}$ are positive integers and without loss of generality we assume that $m > m^{\prime}$.  $m+m^{\prime}=N$ is the total number of photons that $\ket{m::m^{\prime}}$ state possesses. The last input state that is investigated here is called the Holland-Burnett (HB) state, which has an even number of $N$ photons and is given by the following formula~\cite{HB, holland_1993_interferometric} (before experiencing the phase shift):

\begin{equation}
	\begin{aligned}
		\ket{HB(N)} = &\sum_{n=0}^{N/2} c_n \ket{2n}_l\ket{N-2n}_u, \\ 
		&c_n = \frac{\sqrt{(2n)!(N-2n)!}}{2^{N/2}n!(\frac{N}{2}-n)!}
		\label{eq-hb}
	\end{aligned}
\end{equation}

In our simulations we take $N = 6$ and the four input states considered are: $\ket{6::0}$, $\ket{5::1}$, $\ket{4::2}$ and $\ket{HB(6)}$. 

\section{Results}
\label{results}

In this work, two different cases for photon loss are considered. Subsection \ref{result_sym} shows the results for the case of symmetric photon loss where the probability of photon loss in the lower and upper arms of the interferometer are equal $(L_0 = L_1 = L, \, t_0=t_1=t)$, and subsection \ref{result_asym} shows the results for the asymmetric photon loss case where there is no photon loss in the lower arm of the interferometer $(L_0 = 0)$.
In our simulation of estimating the unknown phase shift, it is assumed that $\phi \in [\phi_i,\phi_f]$, where $\phi_f - \phi_i = \frac{\pi}{6}$; this is done to avoid the ambiguity in estimating phase. The actual value of the unknown phase is $\phi^* = \frac{\phi_i + \phi_f}{2}$, and it is chosen in a way that minimizes $\delta \phi$ for each input state. For $\ket{6::0}$, $\ket{5::1}$ and $\ket{4::2}$, $\phi^*$ is $\pi/12$, $\pi/8$ and $\pi/4$, respectively~\cite{dowling_2013}. For $\ket{HB(6)}$, approximate value of the $\phi^*$ is found by simulating the quantum circuits with $\ket{HB(6)}$ as input state for different values of phase shifts, and comparing their   
estimated average standard deviations.
\subsection{Symmetric photon dissipation}
\label{result_sym}
		\begin{table*}
	\caption{Unnormalized likelihood functions for different input states in symmetric photon loss when $t_0=t_1=0.5$. The vertical axis in each row shows $P_{n_0,n_1}(\phi)/\max(P_{n_0,n_1}(\phi))$ for the measurement result $D(n_0,n_1)$, which is shown in the left side of each row. Each column in the table corresponds to one of the input states, which is shown at the top of the column. By moving down the table, the total number of detected photons $(n_0 + n_1)$ generally decreases.} \label{tab_interference_pattern}
	\centering
	
	\begin{tabular}{c c c c c}
		&$\ket{6::0}$ & $\ket{5::1}$ & $\ket{4::2}$ & $\ket{HB(6)}$\\
		
		$D(6,0)$
		&\IncG[width=.23\textwidth,height=.12\textheight]{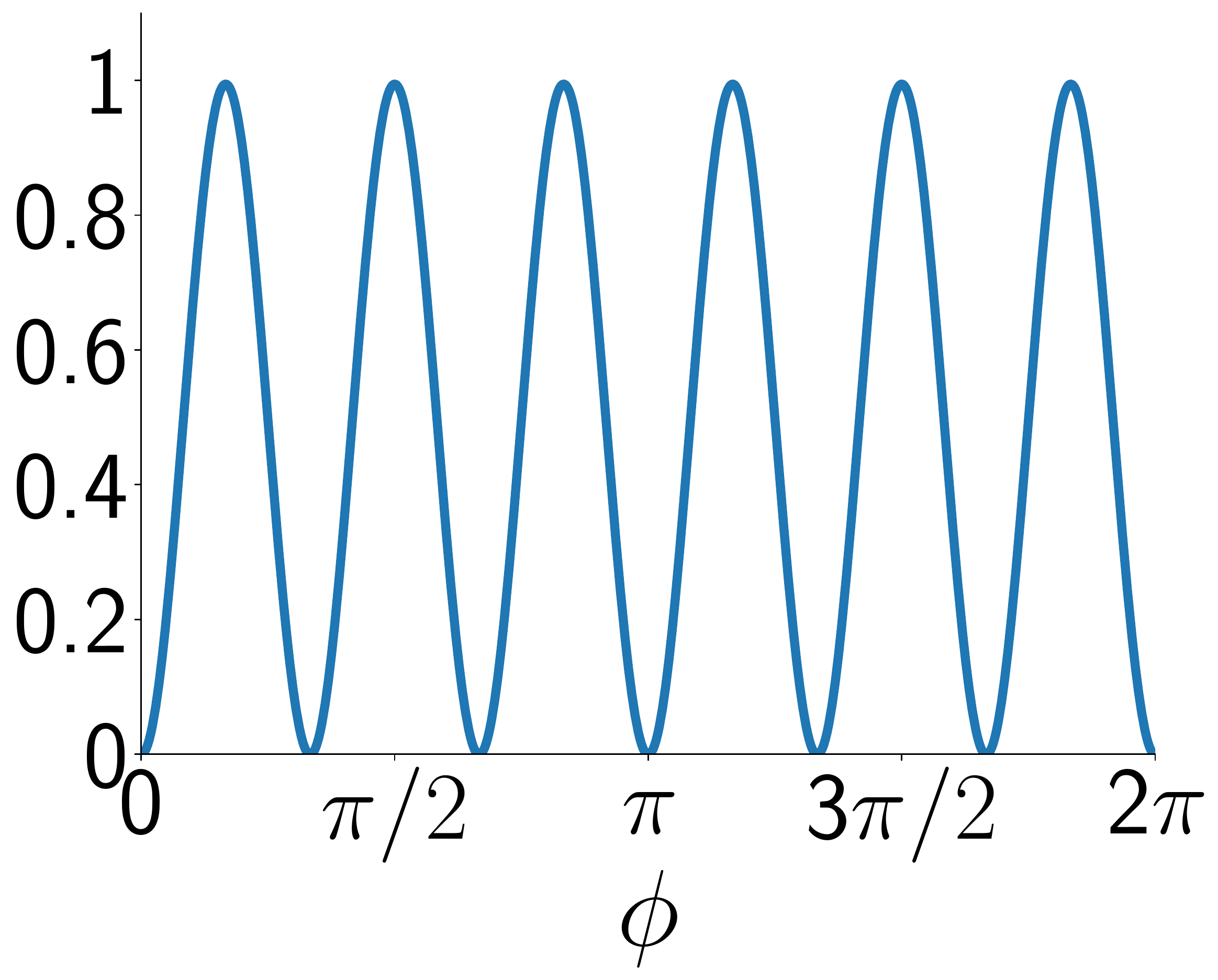}
		&\IncG[width=.23\textwidth,height=.12\textheight]{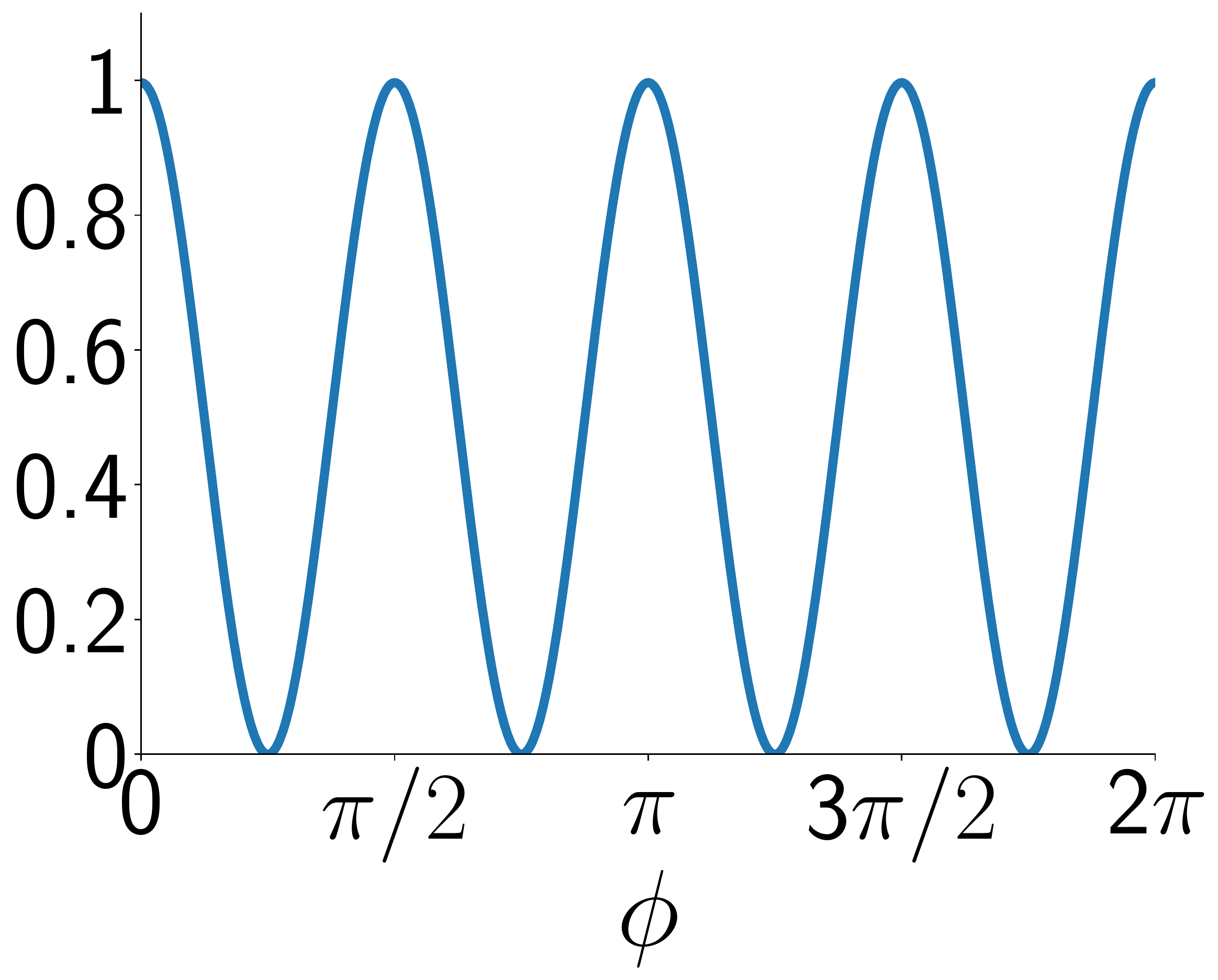}
		&\IncG[width=.23\textwidth,height=.12\textheight]{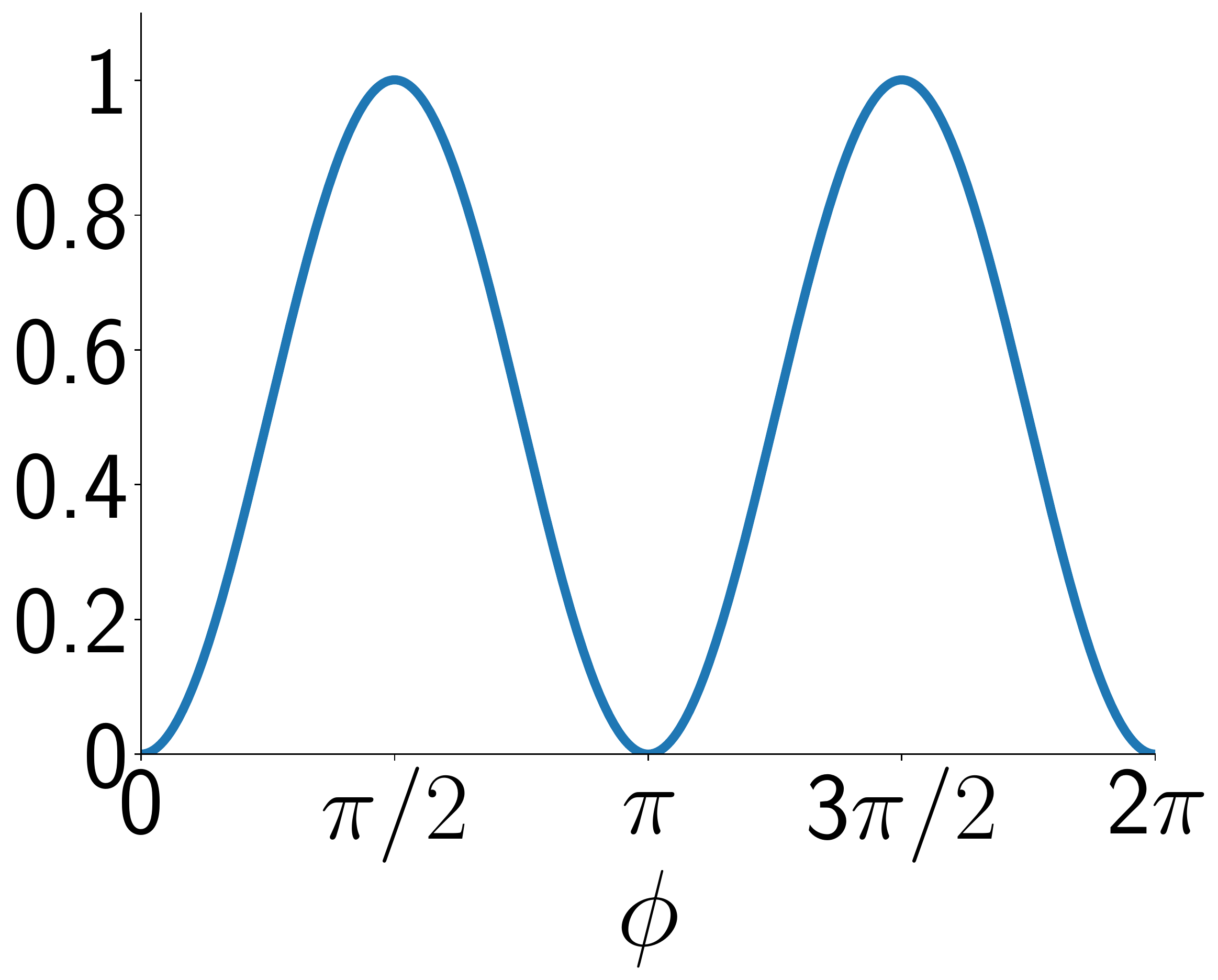}
		&\IncG[width=.23\textwidth,height=.12\textheight]{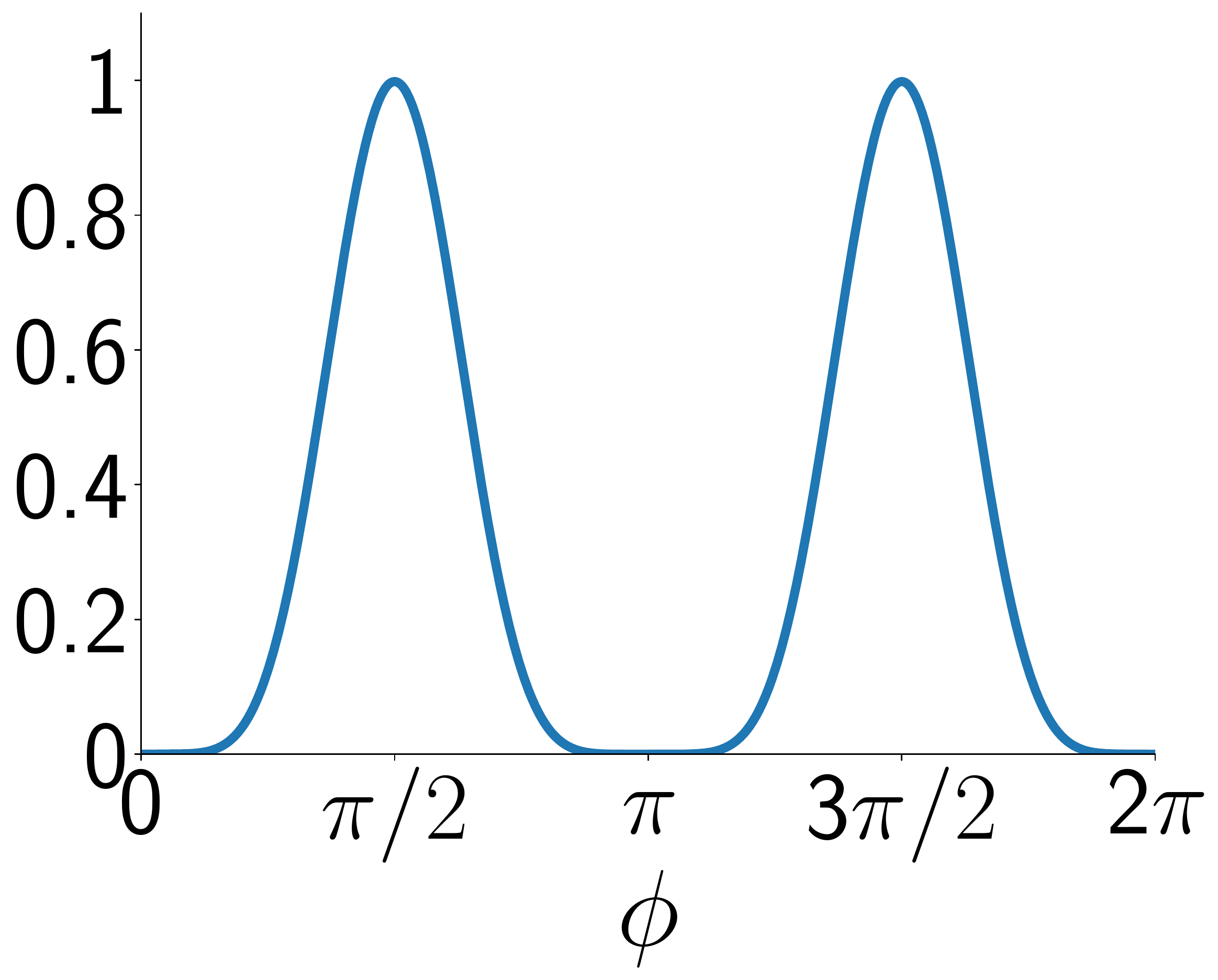}\\
		
		$D(5,1)$
		&\IncG[width=.23\textwidth,height=.12\textheight]{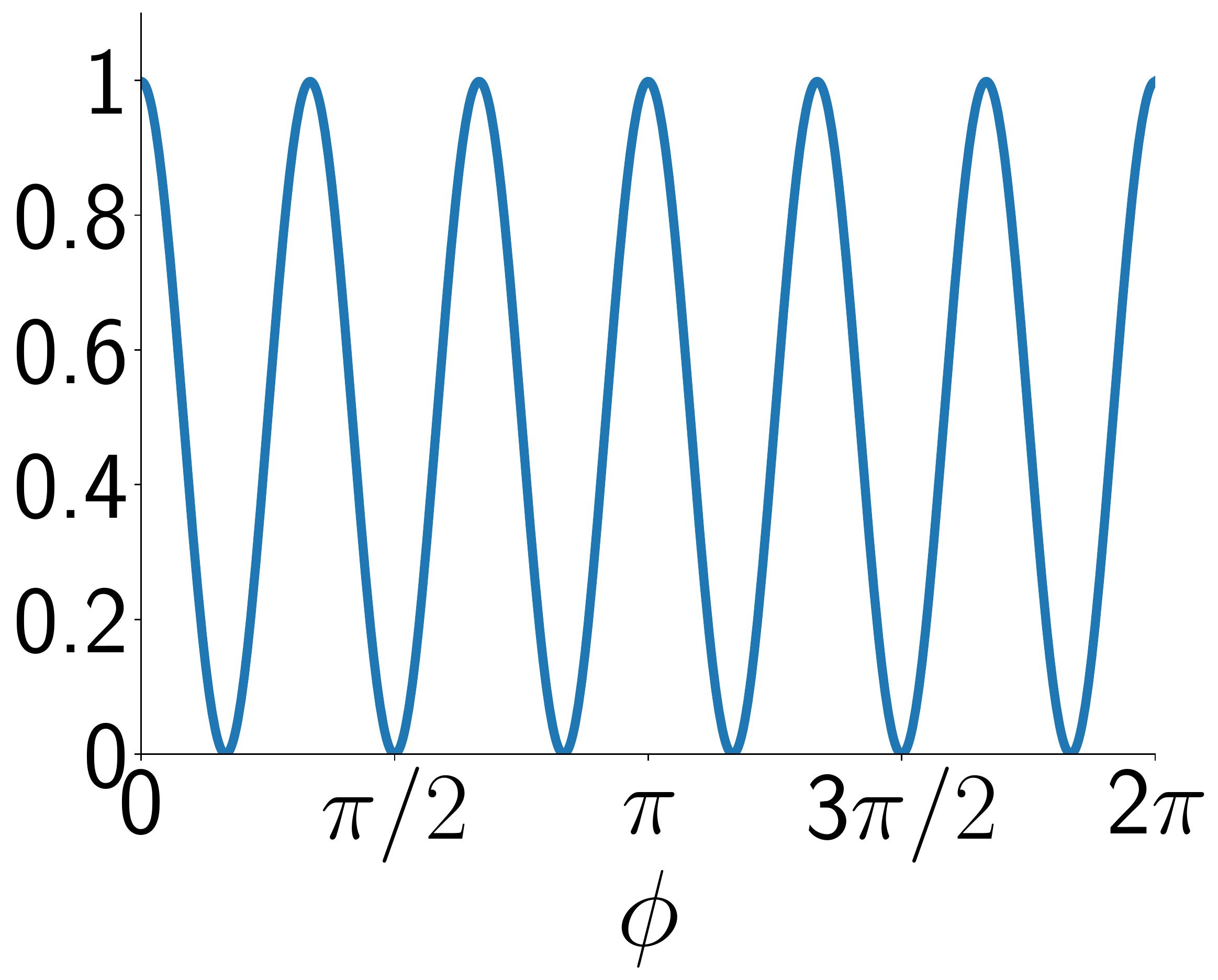}
		&\IncG[width=.23\textwidth,height=.12\textheight]{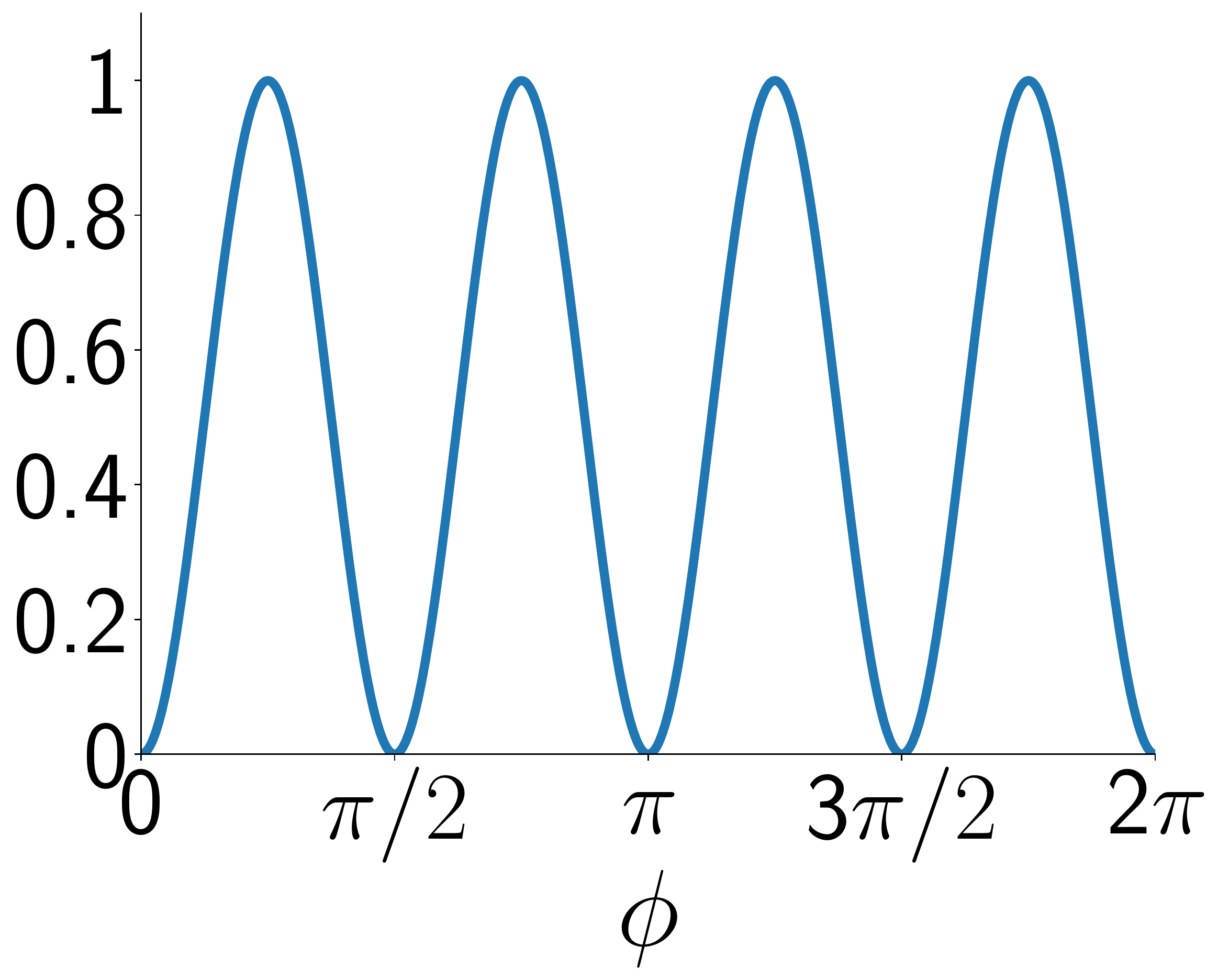}
		&\IncG[width=.23\textwidth,height=.12\textheight]{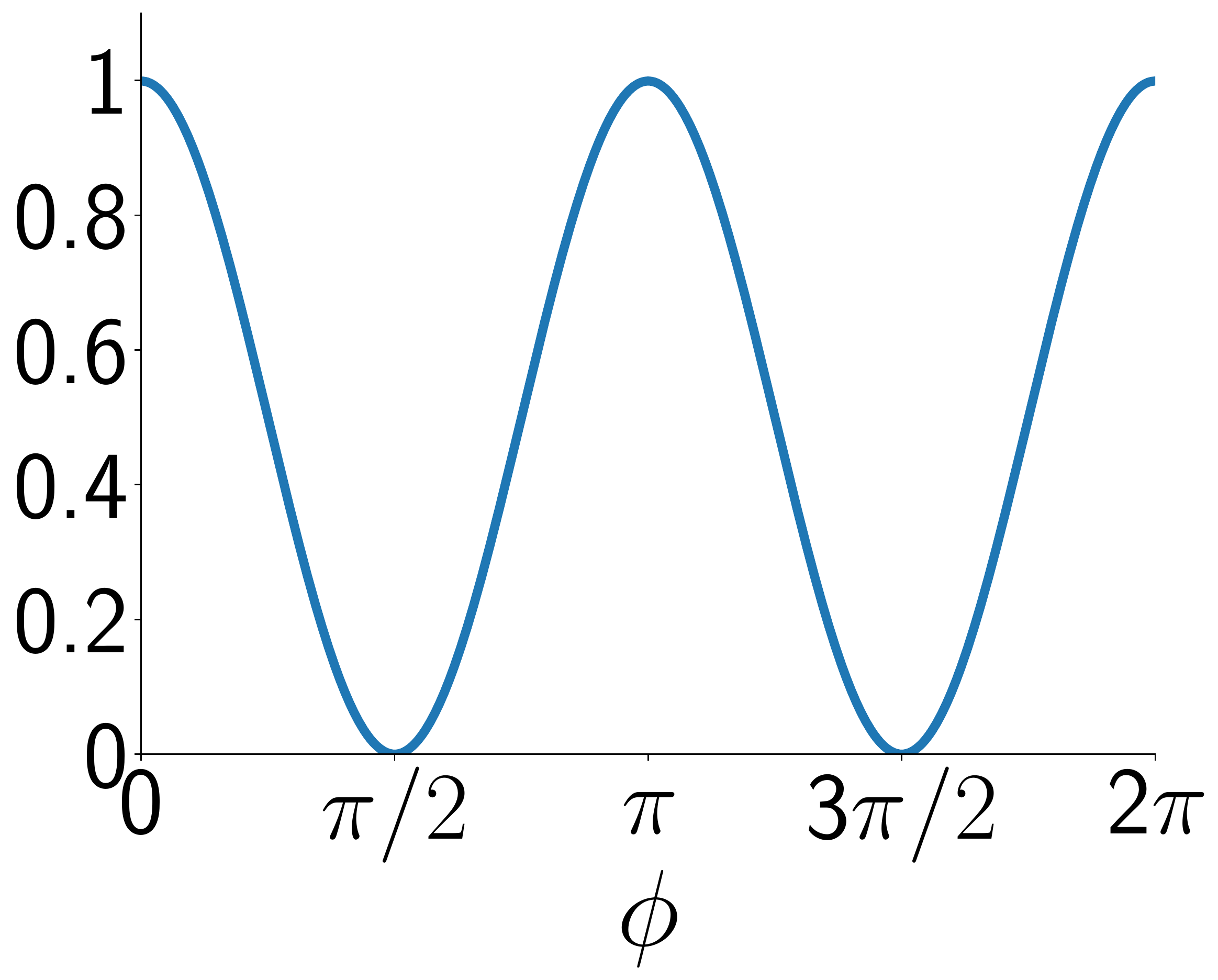}
		&\IncG[width=.23\textwidth,height=.12\textheight]{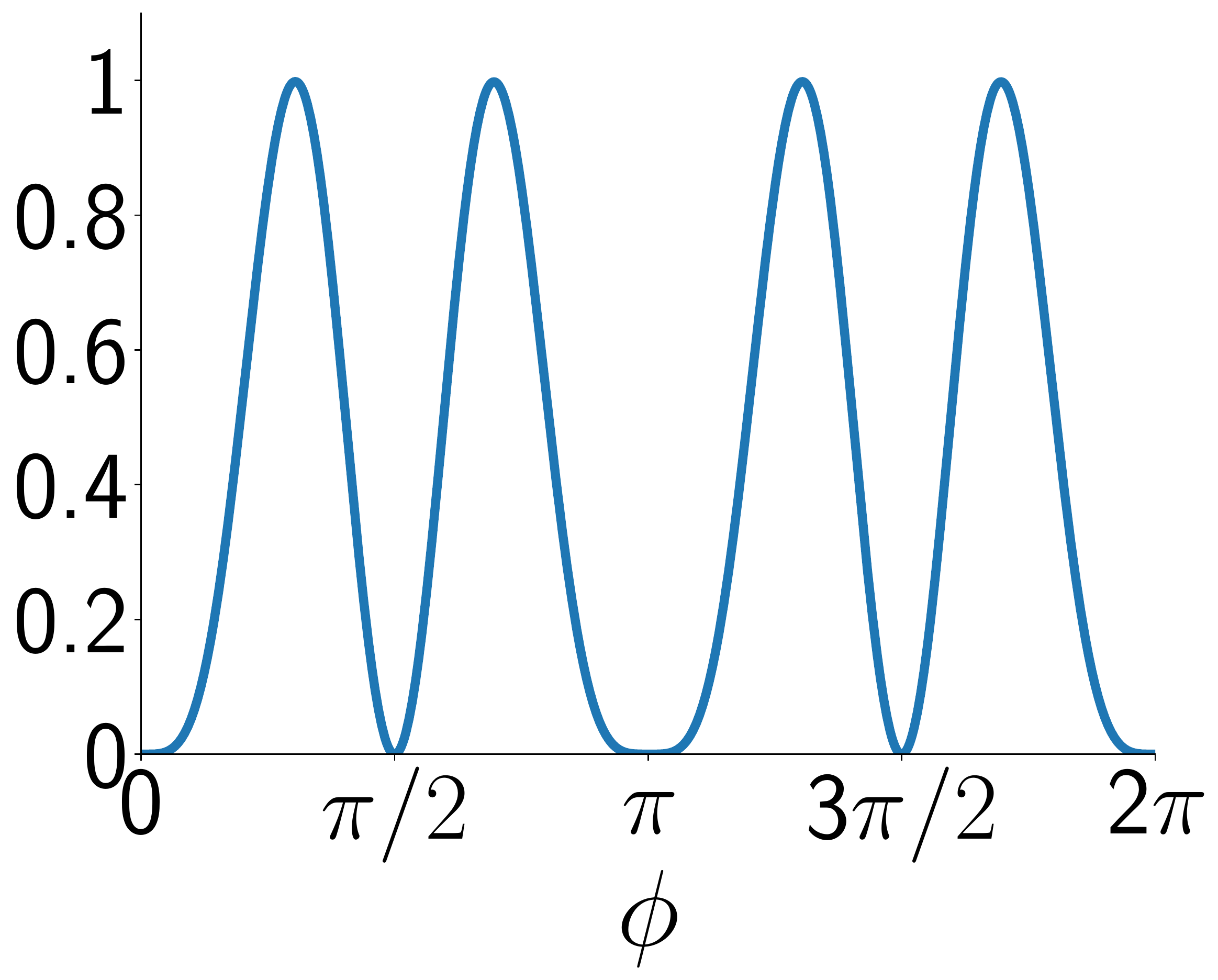}\\

		$D(4,2)$
		&\IncG[width=.23\textwidth,height=.12\textheight]{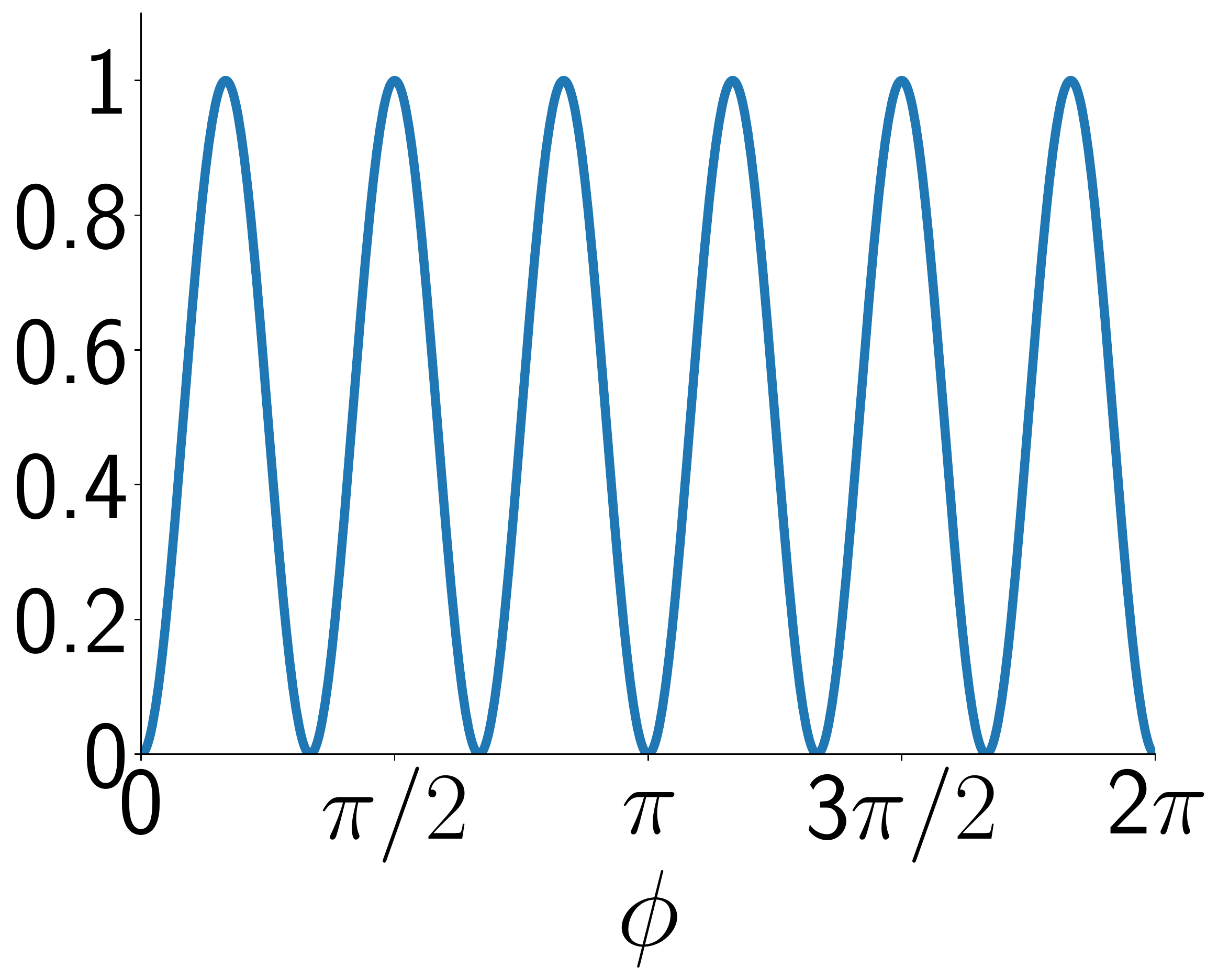}
		&\IncG[width=.23\textwidth,height=.12\textheight]{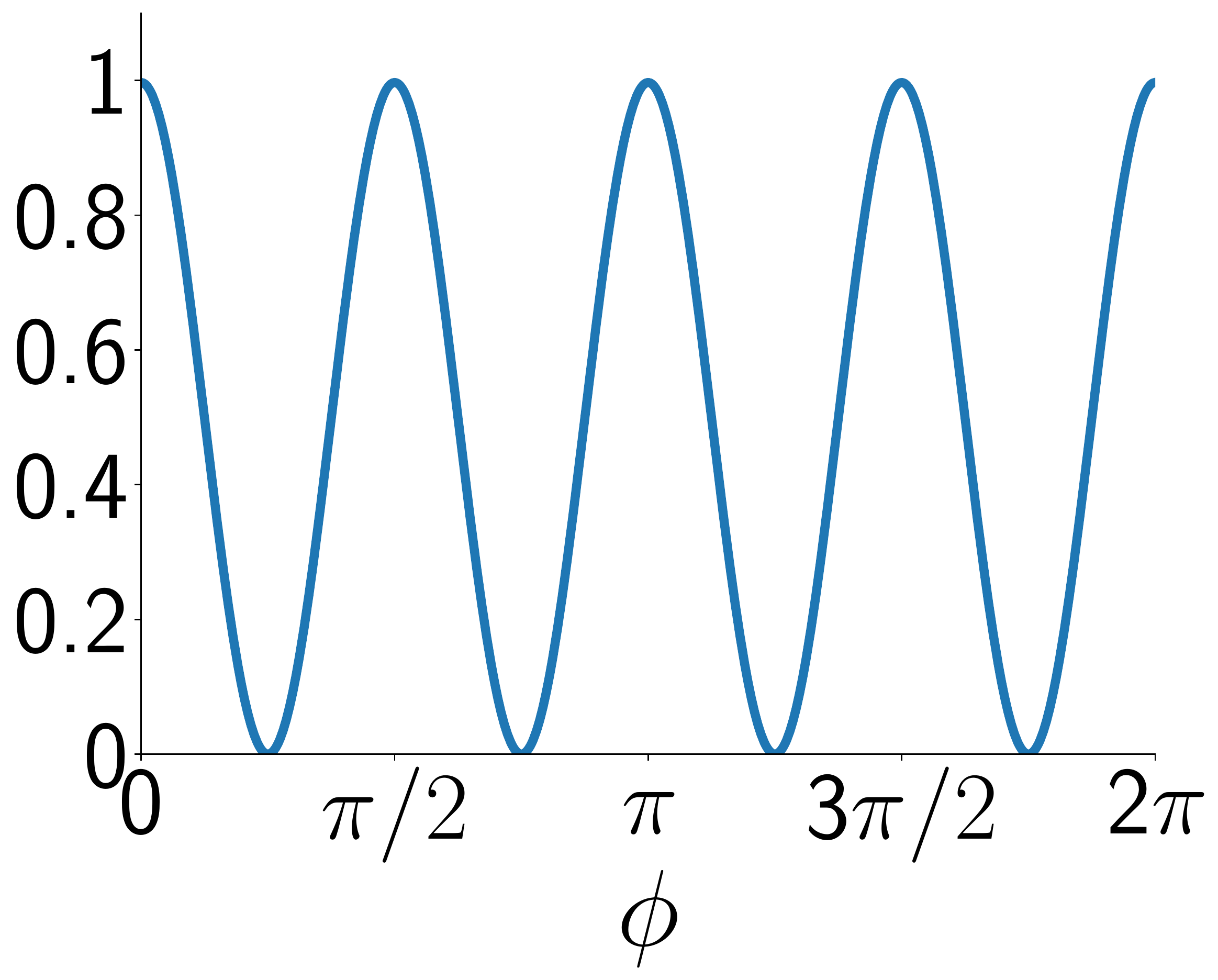}
		&\IncG[width=.23\textwidth,height=.12\textheight]{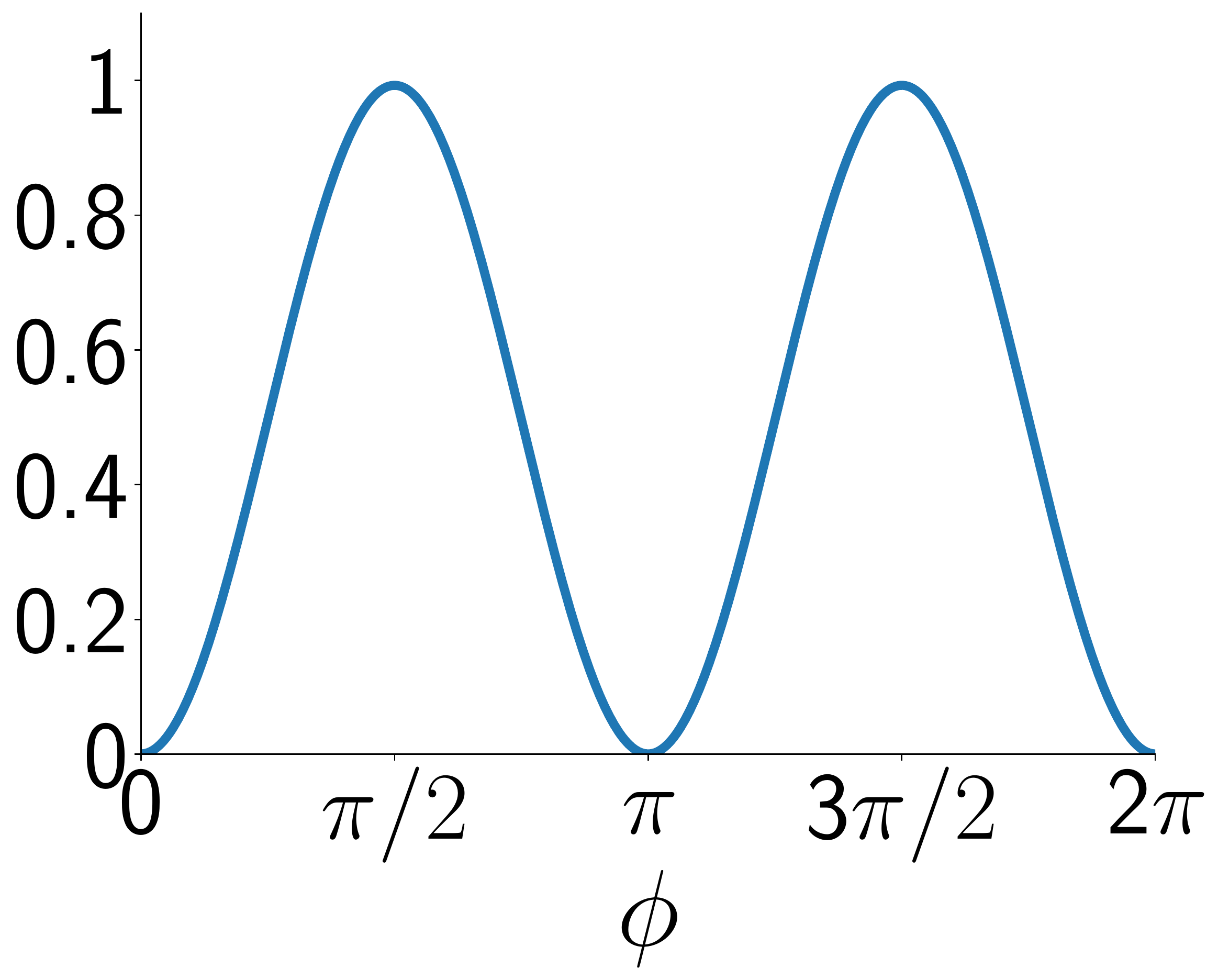}
		&\IncG[width=.23\textwidth,height=.12\textheight]{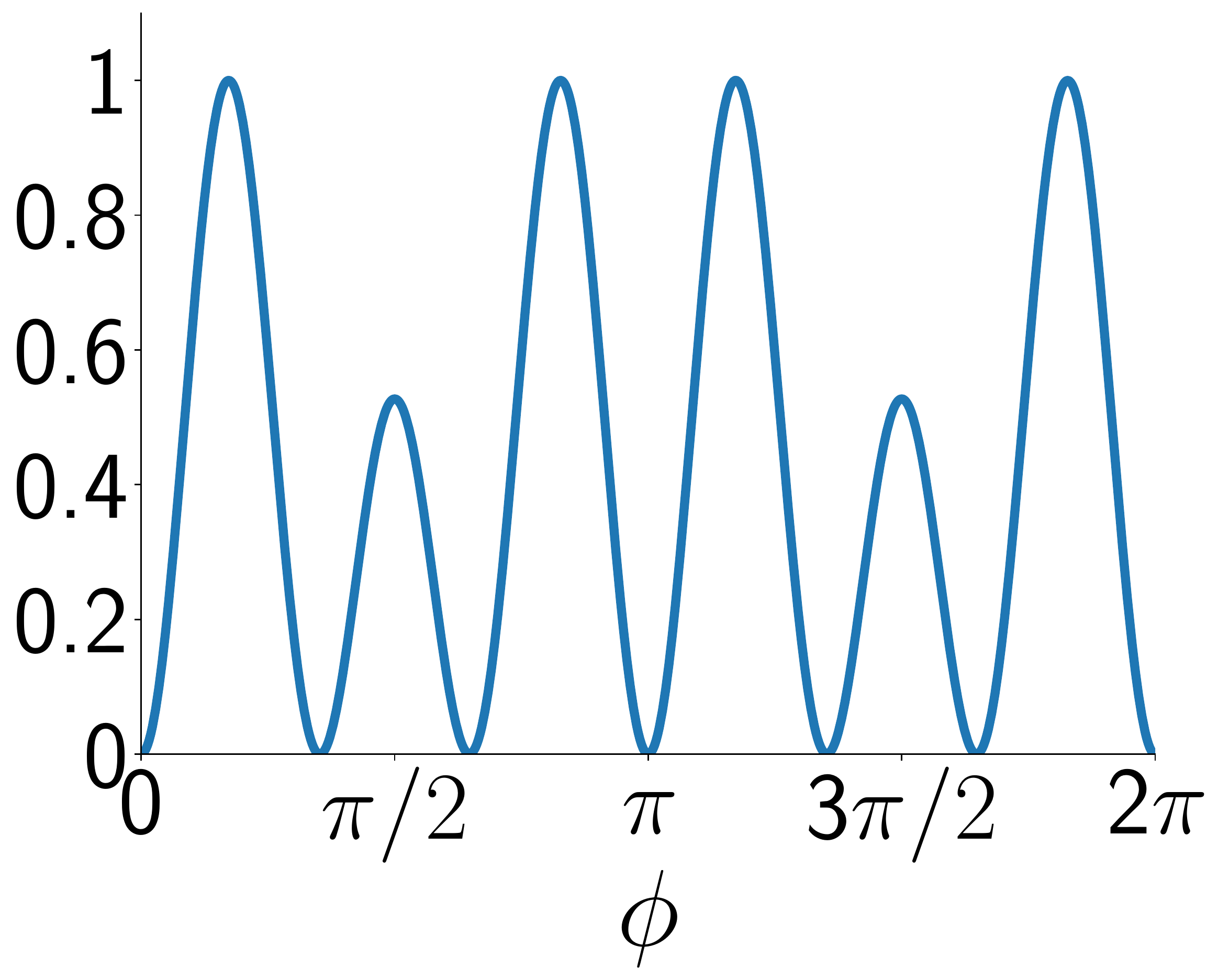}\\

		$D(3,3)$
		&\IncG[width=.23\textwidth,height=.12\textheight]{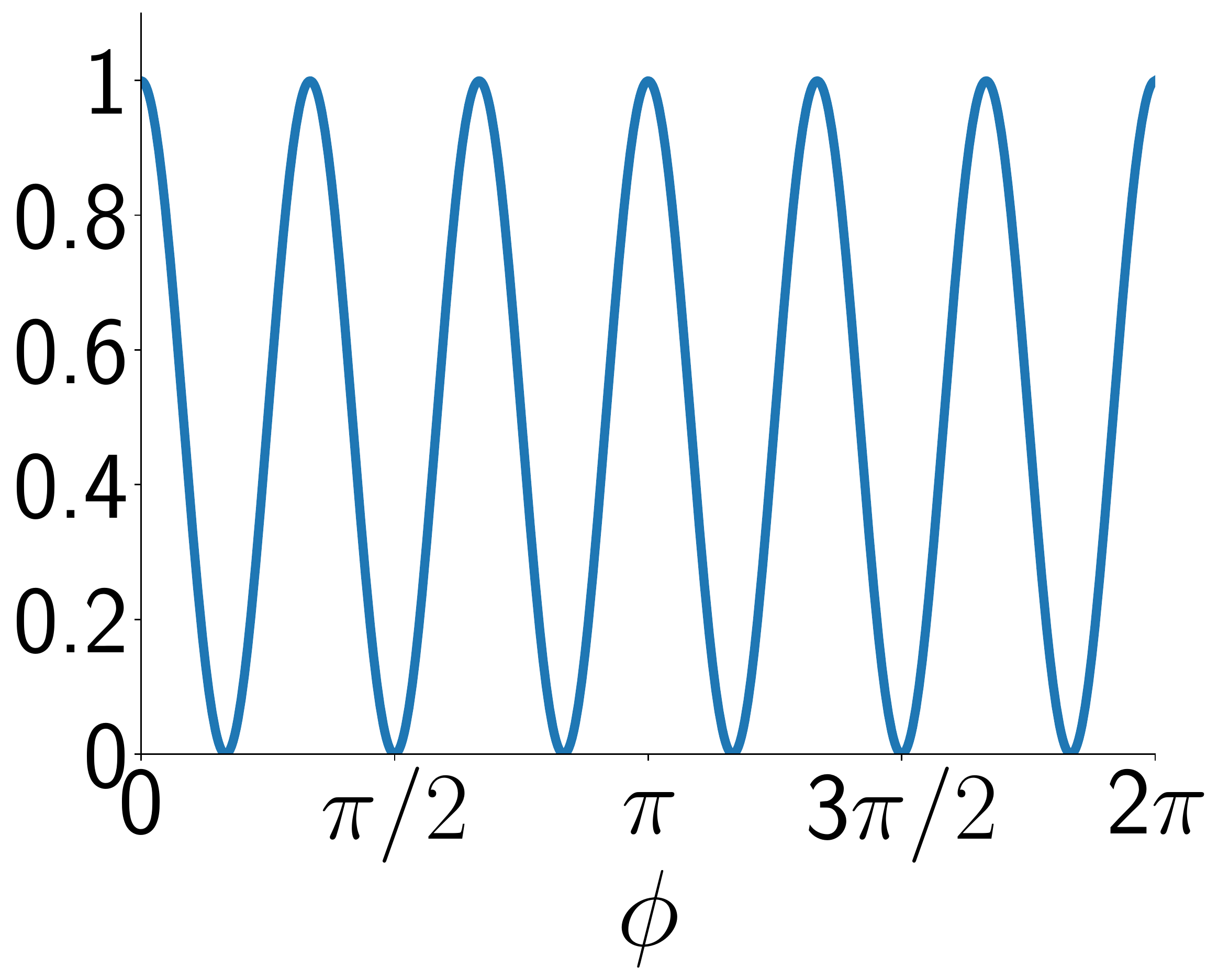}
		&\IncG[width=.23\textwidth,height=.12\textheight]{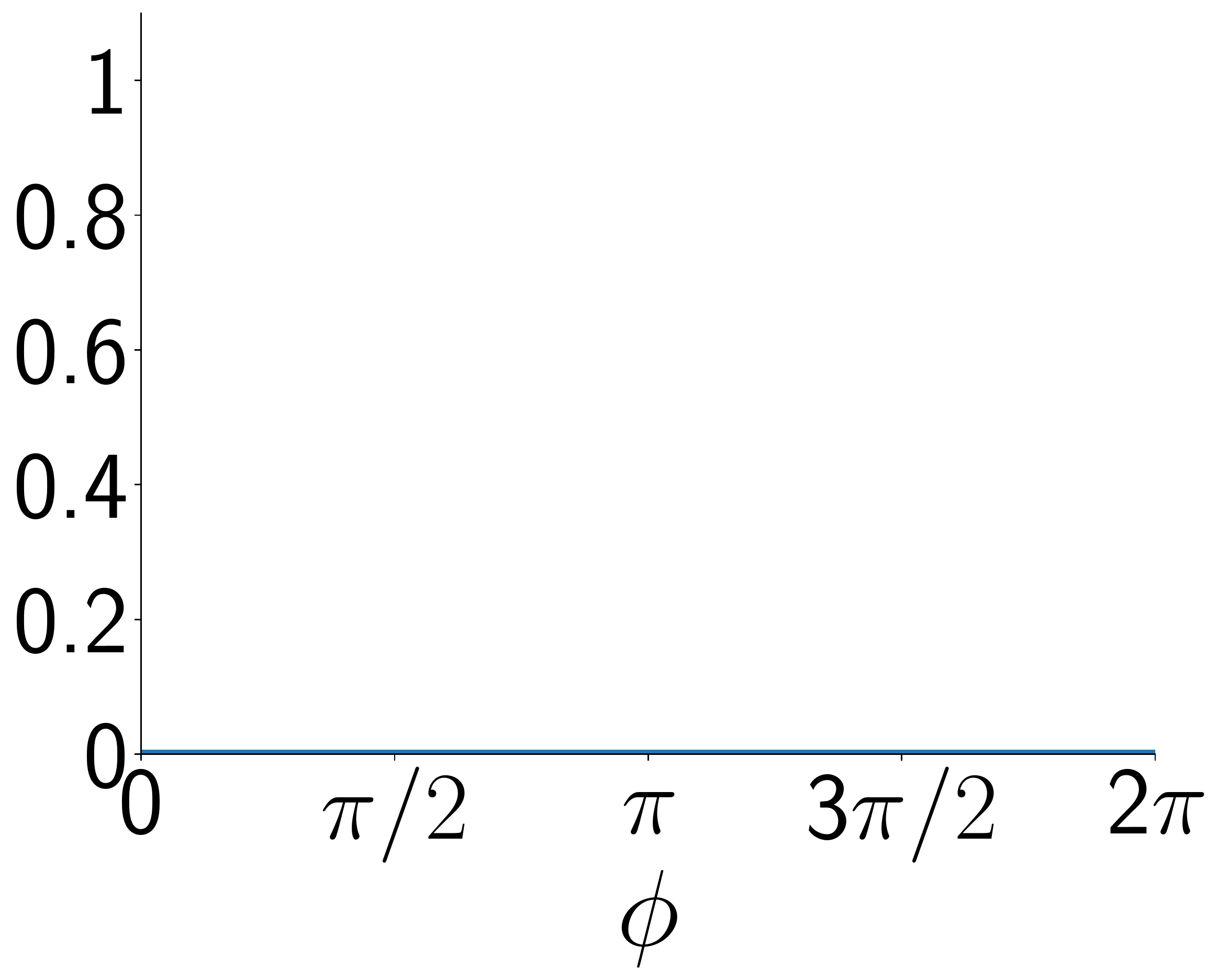}
		&\IncG[width=.23\textwidth,height=.12\textheight]{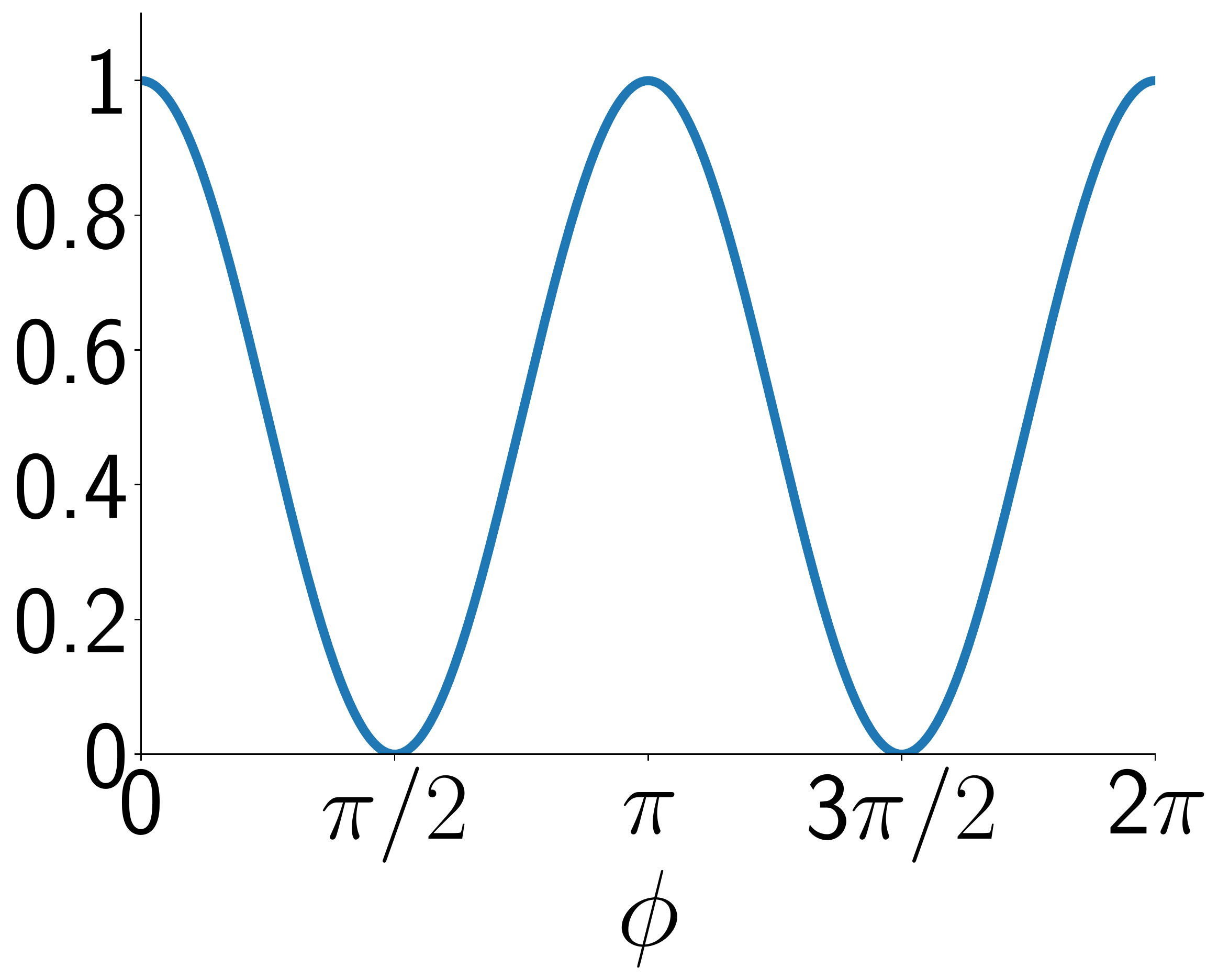}
		&\IncG[width=.23\textwidth,height=.12\textheight]{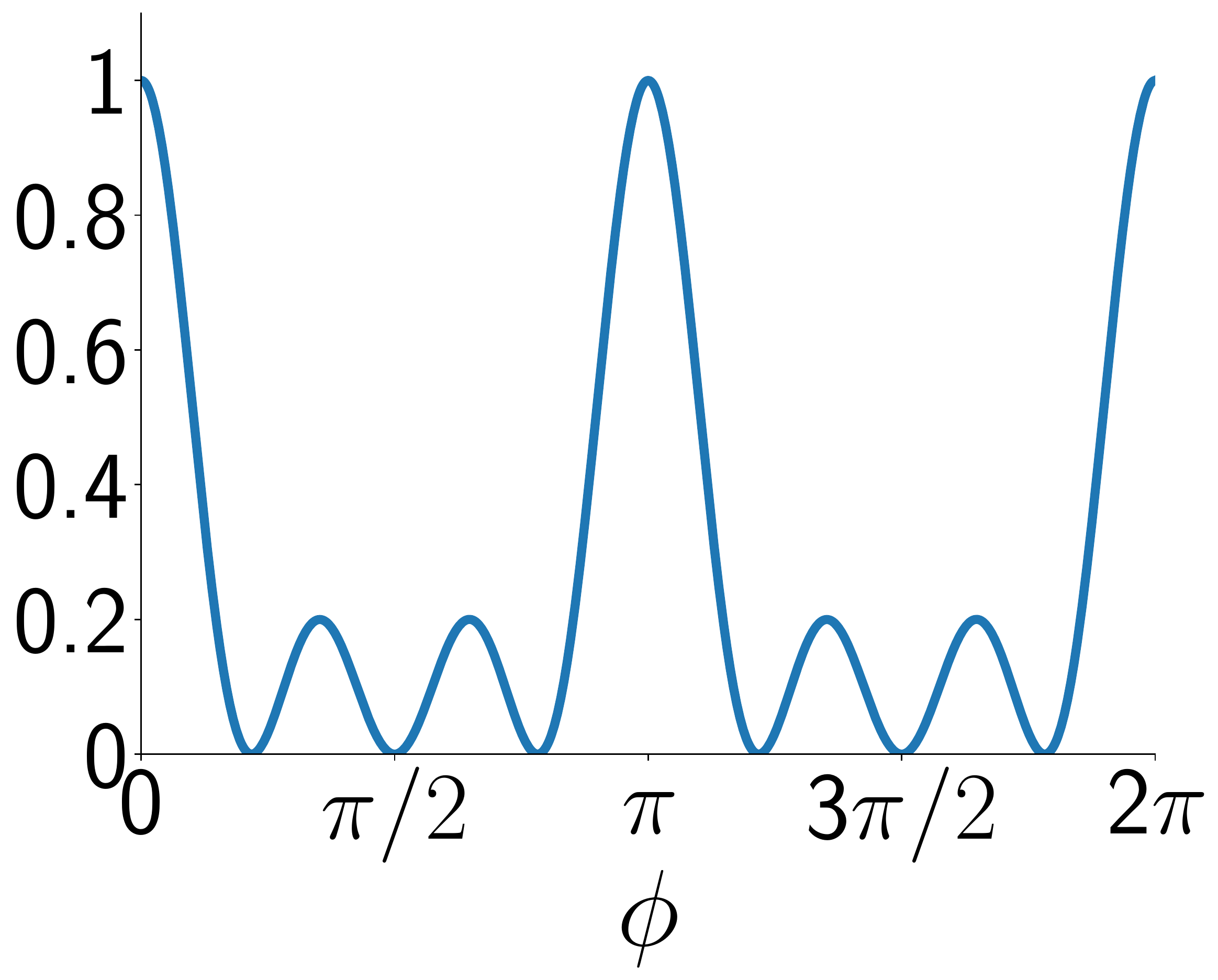}\\
		
		$D(5,0)$
		&\IncG[width=.23\textwidth,height=.12\textheight]{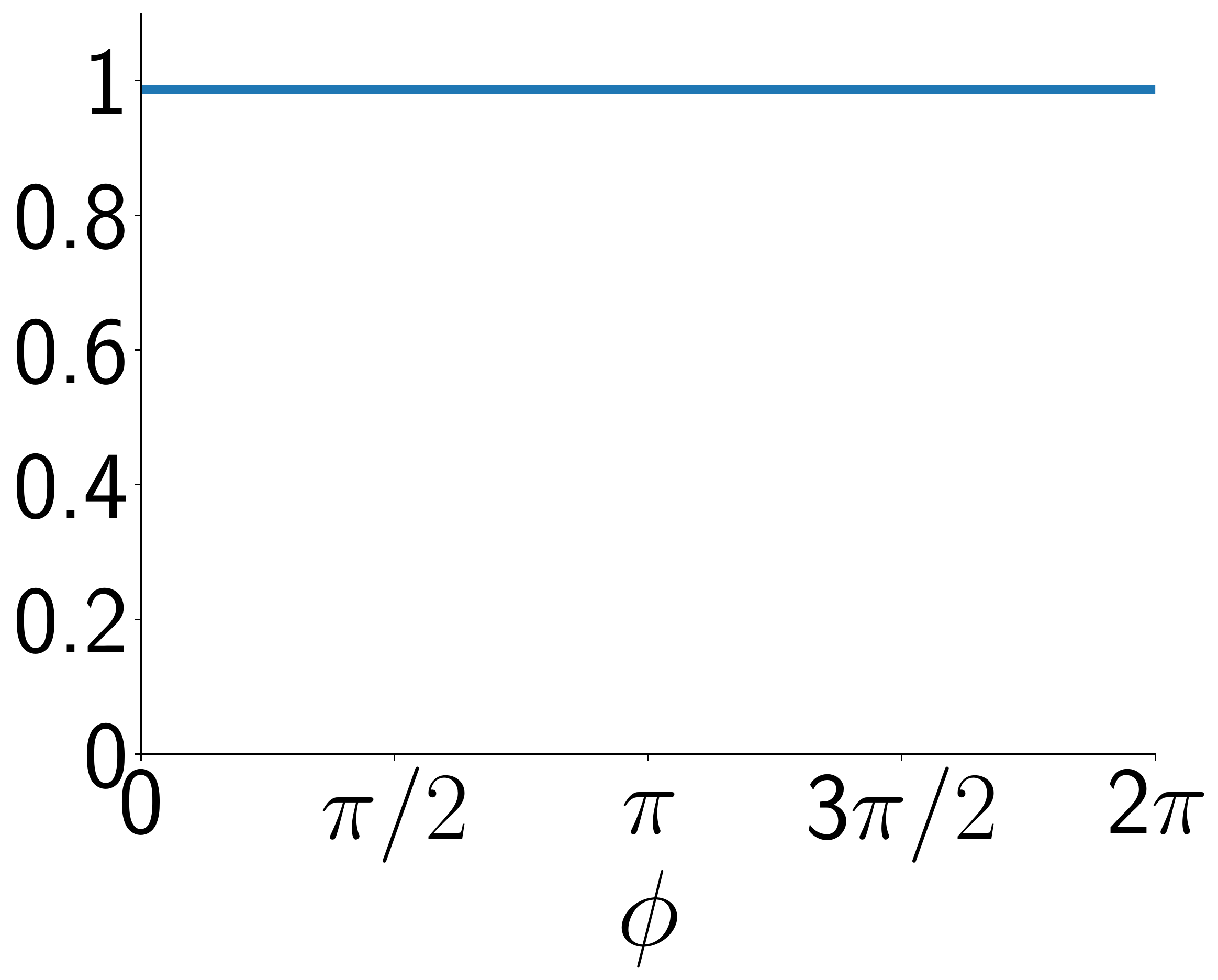}
		&\IncG[width=.23\textwidth,height=.12\textheight]{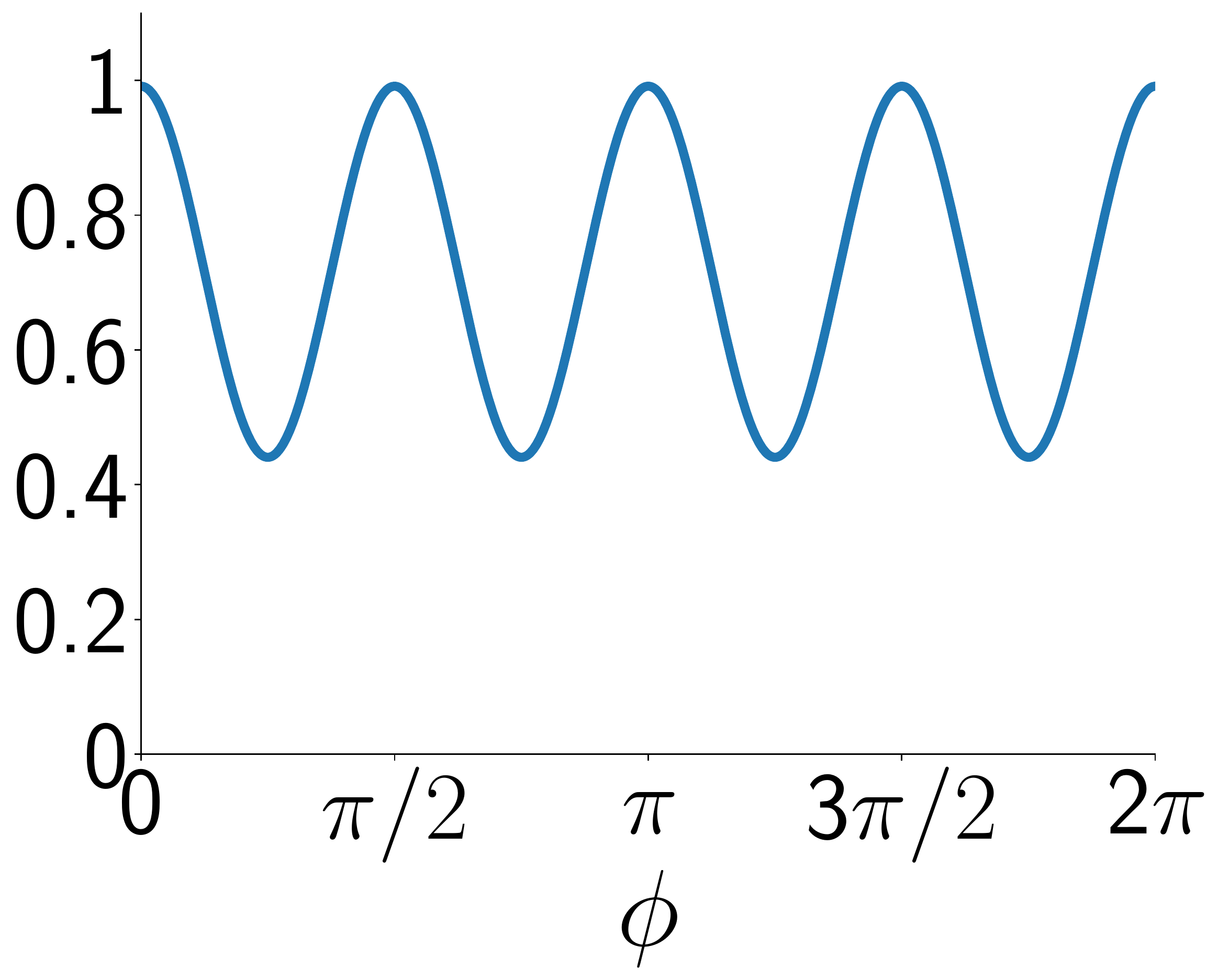}
		&\IncG[width=.23\textwidth,height=.12\textheight]{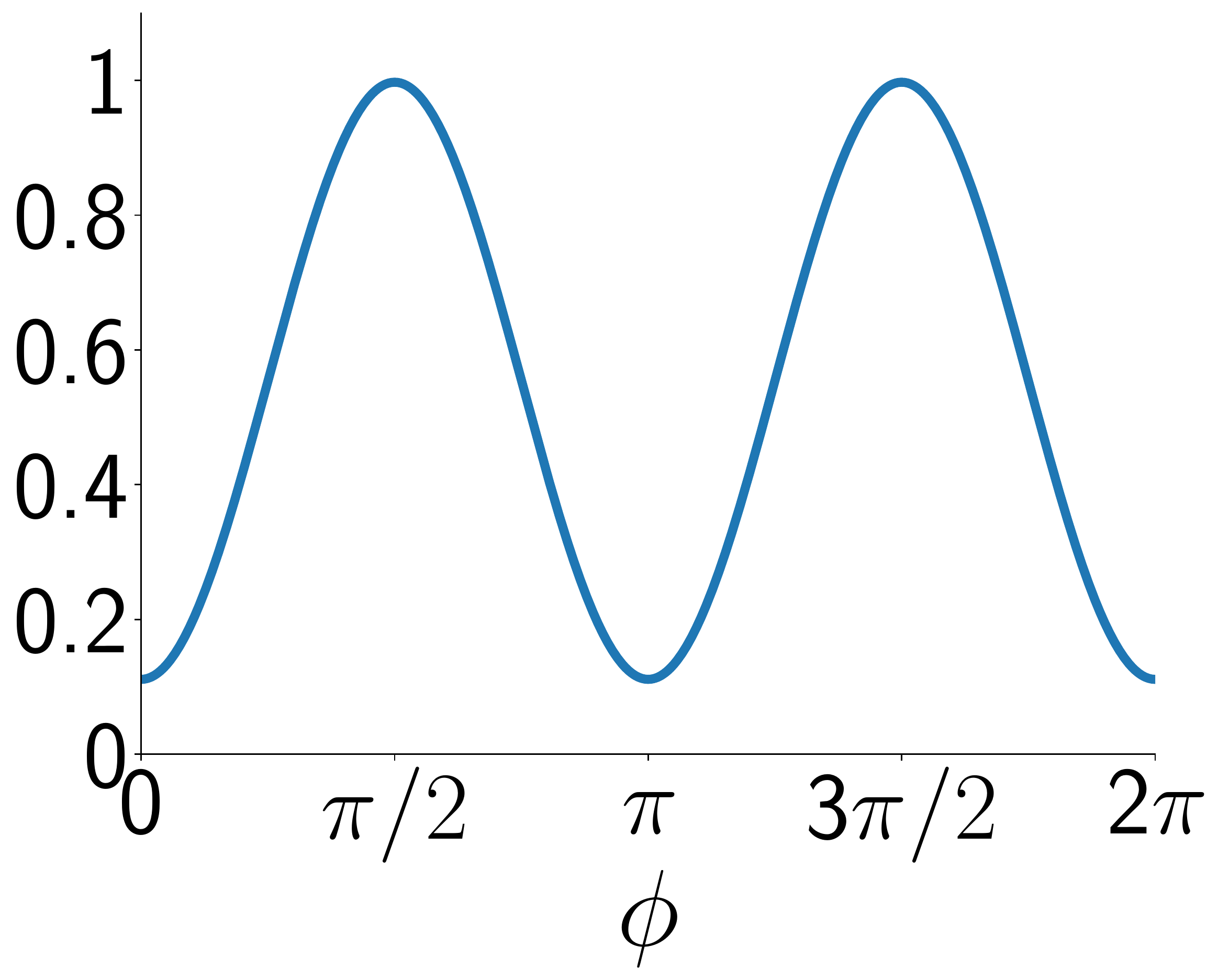}
		&\IncG[width=.23\textwidth,height=.12\textheight]{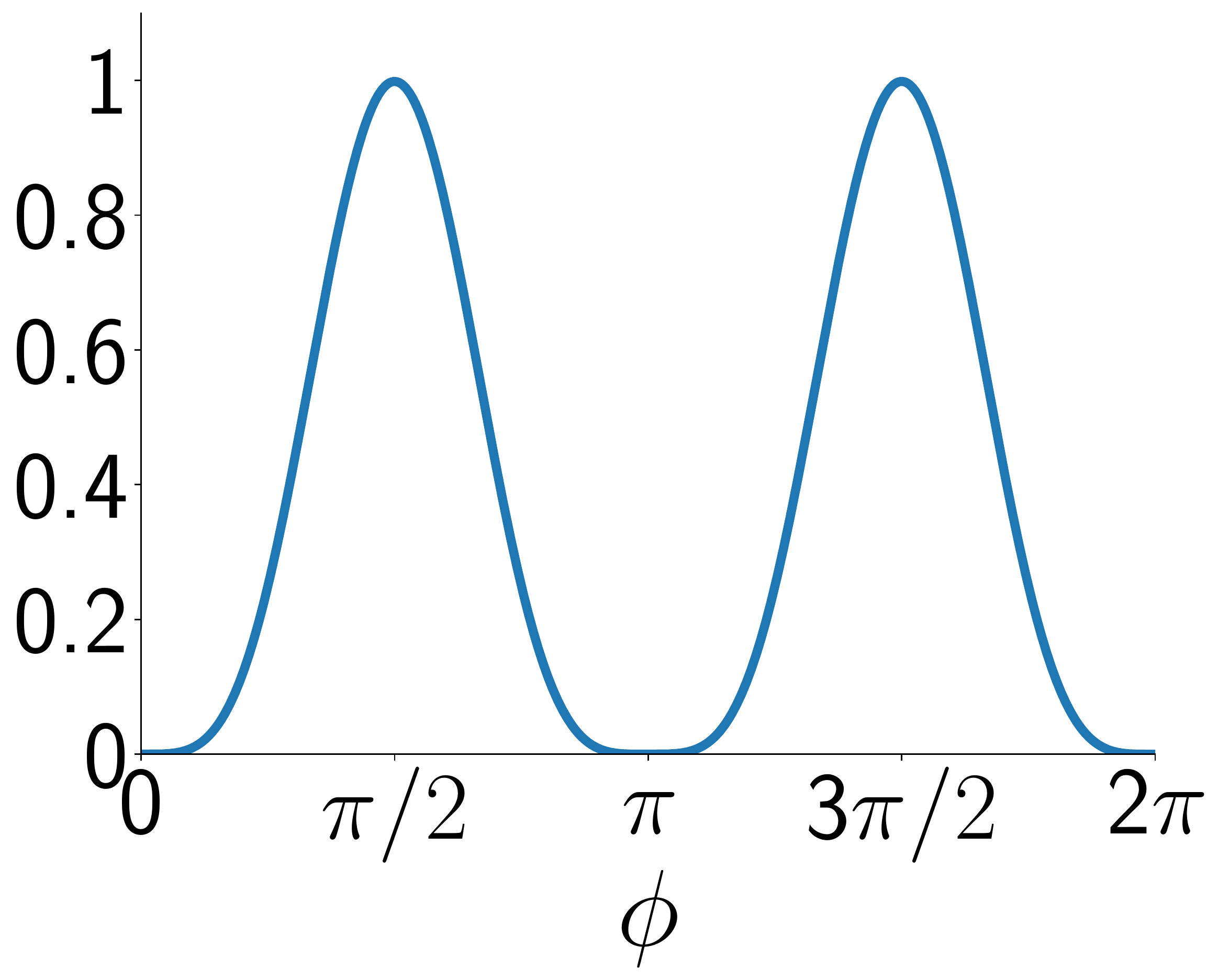}\\
		
		$D(4,1)$
		&\IncG[width=.23\textwidth,height=.12\textheight]{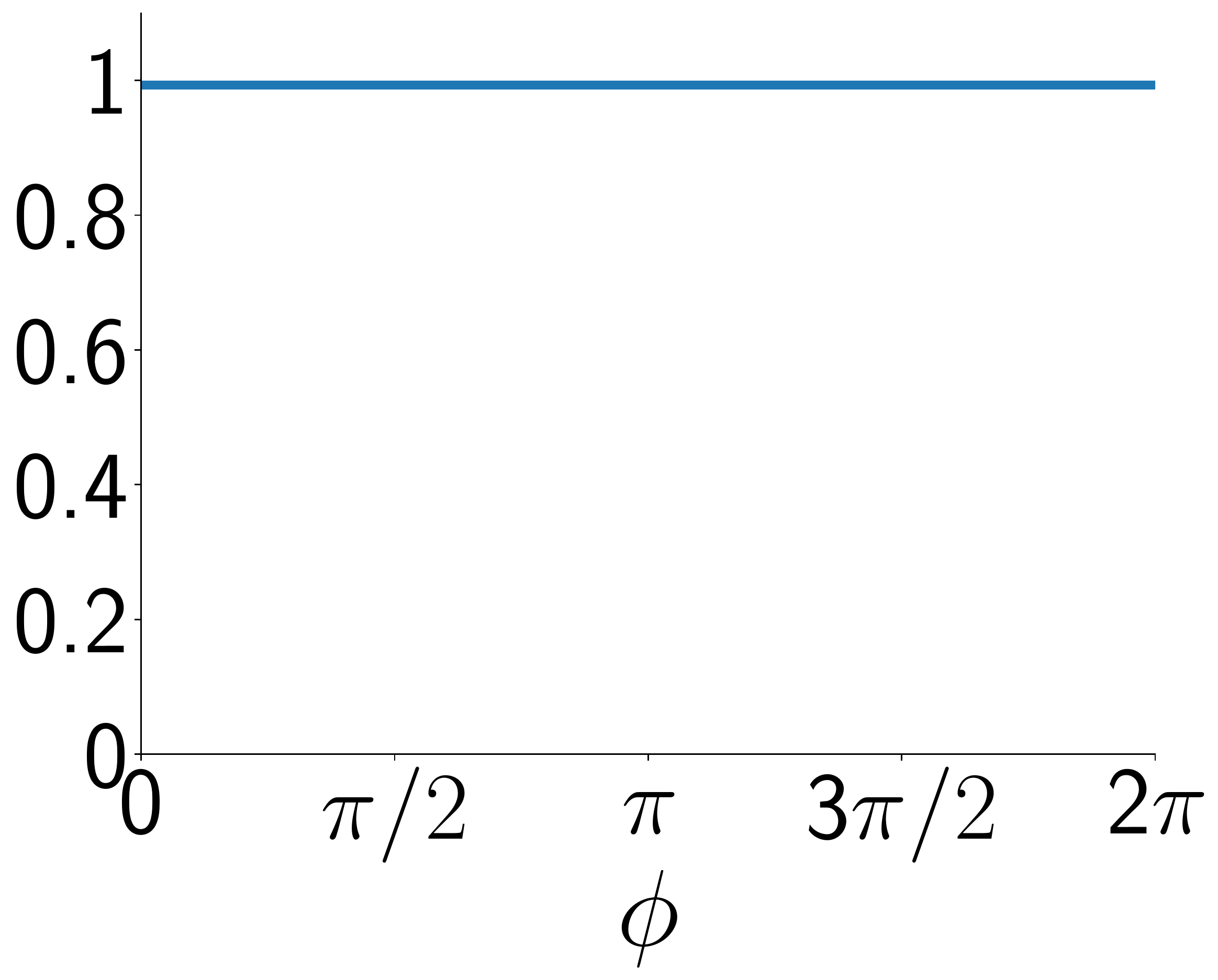}
		&\IncG[width=.23\textwidth,height=.12\textheight]{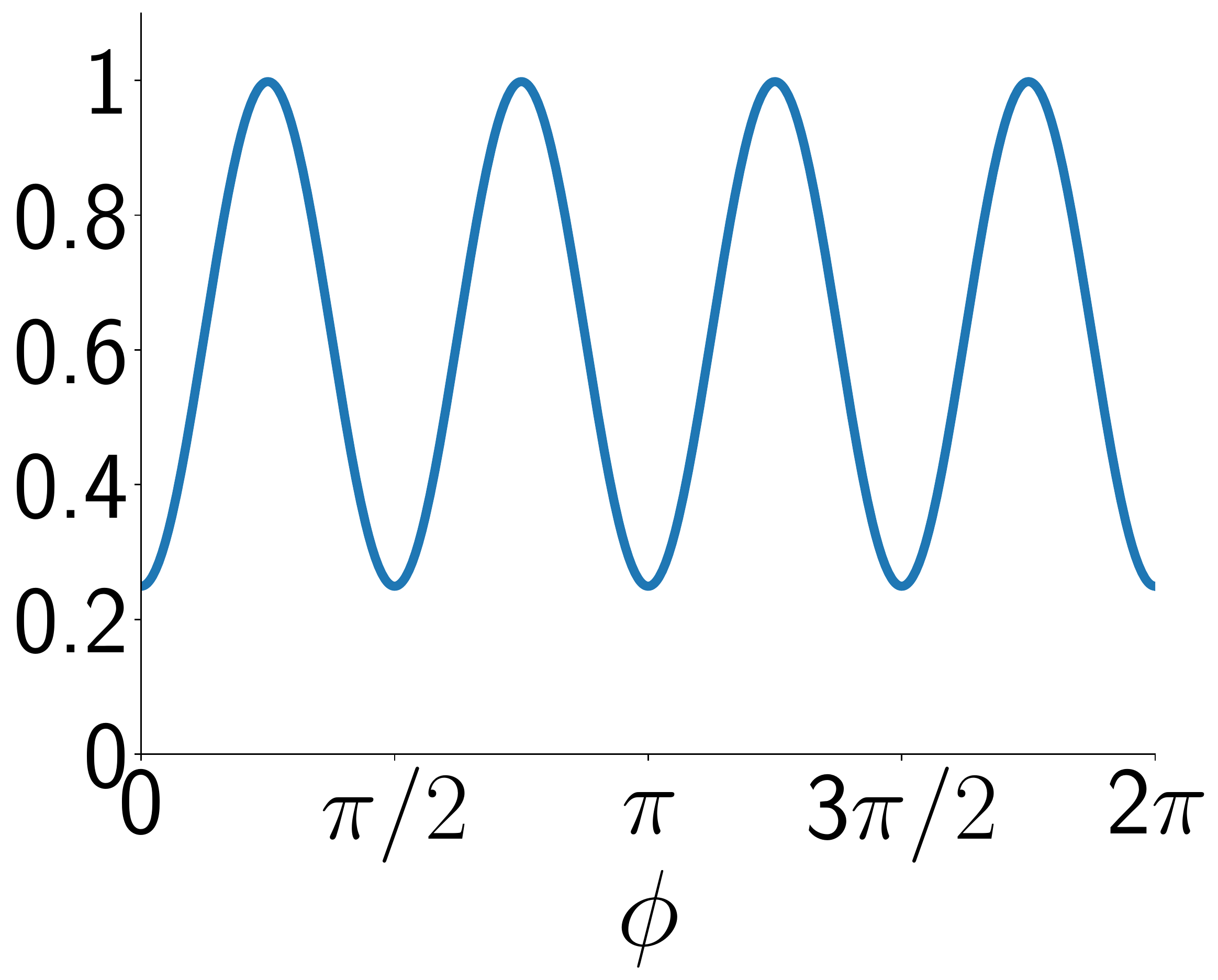}
		&\IncG[width=.23\textwidth,height=.12\textheight]{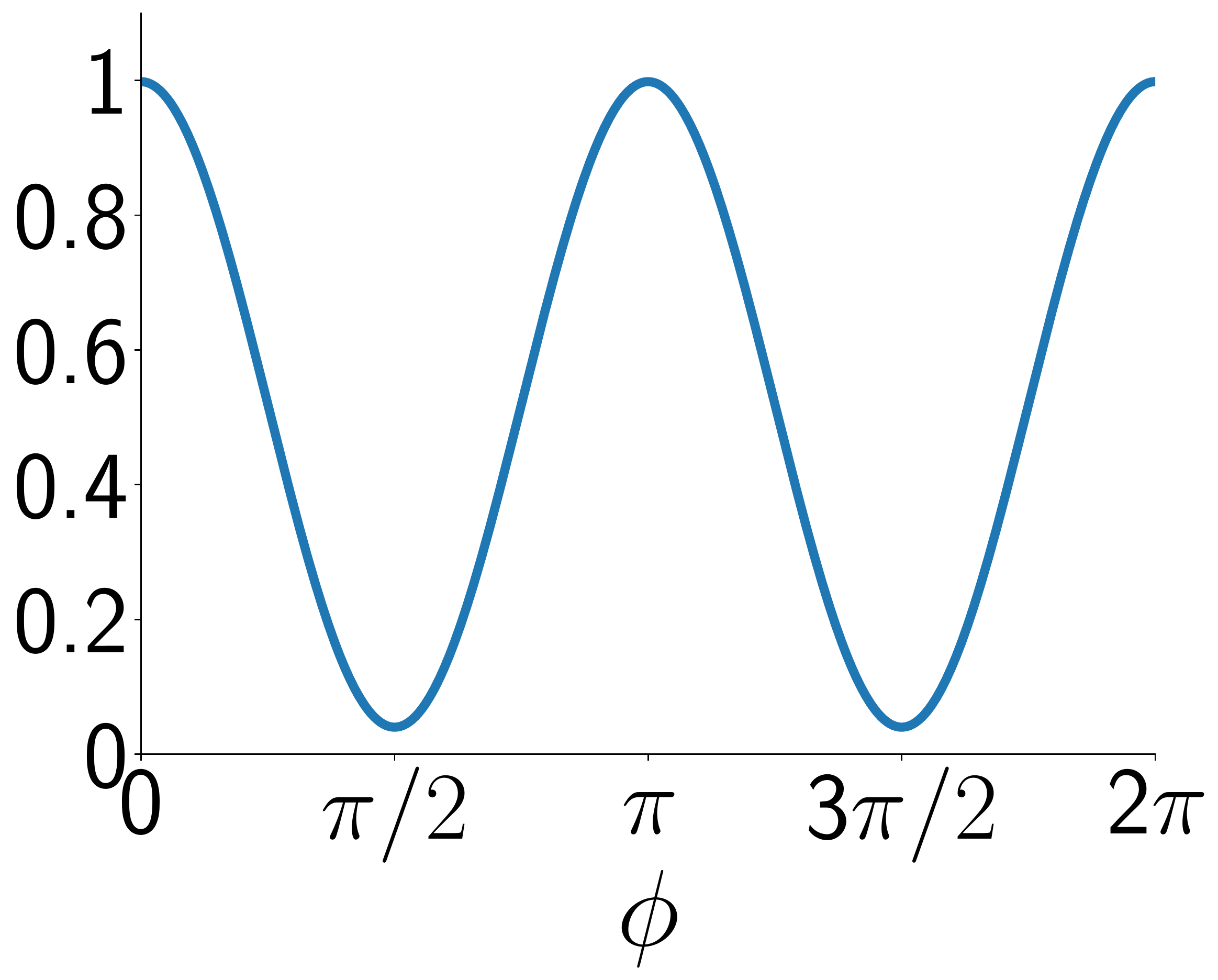}
		&\IncG[width=.23\textwidth,height=.12\textheight]{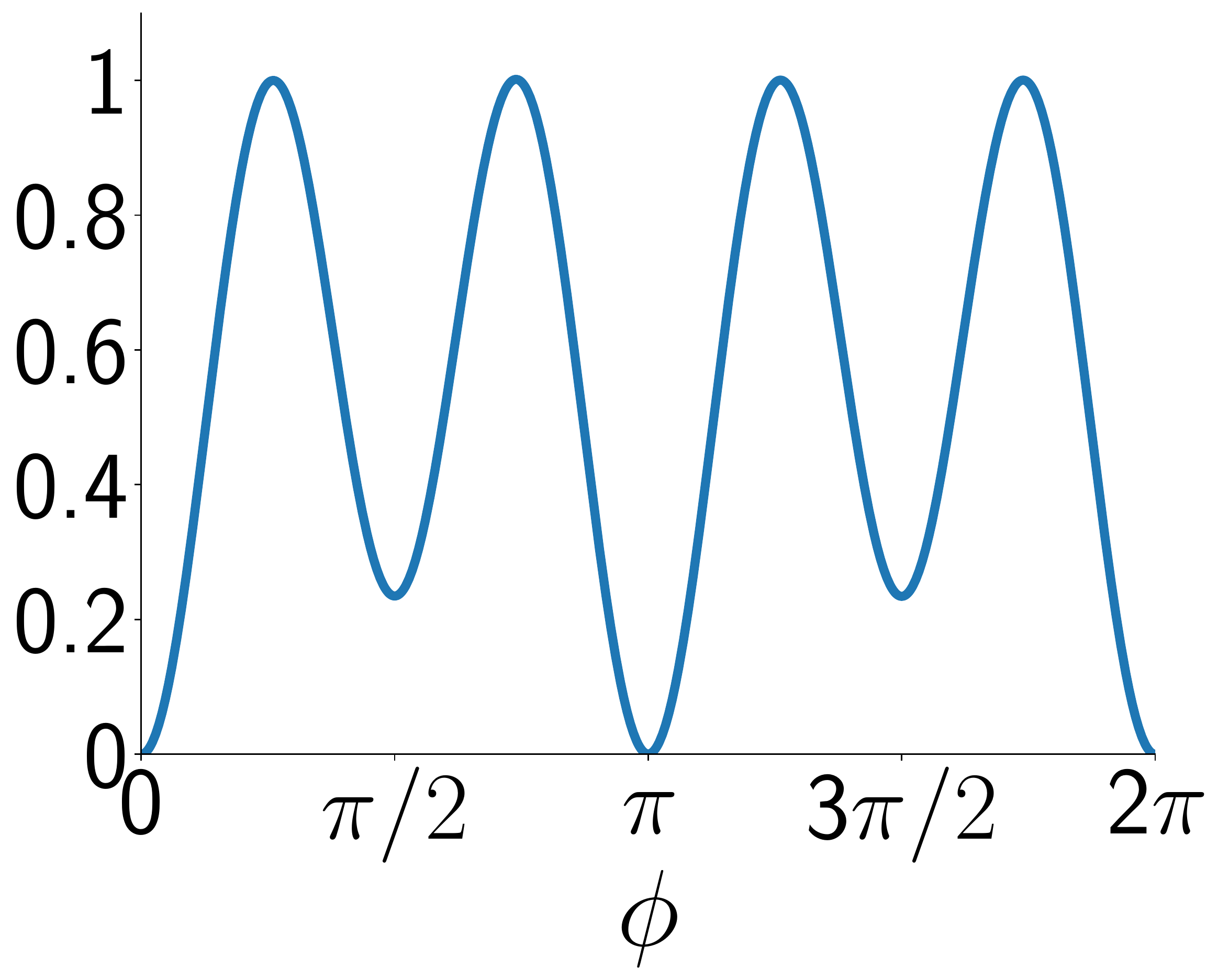}\\
		
		$D(3,2)$
		&\IncG[width=.23\textwidth,height=.12\textheight]{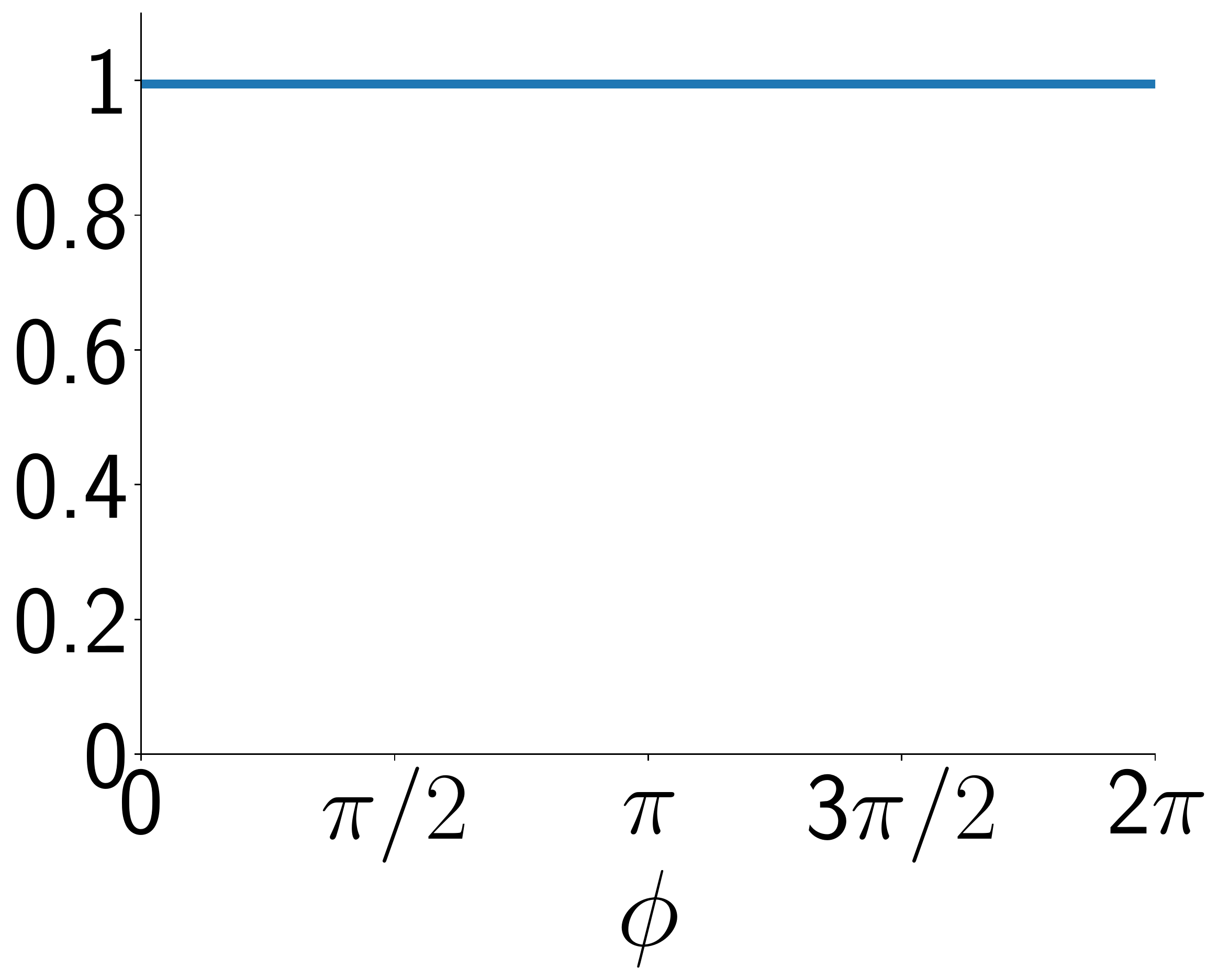}
		&\IncG[width=.23\textwidth,height=.12\textheight]{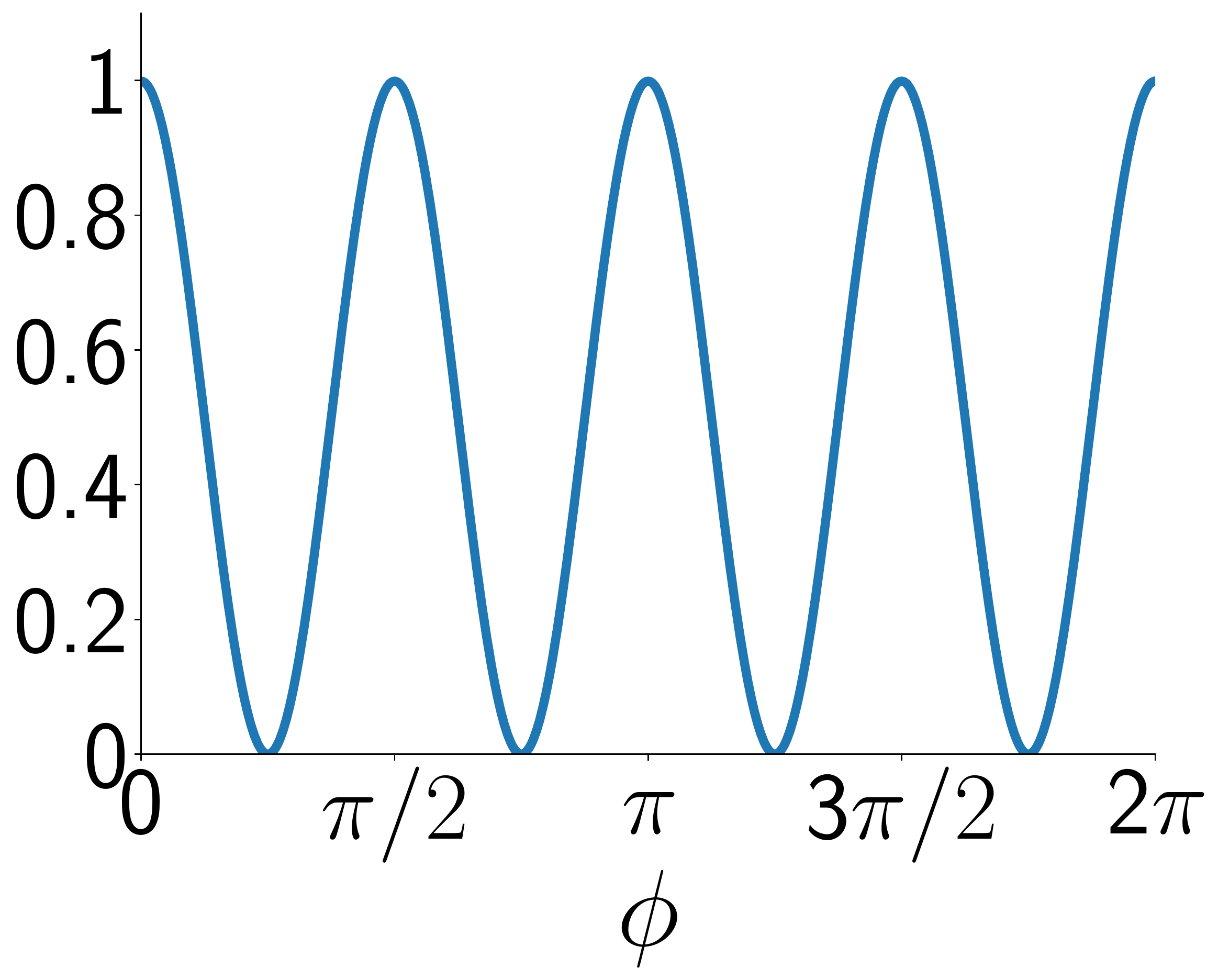}
		&\IncG[width=.23\textwidth,height=.12\textheight]{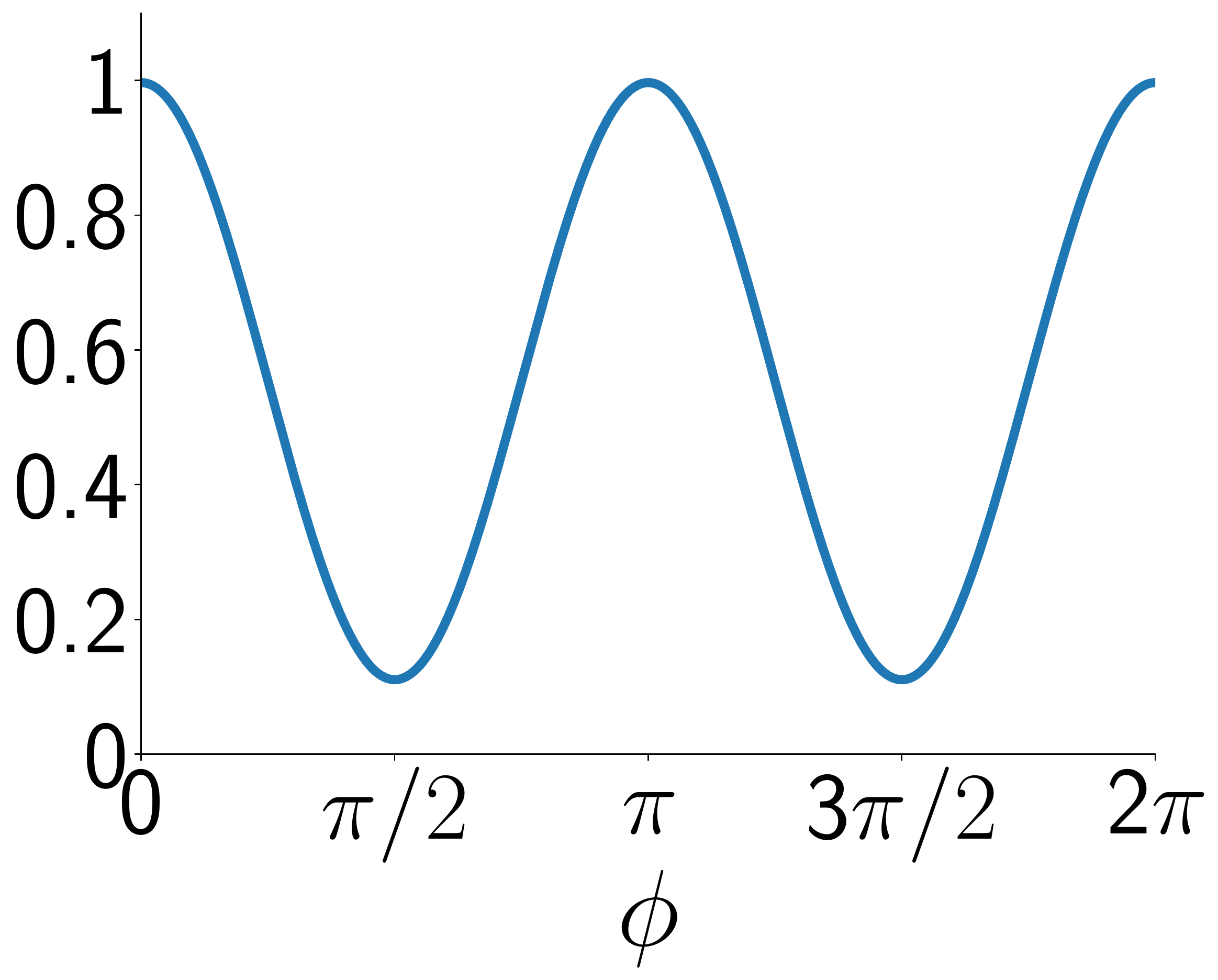}
		&\IncG[width=.23\textwidth,height=.12\textheight]{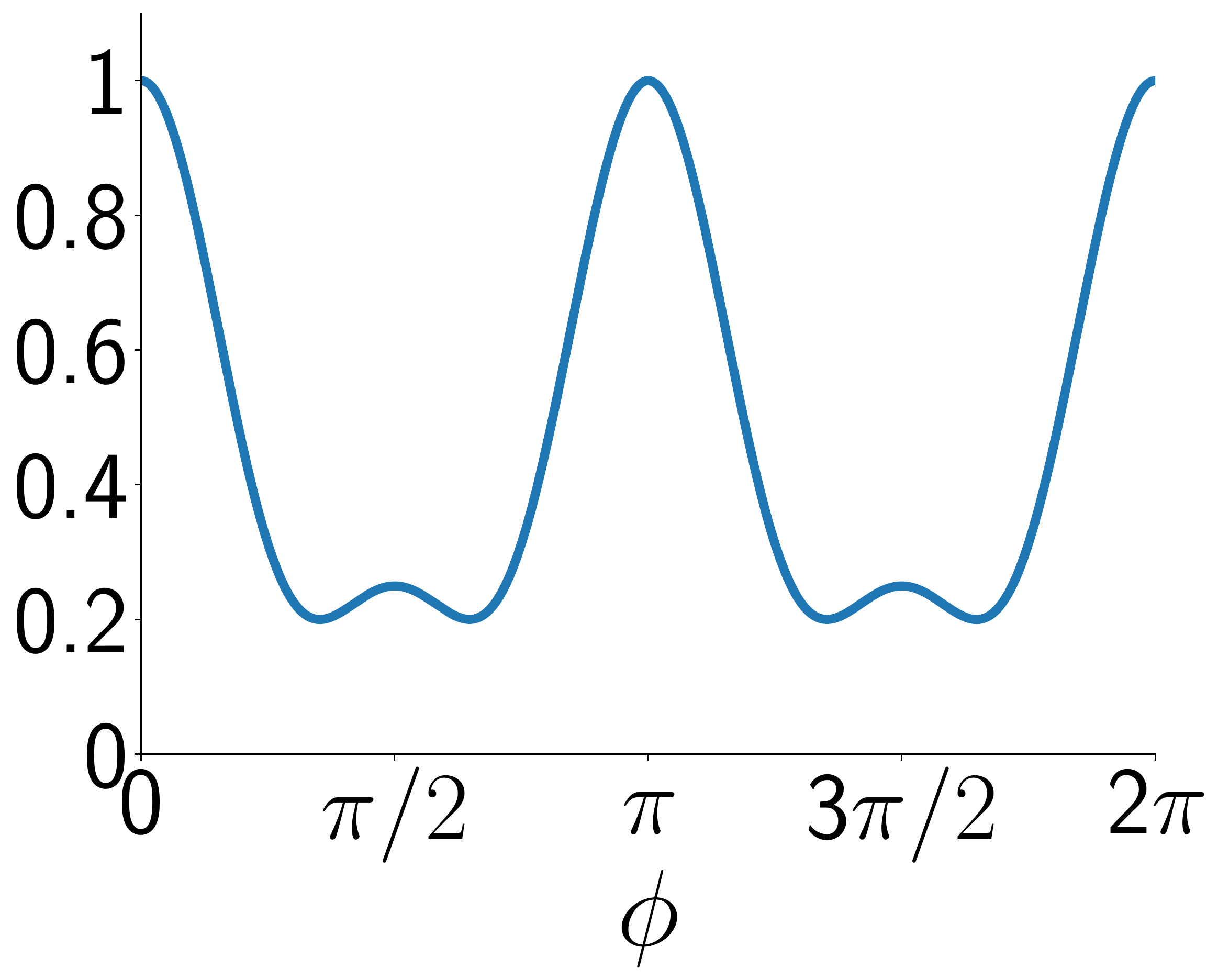}\\
		
	\end{tabular} 
	
\end{table*}

\begin{table*}
	\ContinuedFloat
	\centering
	\begin{tabular}{c c c c c}
		&$\ket{6::0}$ & $\ket{5::1}$ & $\ket{4::2}$ & $\ket{HB(6)}$\\
		
		$D(4,0)$
		&\IncG[width=.23\textwidth,height=.12\textheight]{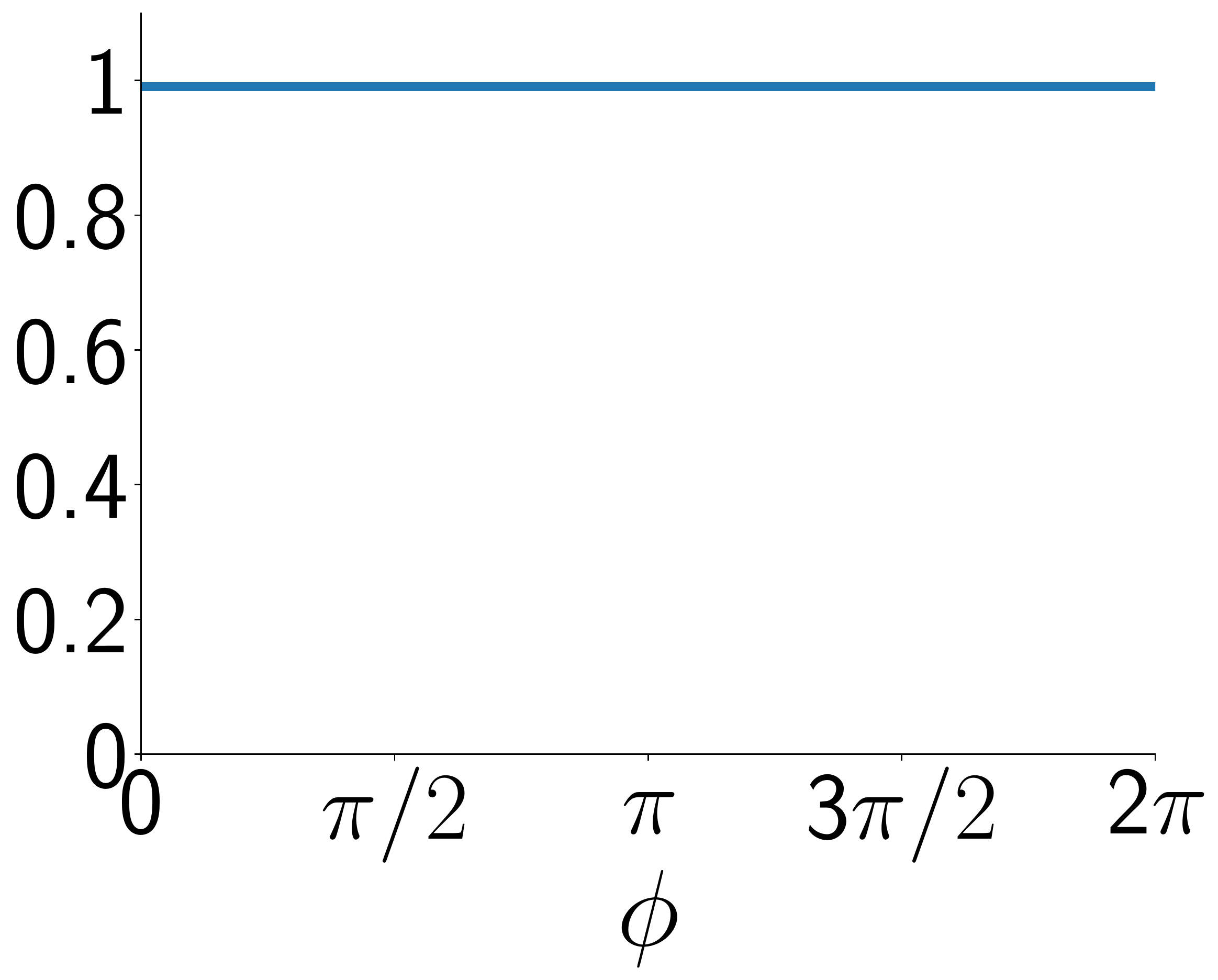}
		&\IncG[width=.23\textwidth,height=.12\textheight]{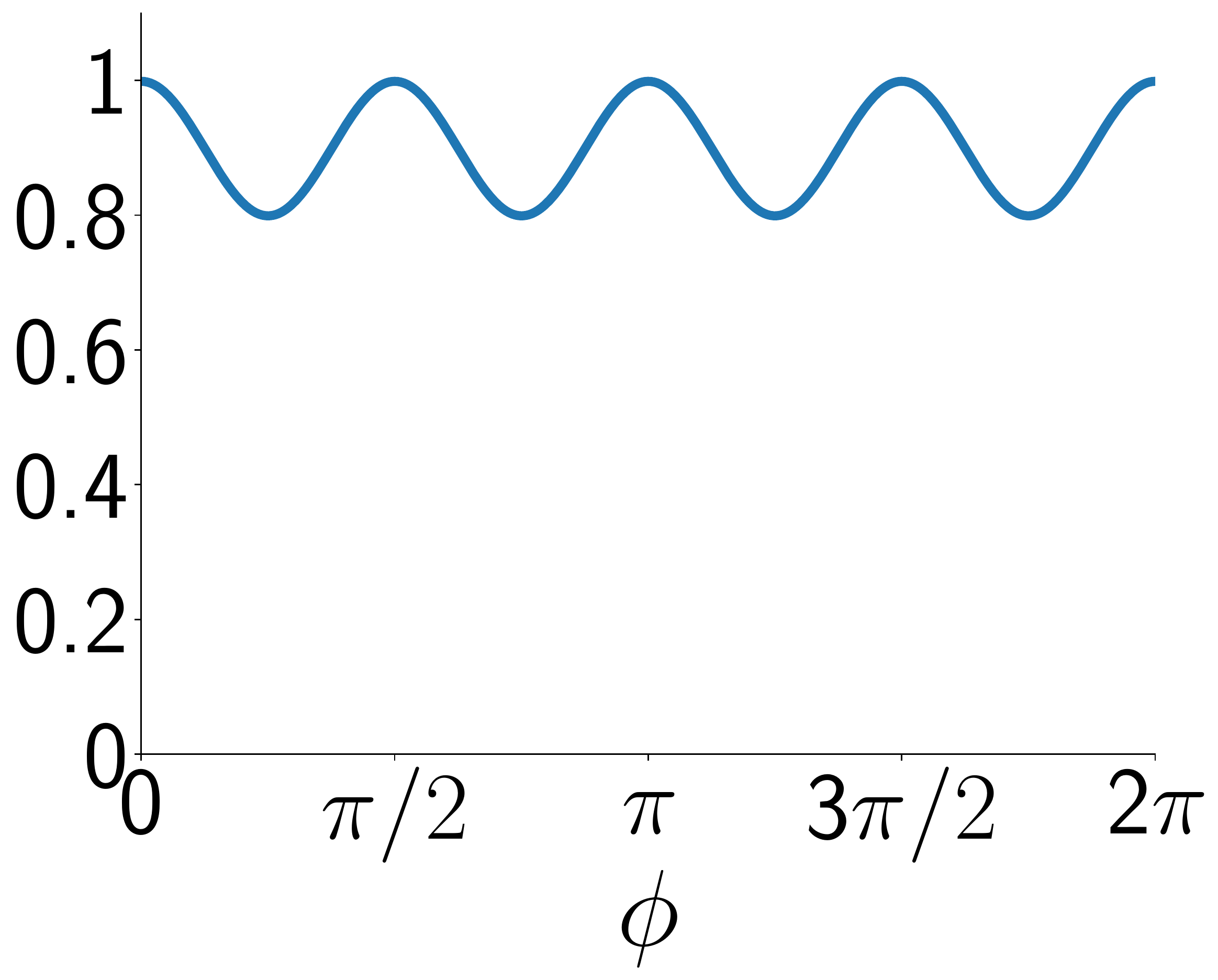}
		&\IncG[width=.23\textwidth,height=.12\textheight]{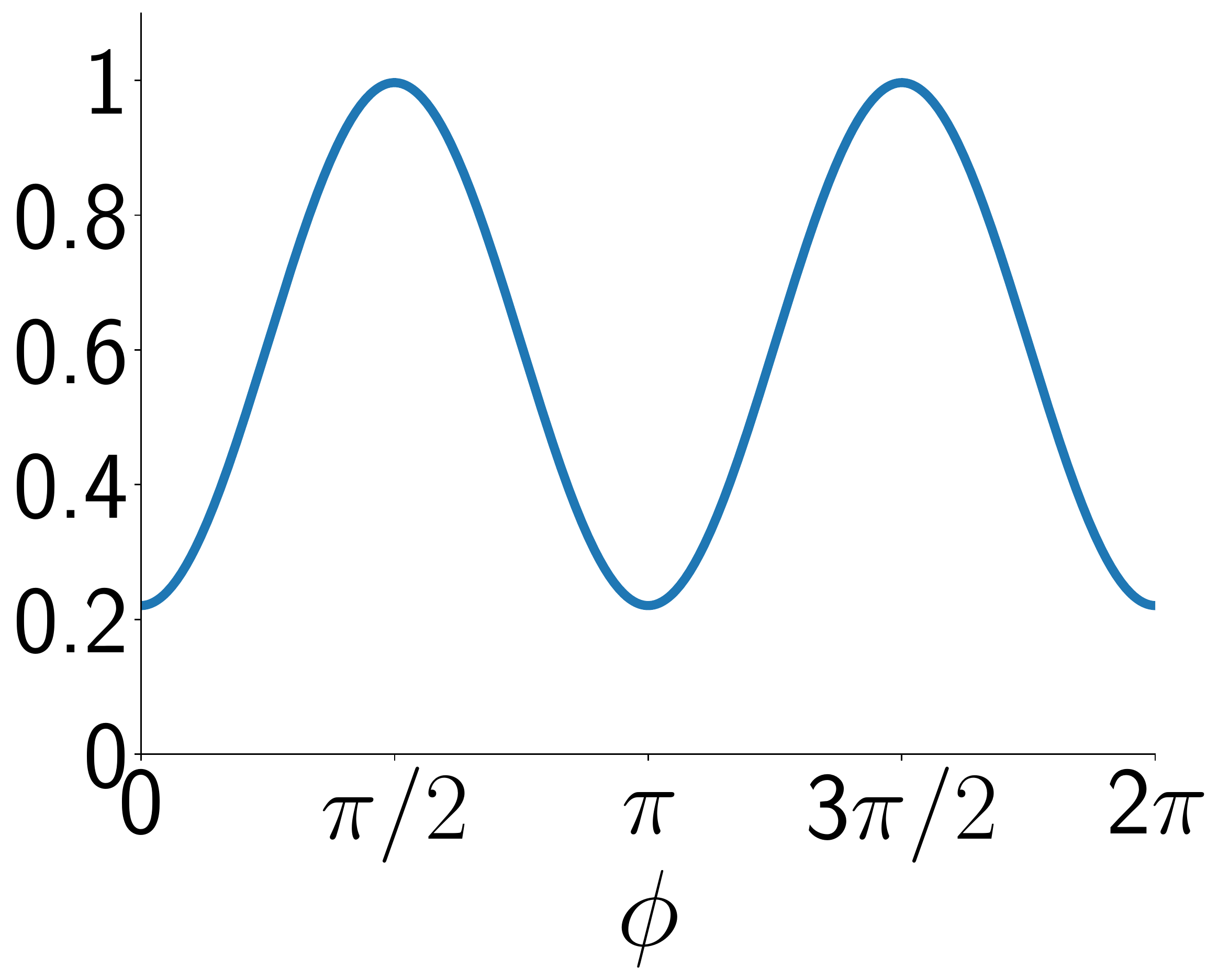}
		&\IncG[width=.23\textwidth,height=.12\textheight]{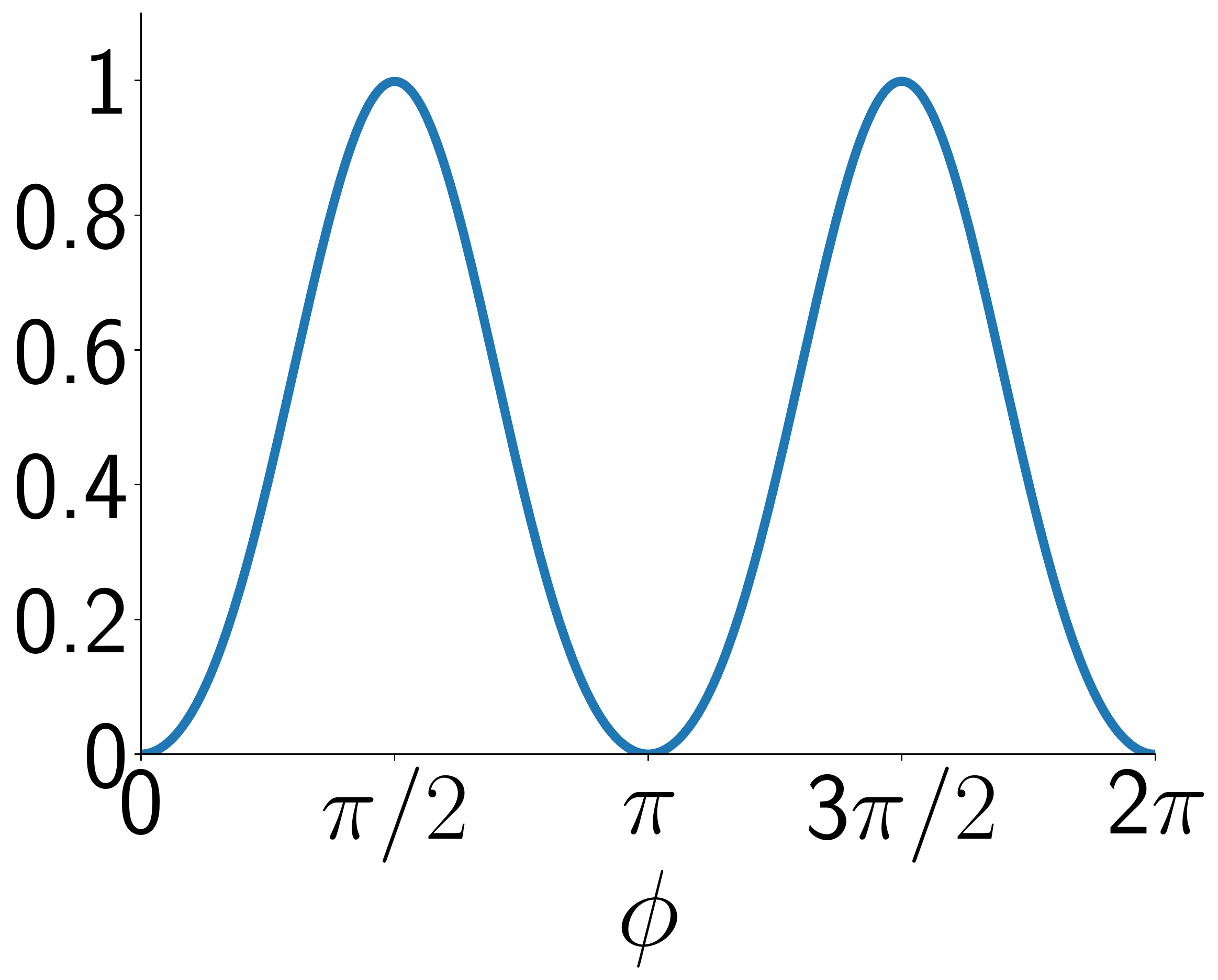}\\
		
		$D(3,1)$
		&\IncG[width=.23\textwidth,height=.12\textheight]{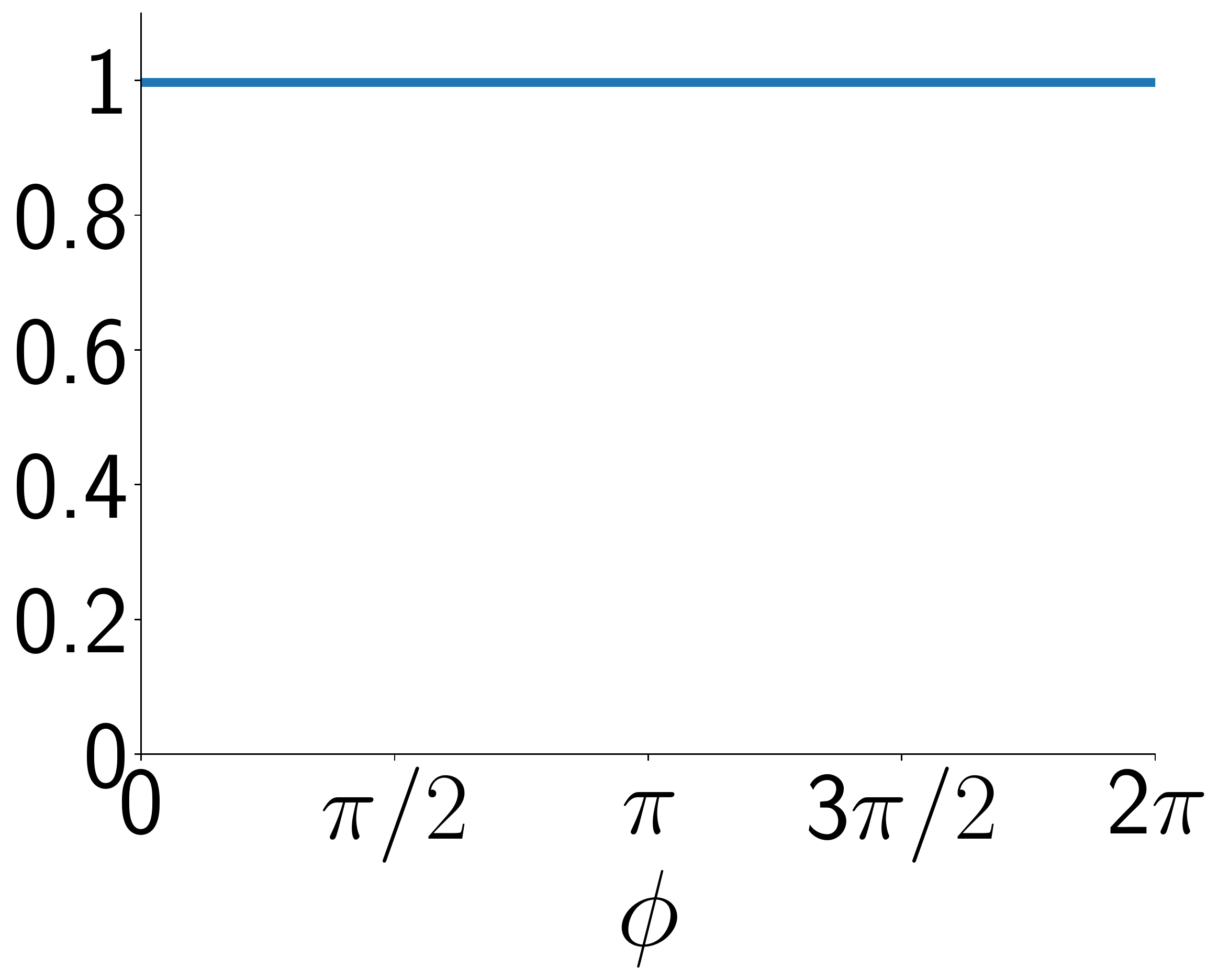}
		&\IncG[width=.23\textwidth,height=.12\textheight]{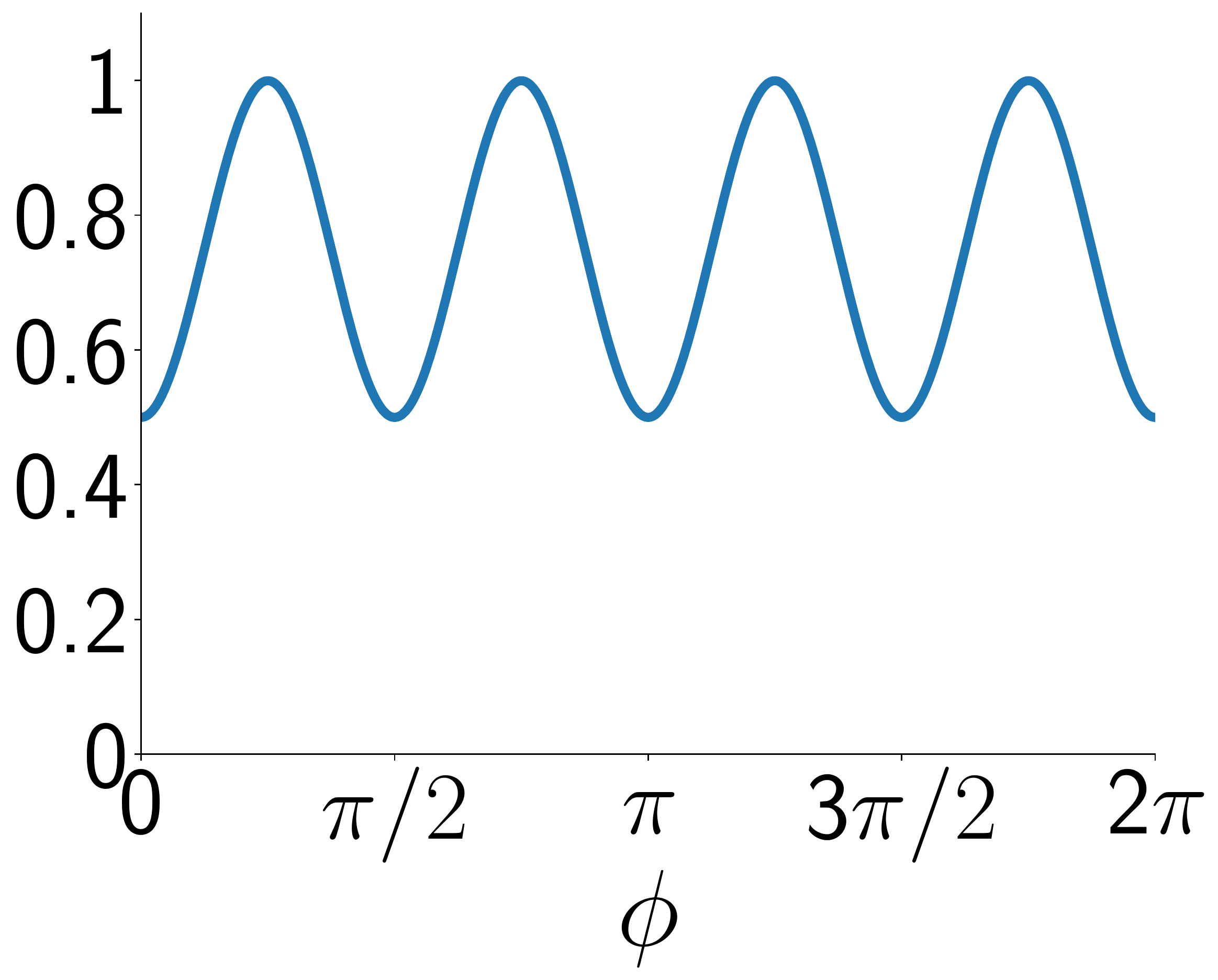}
		&\IncG[width=.23\textwidth,height=.12\textheight]{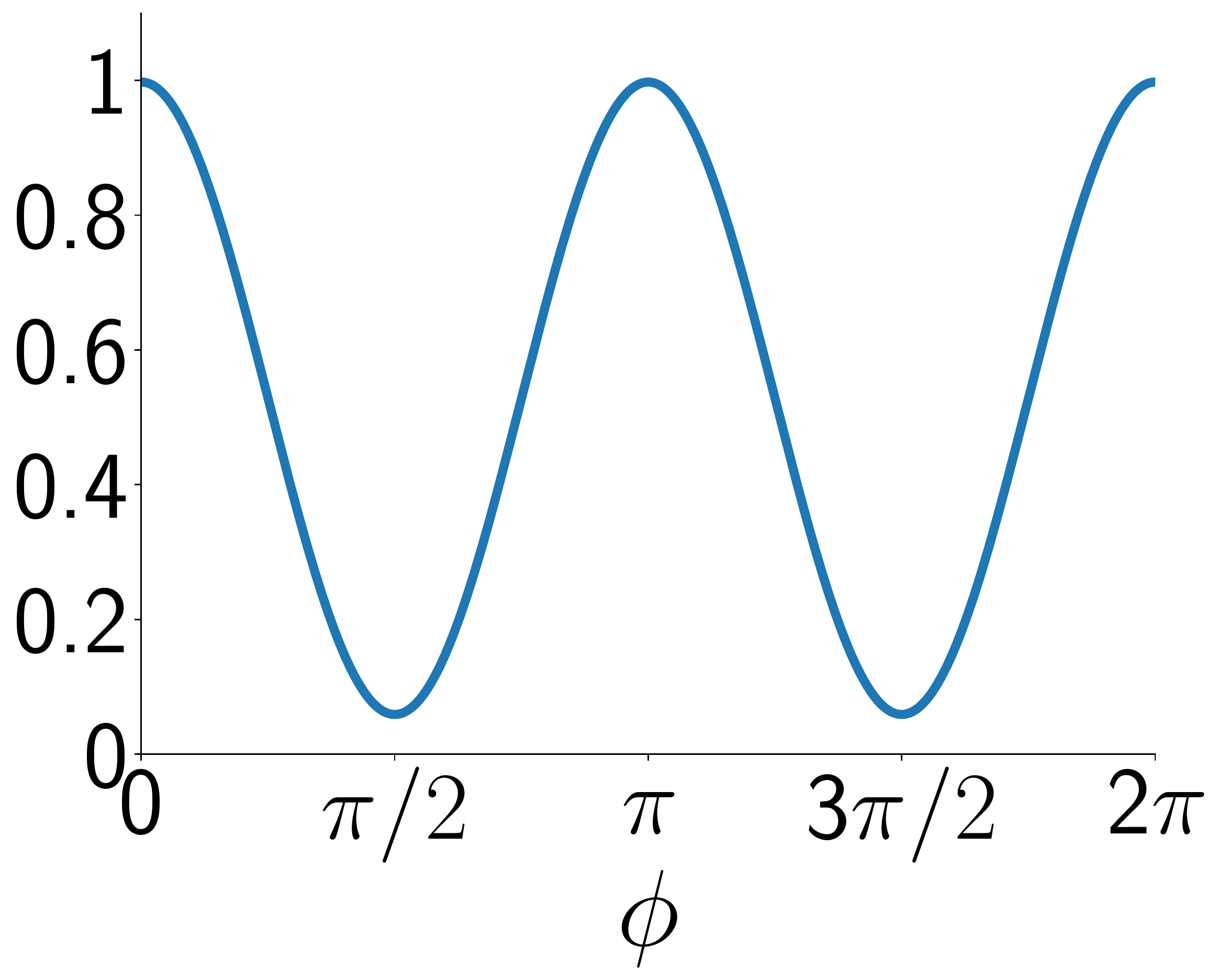}
		&\IncG[width=.23\textwidth,height=.12\textheight]{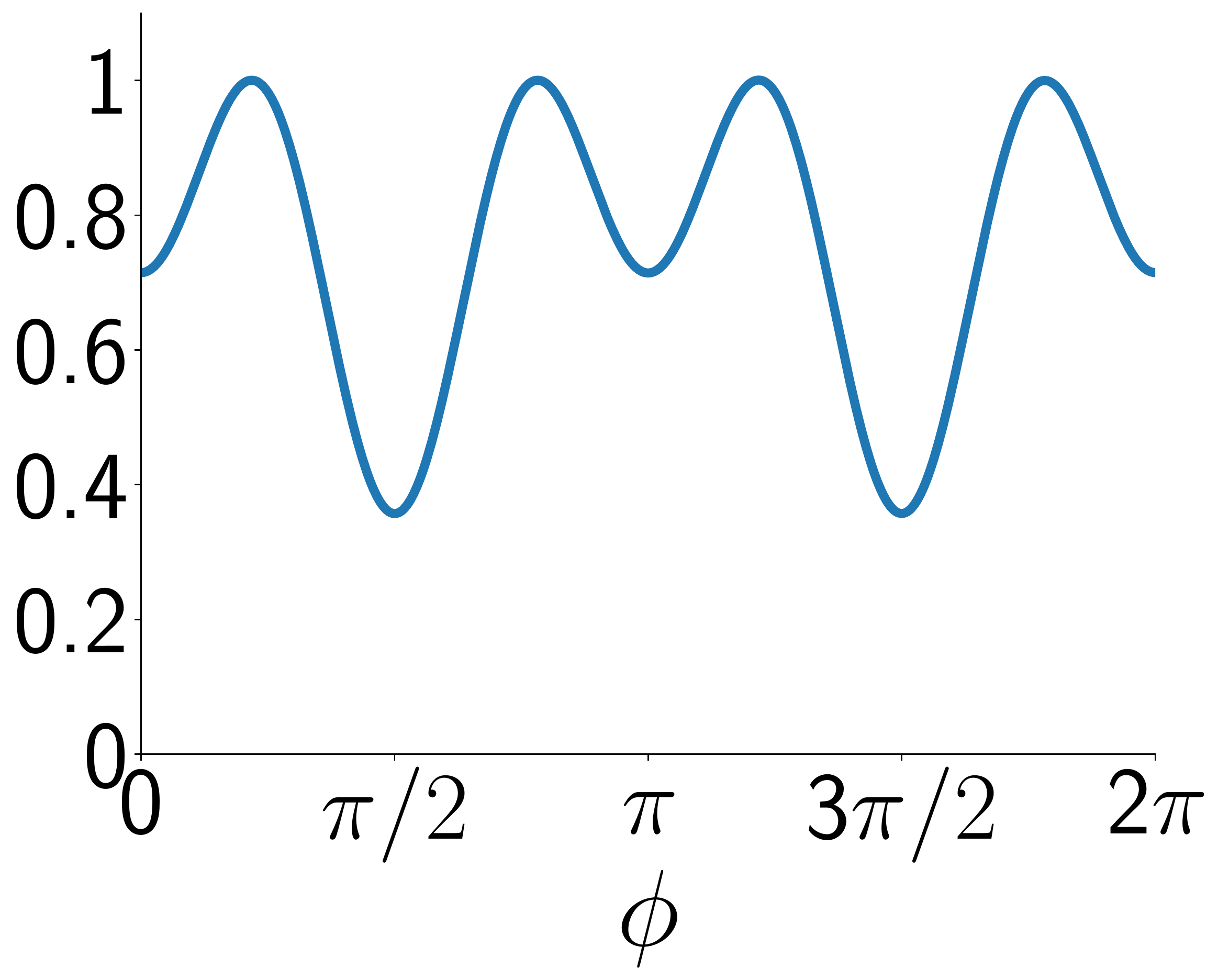}\\

		$D(2,2)$
		&\IncG[width=.23\textwidth,height=.12\textheight]{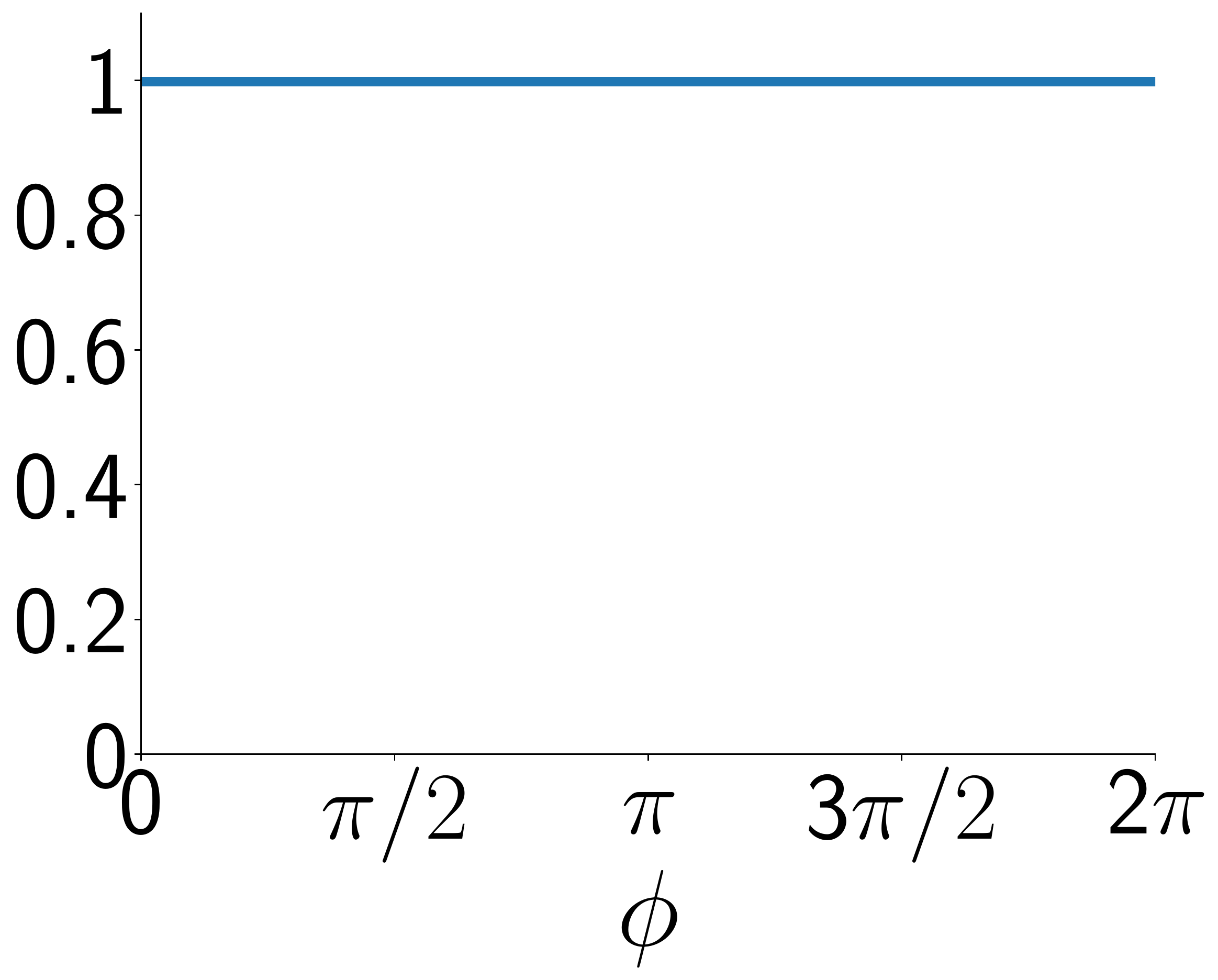}
		&\IncG[width=.23\textwidth,height=.12\textheight]{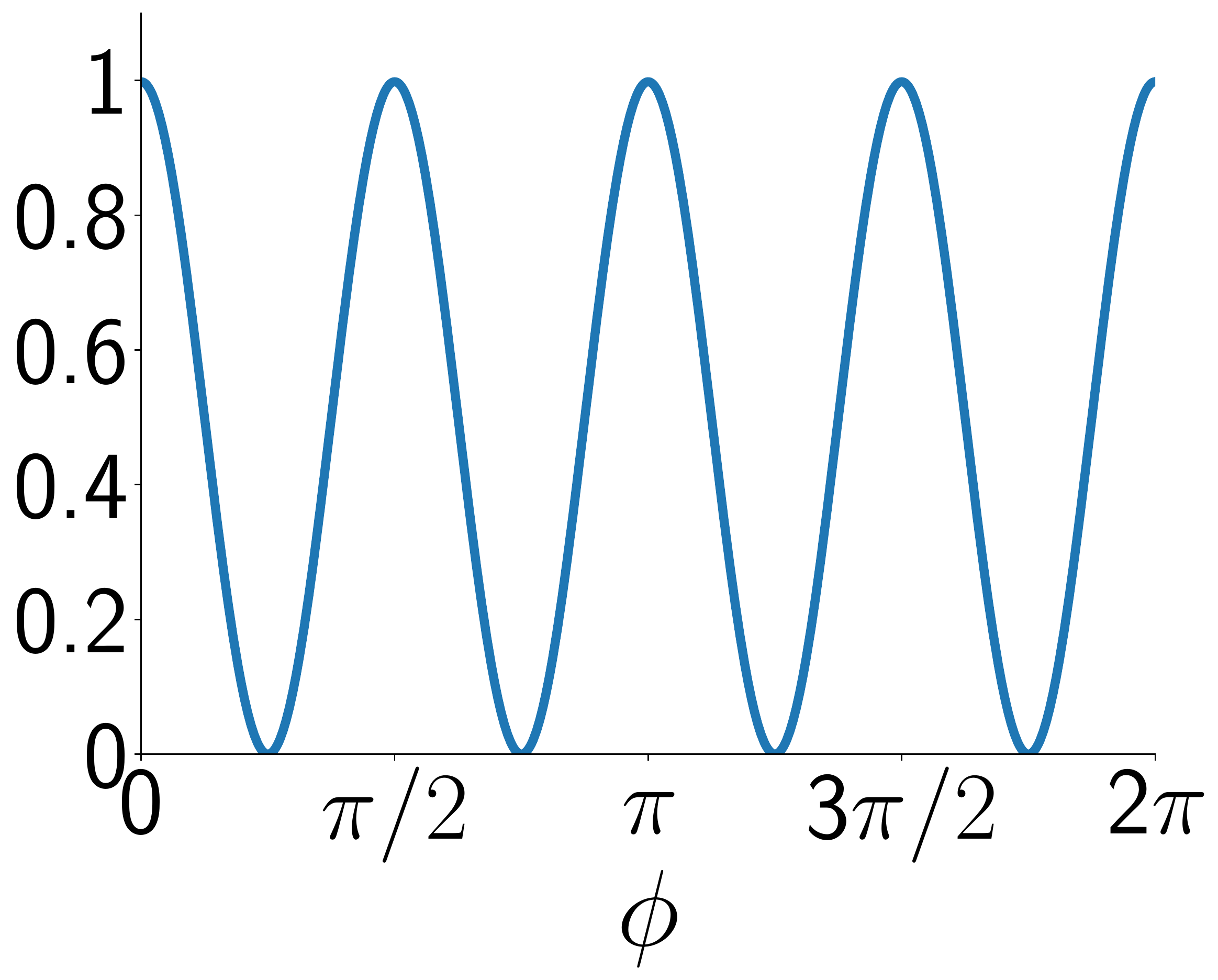}
		&\IncG[width=.23\textwidth,height=.12\textheight]{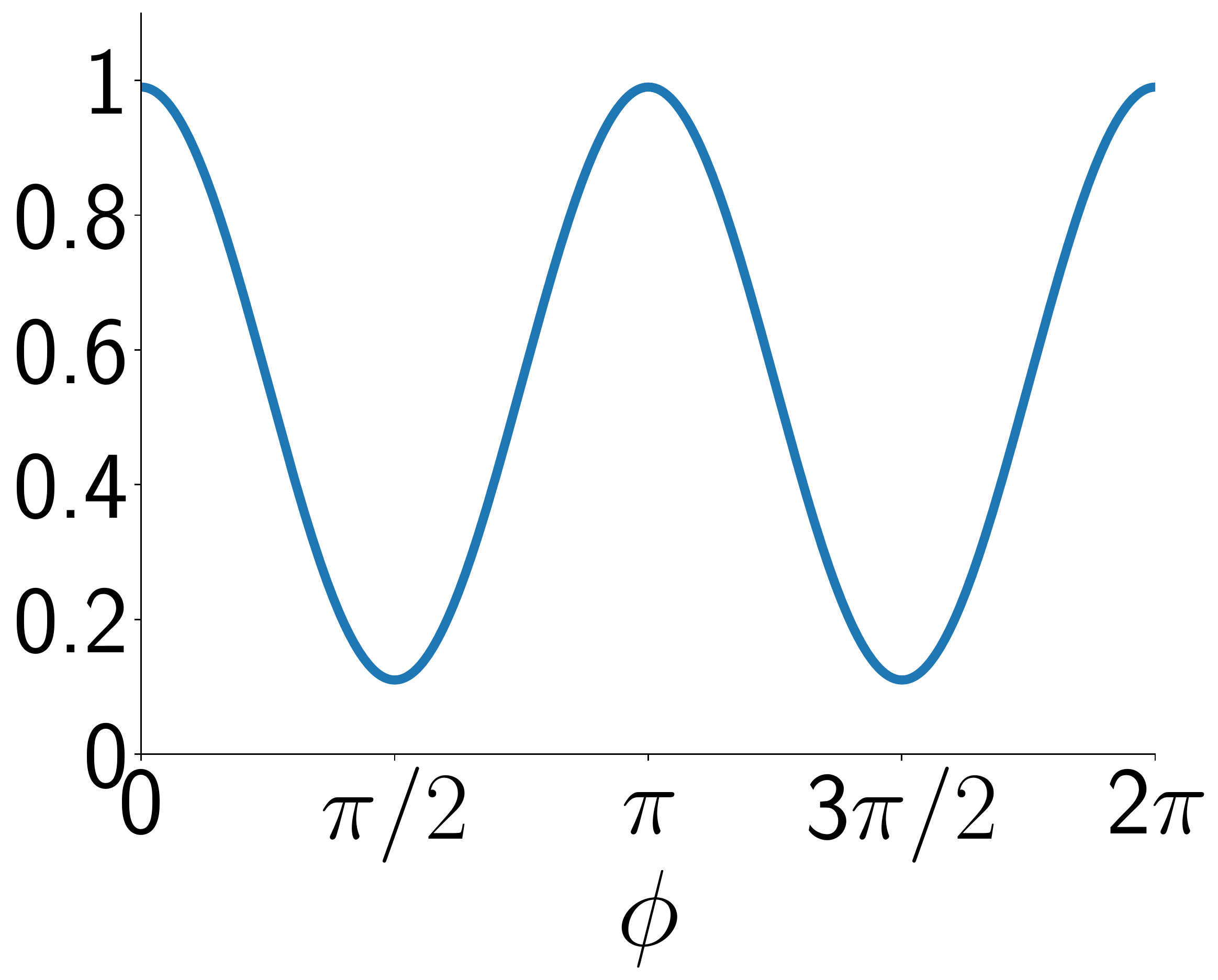}
		&\IncG[width=.23\textwidth,height=.12\textheight]{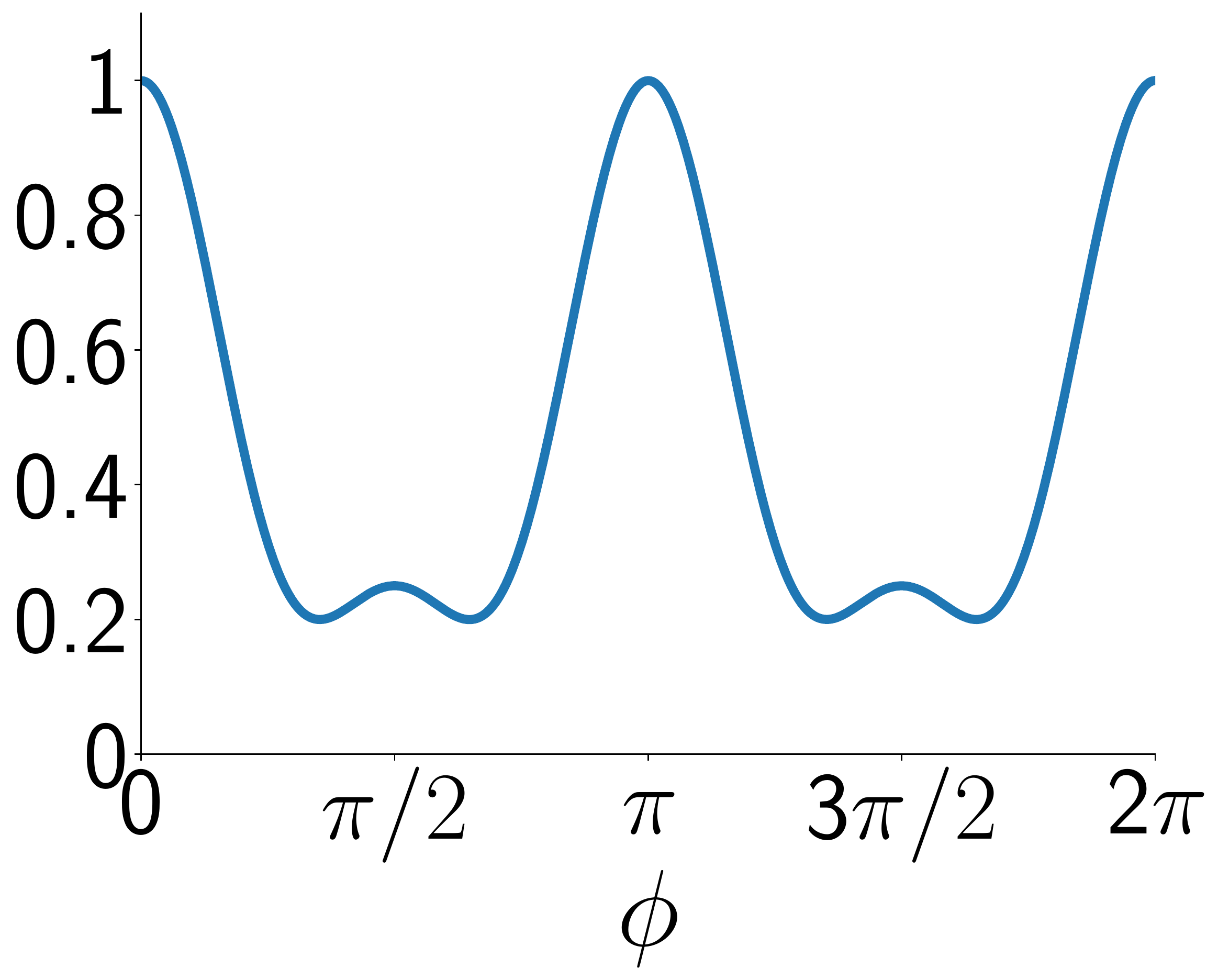}\\	
		
		$D(3,0)$
		&\IncG[width=.23\textwidth,height=.12\textheight]{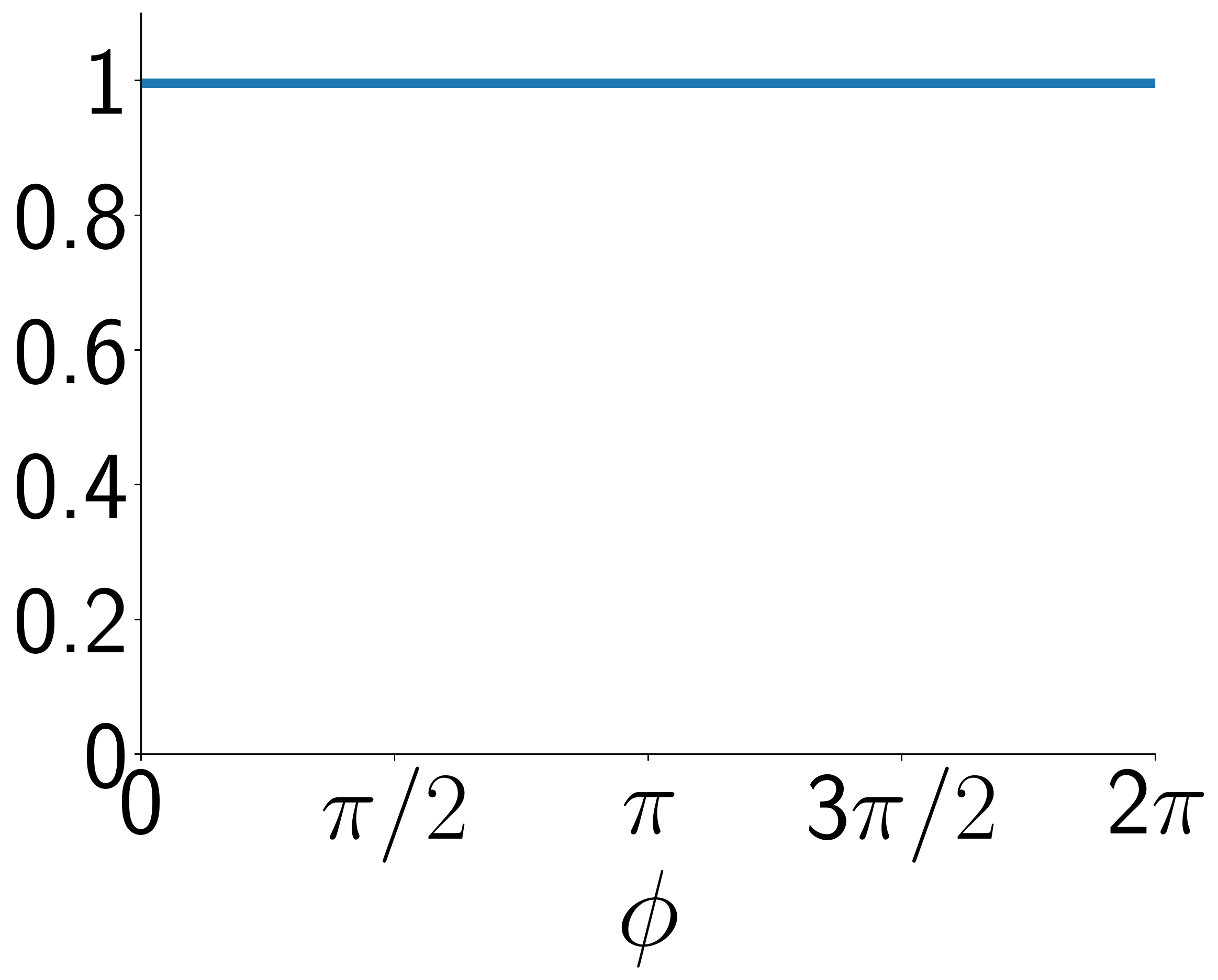}
		&\IncG[width=.23\textwidth,height=.12\textheight]{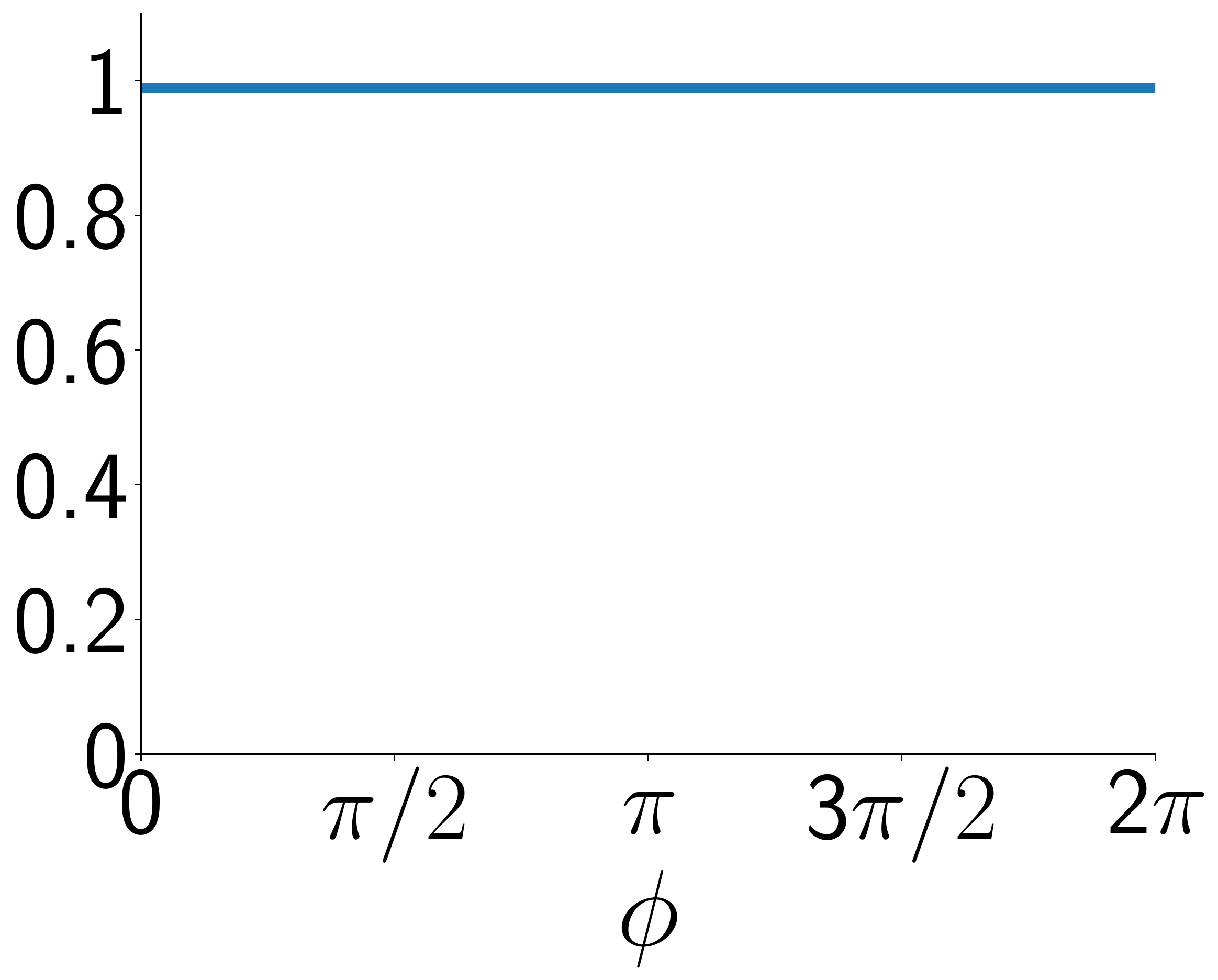}
		&\IncG[width=.23\textwidth,height=.12\textheight]{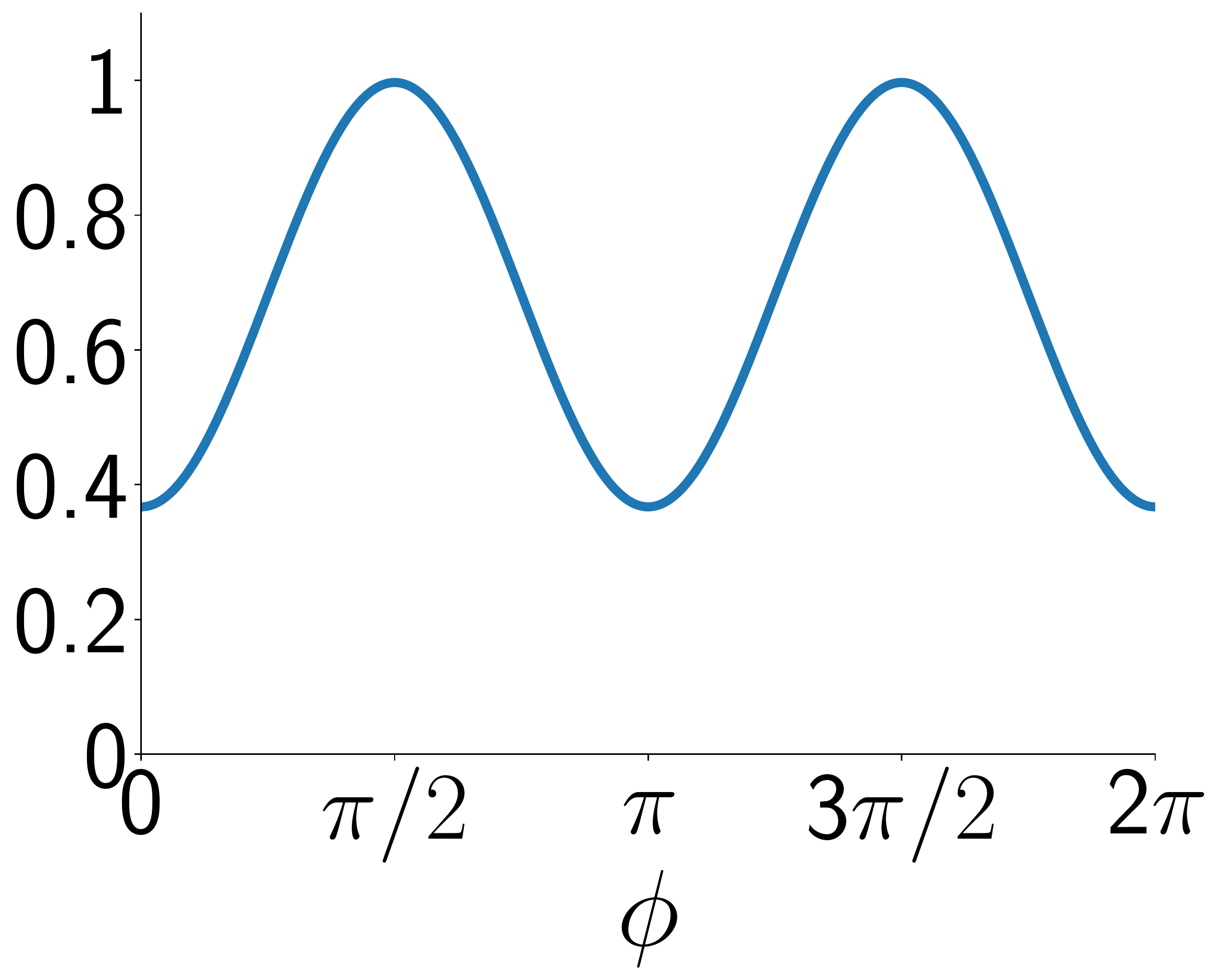}
		&\IncG[width=.23\textwidth,height=.12\textheight]{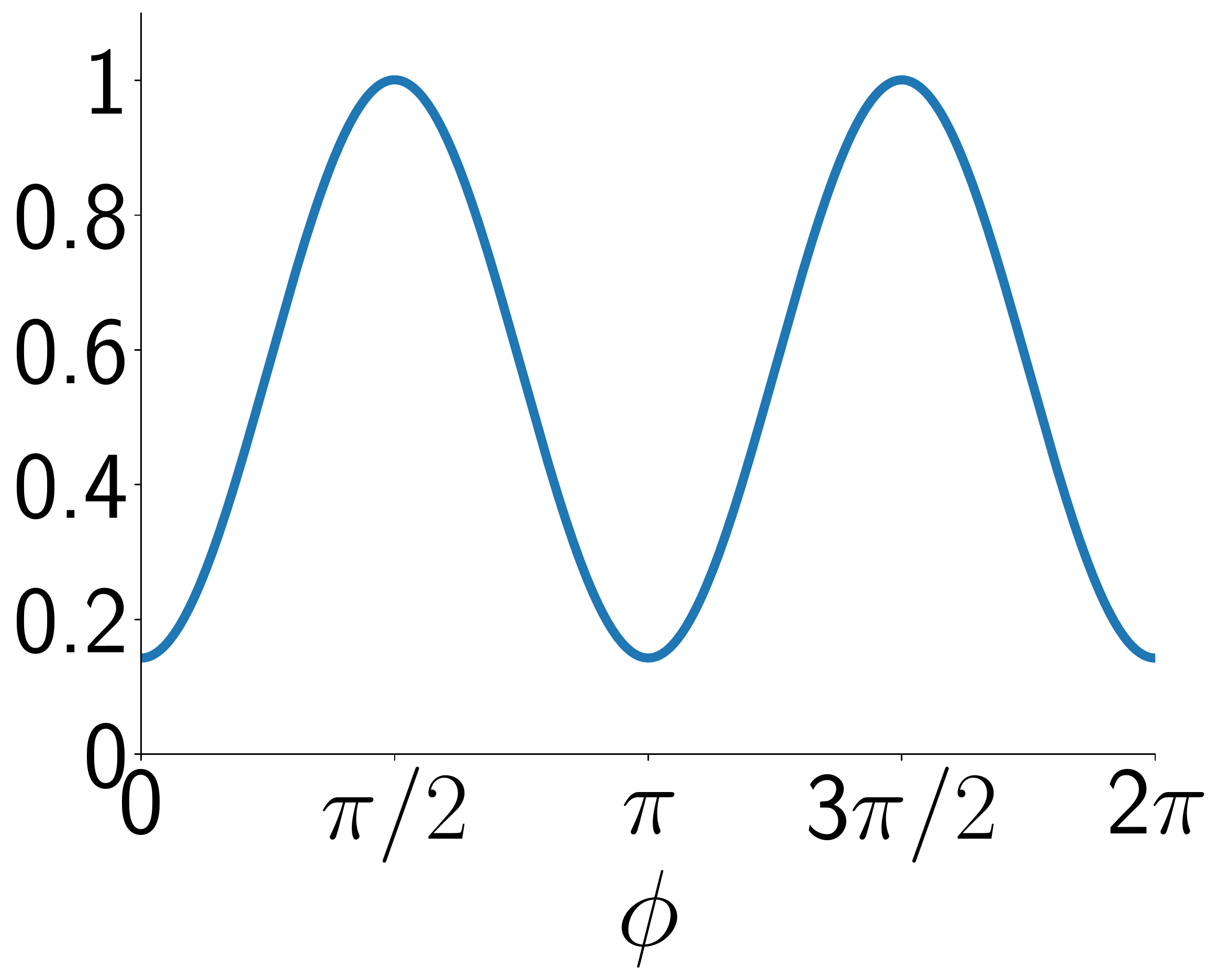}\\
		
		$D(2,1)$
		&\IncG[width=.23\textwidth,height=.12\textheight]{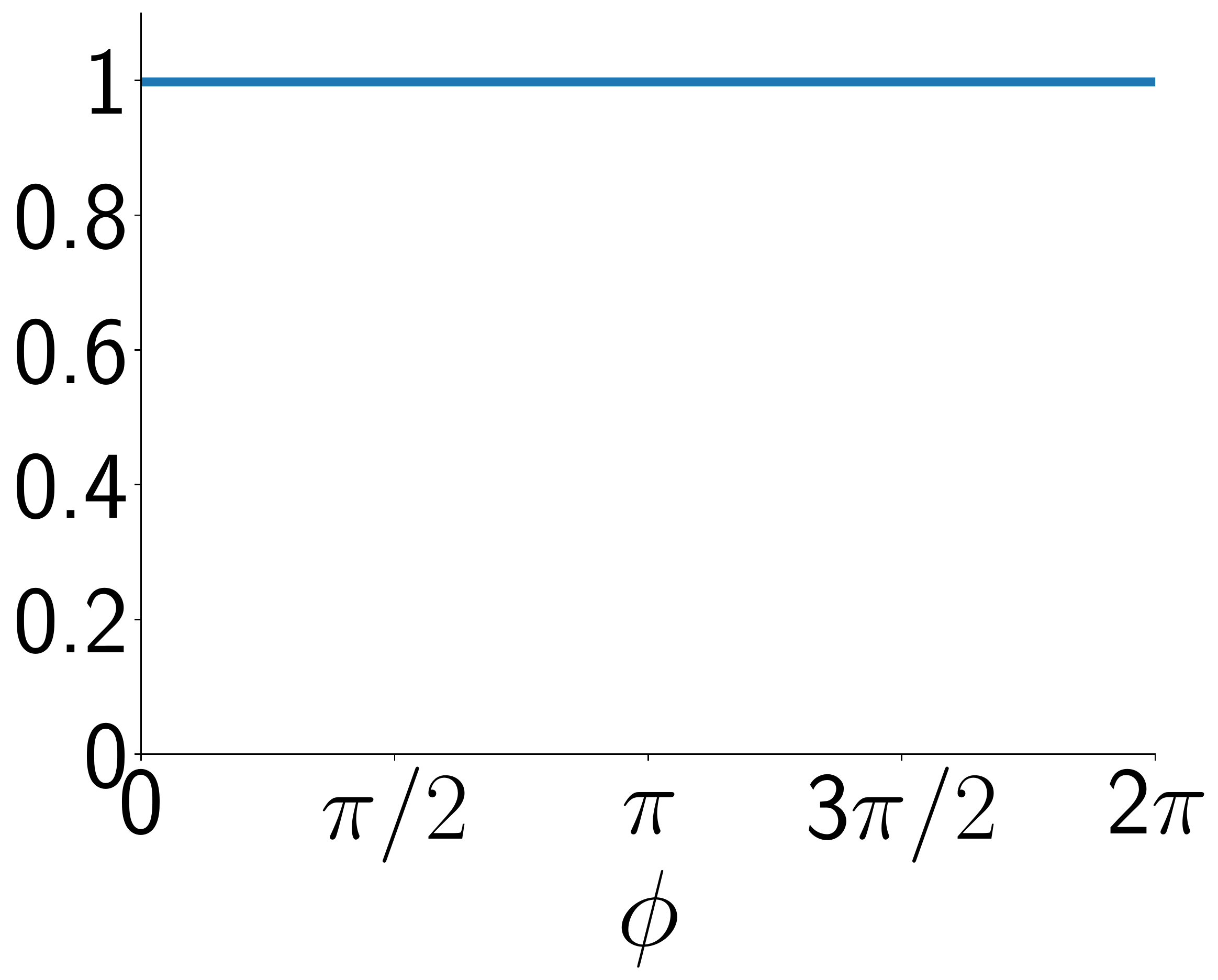}
		&\IncG[width=.23\textwidth,height=.12\textheight]{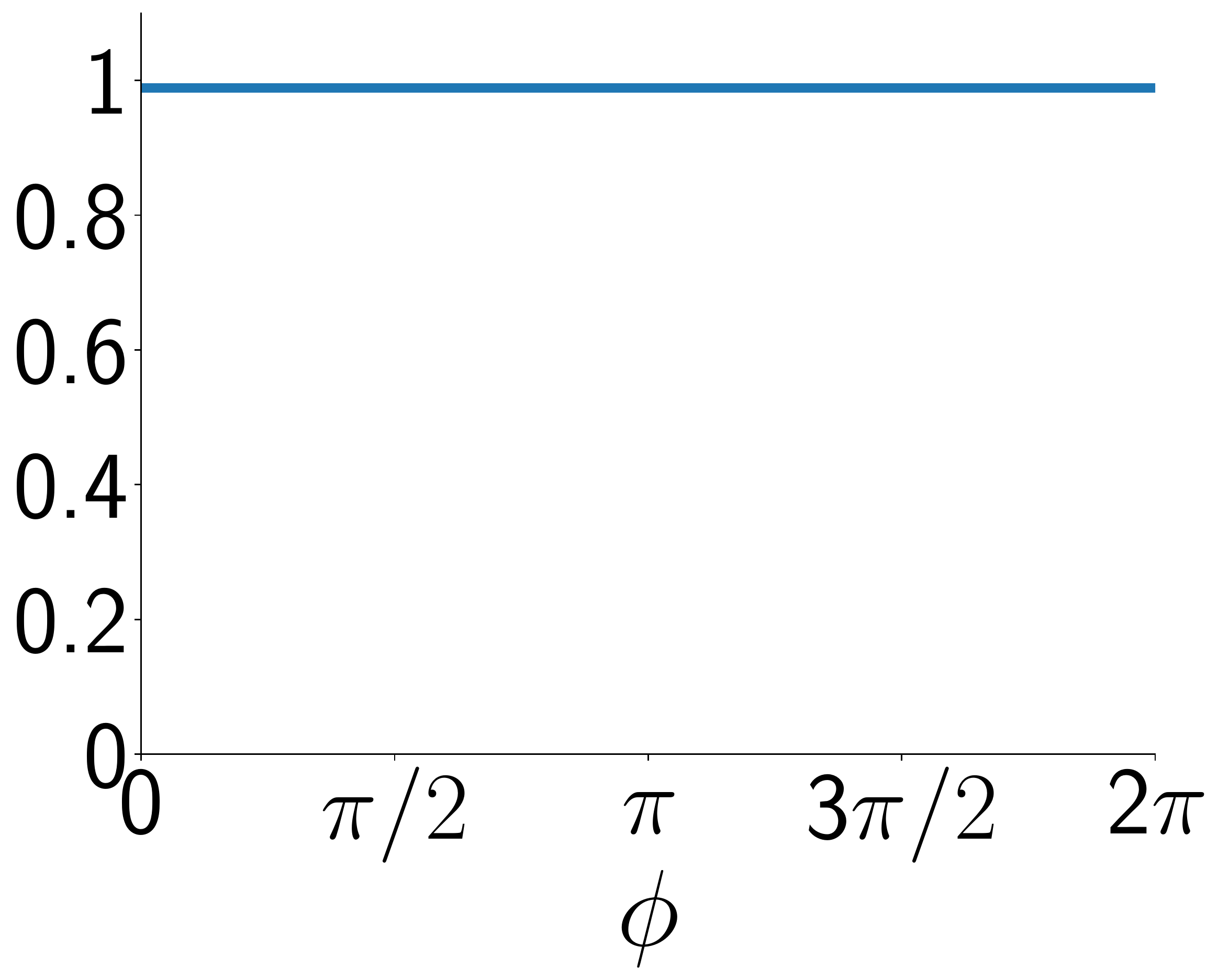}
		&\IncG[width=.23\textwidth,height=.12\textheight]{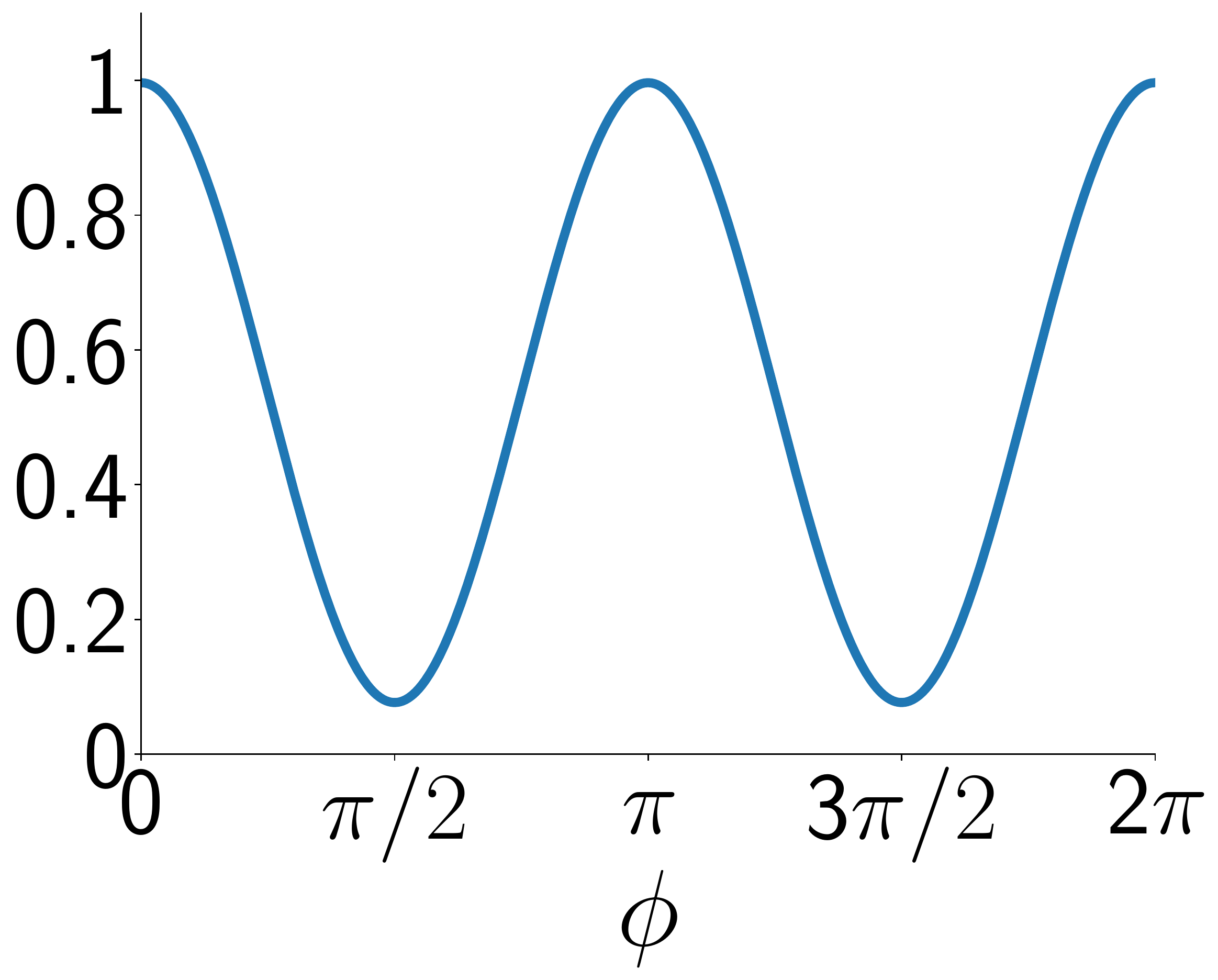}
		&\IncG[width=.23\textwidth,height=.12\textheight]{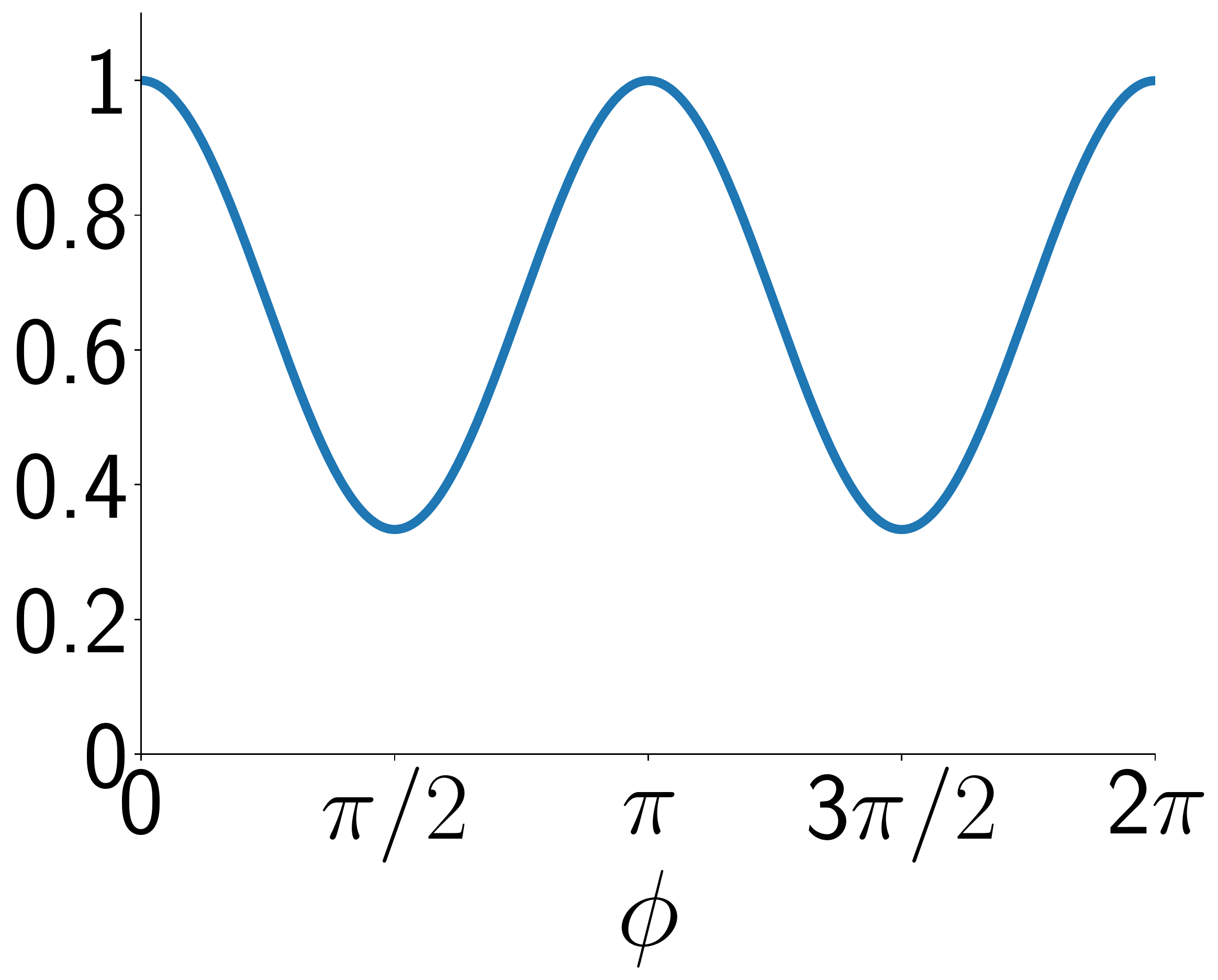}\\
		
		$D(2,0)$
		&\IncG[width=.23\textwidth,height=.12\textheight]{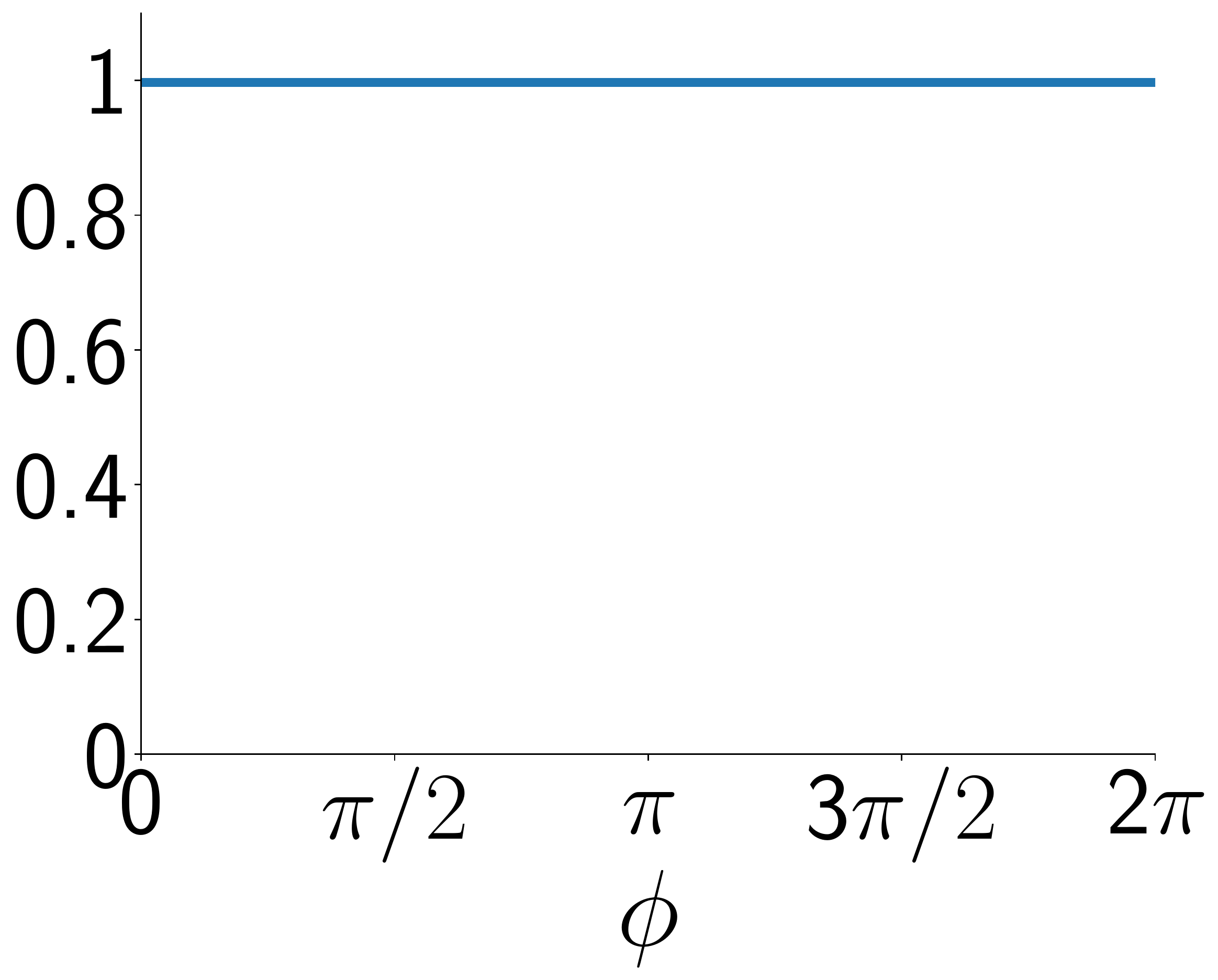}
		&\IncG[width=.23\textwidth,height=.12\textheight]{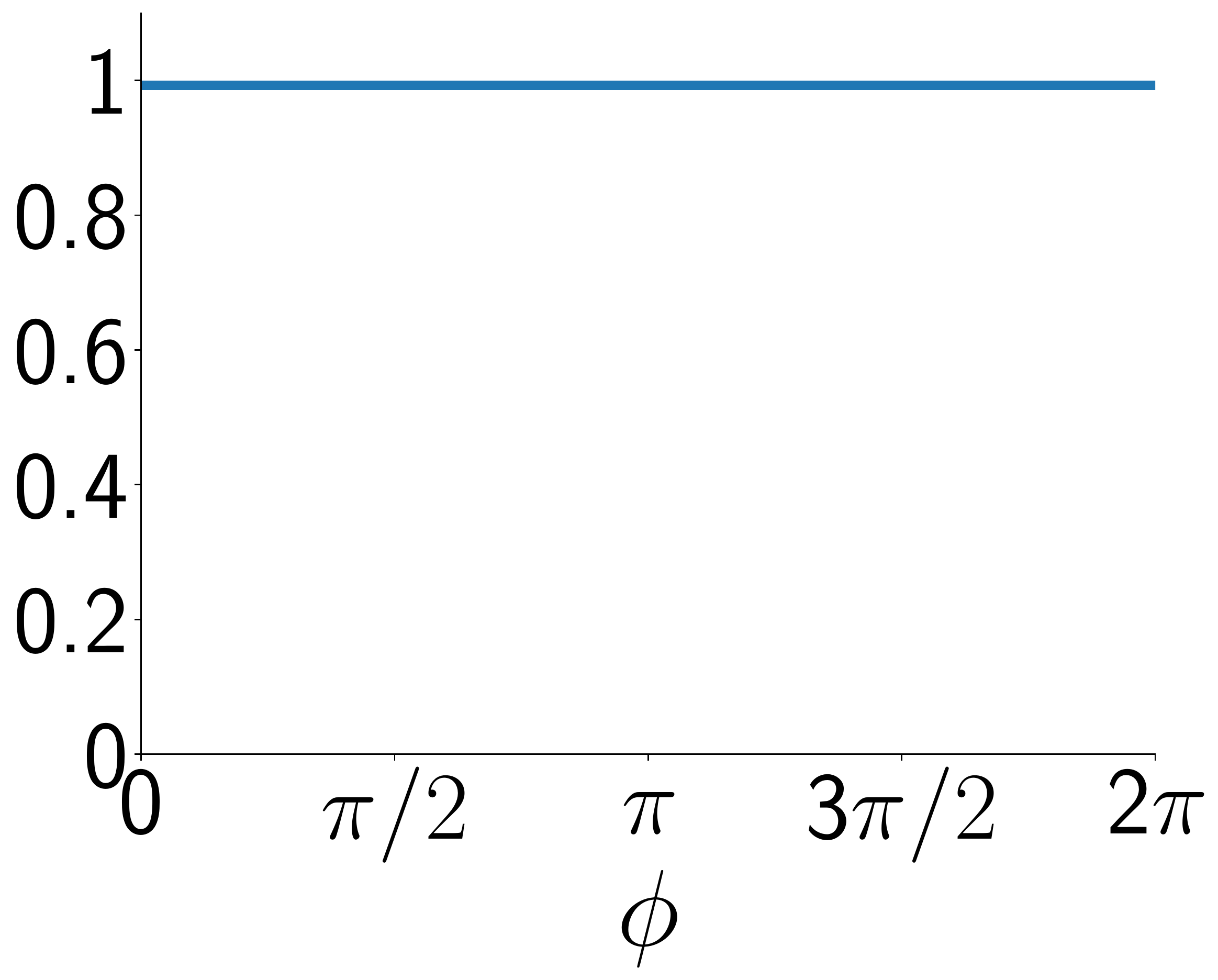}
		&\IncG[width=.23\textwidth,height=.12\textheight]{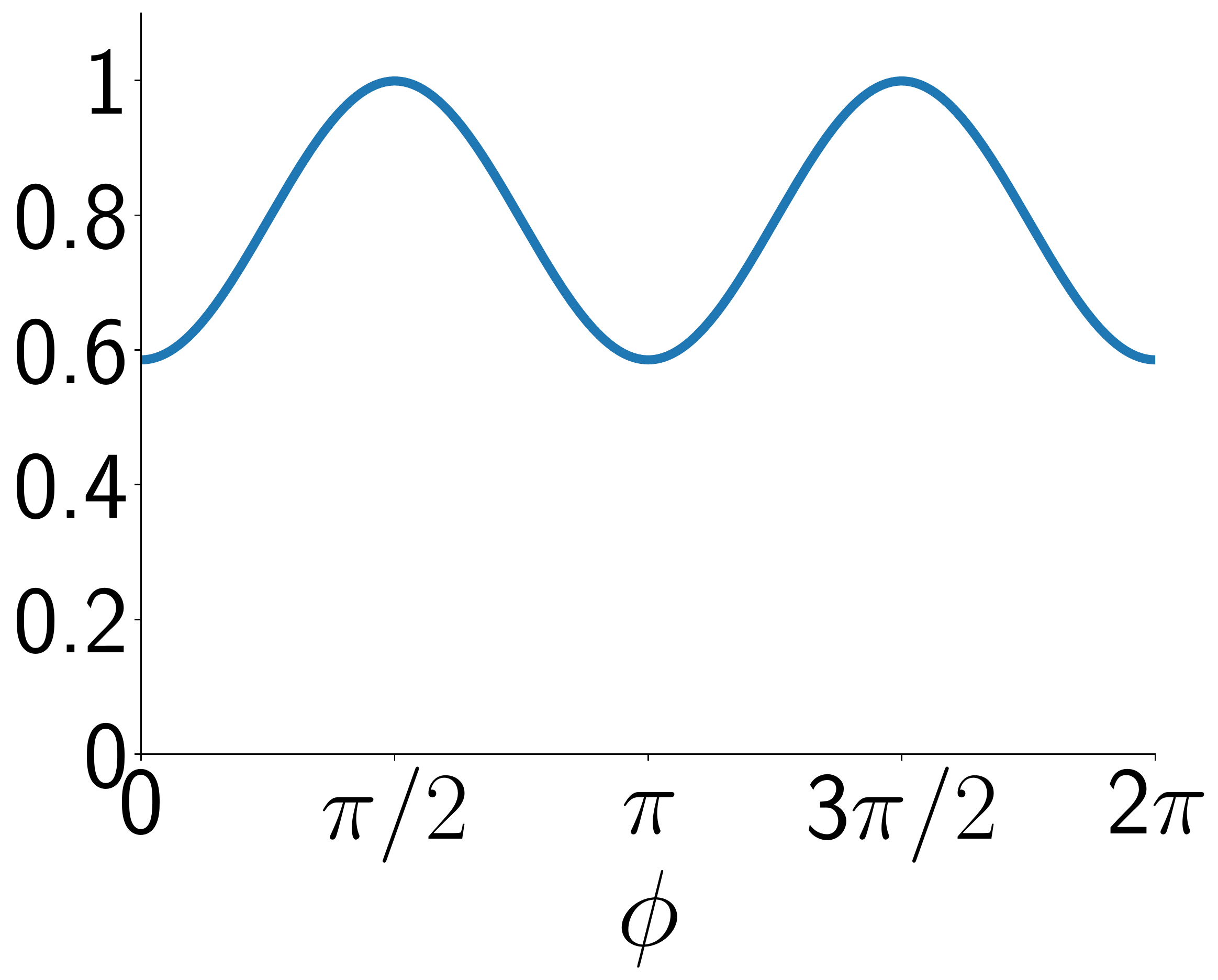}
		&\IncG[width=.23\textwidth,height=.12\textheight]{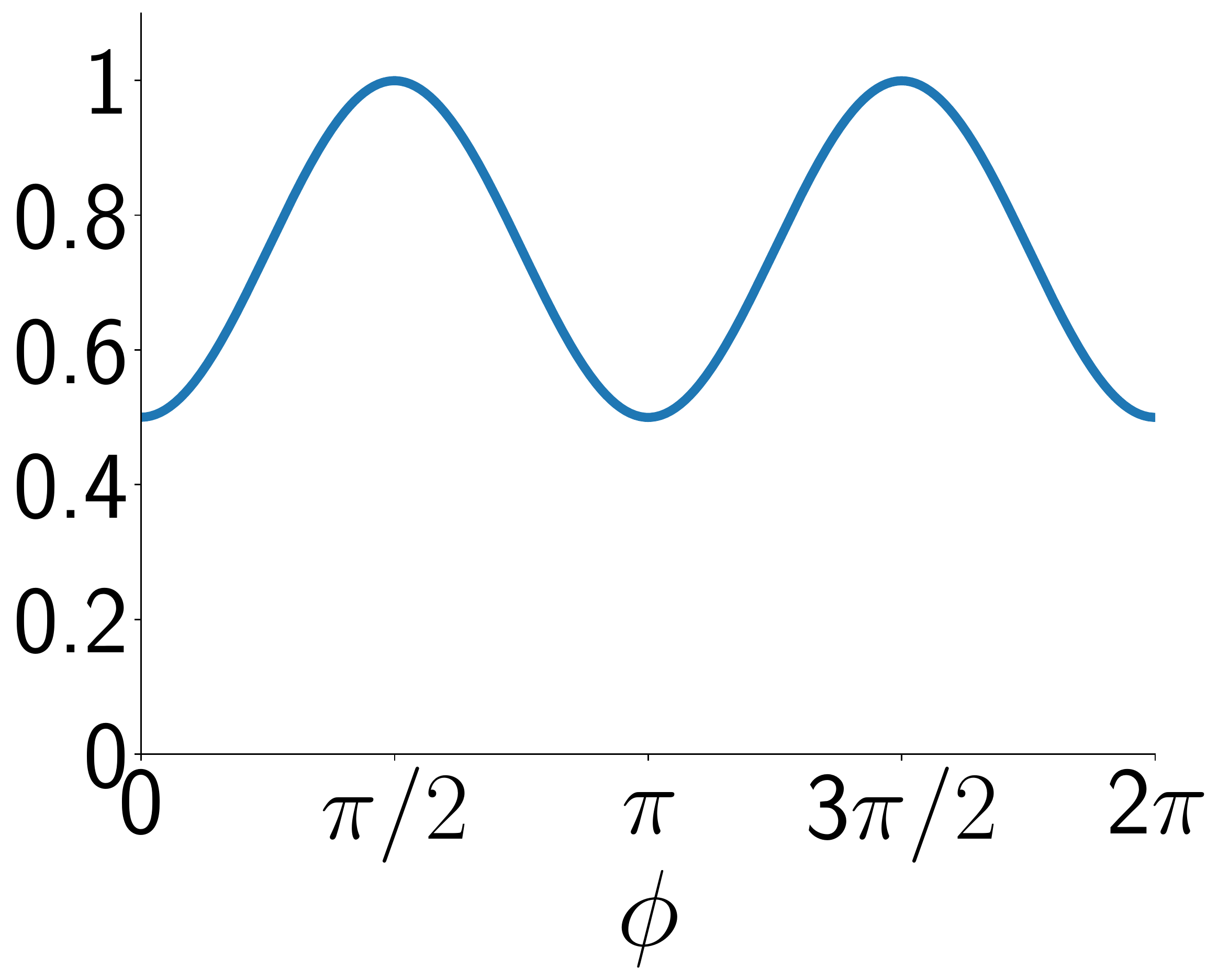}\\
		
		$D(1,1)$
		&\IncG[width=.23\textwidth,height=.12\textheight]{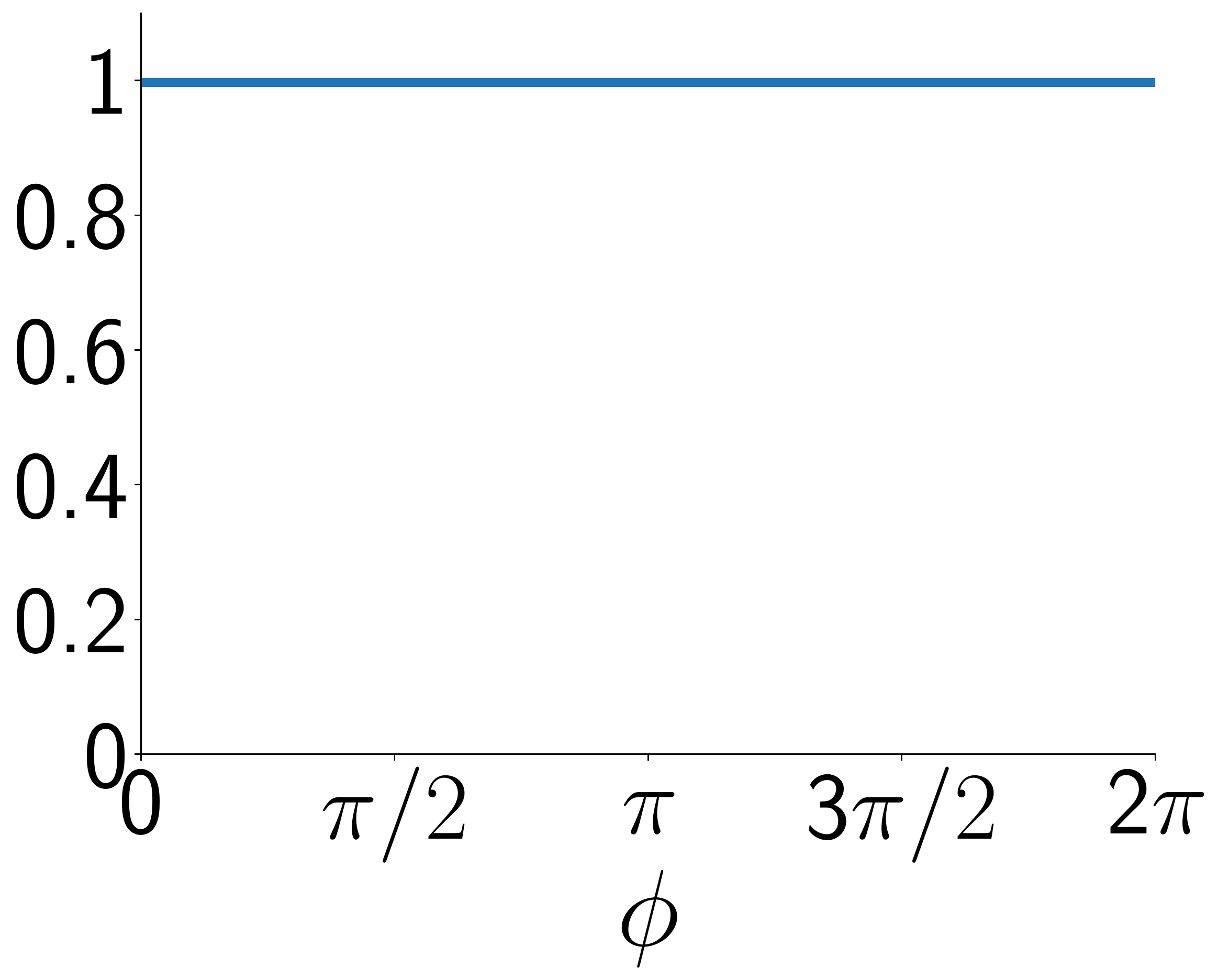}
		&\IncG[width=.23\textwidth,height=.12\textheight]{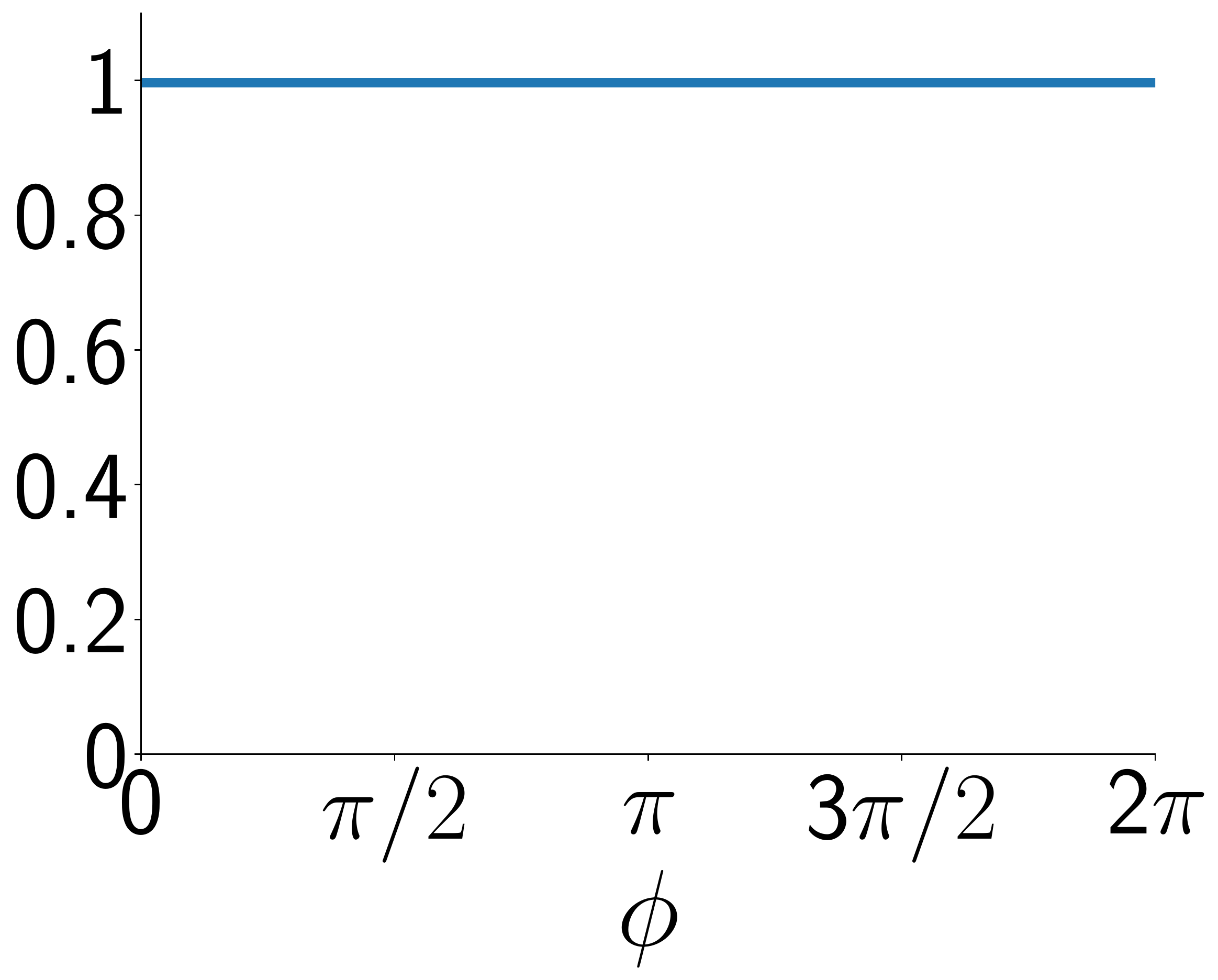}
		&\IncG[width=.23\textwidth,height=.12\textheight]{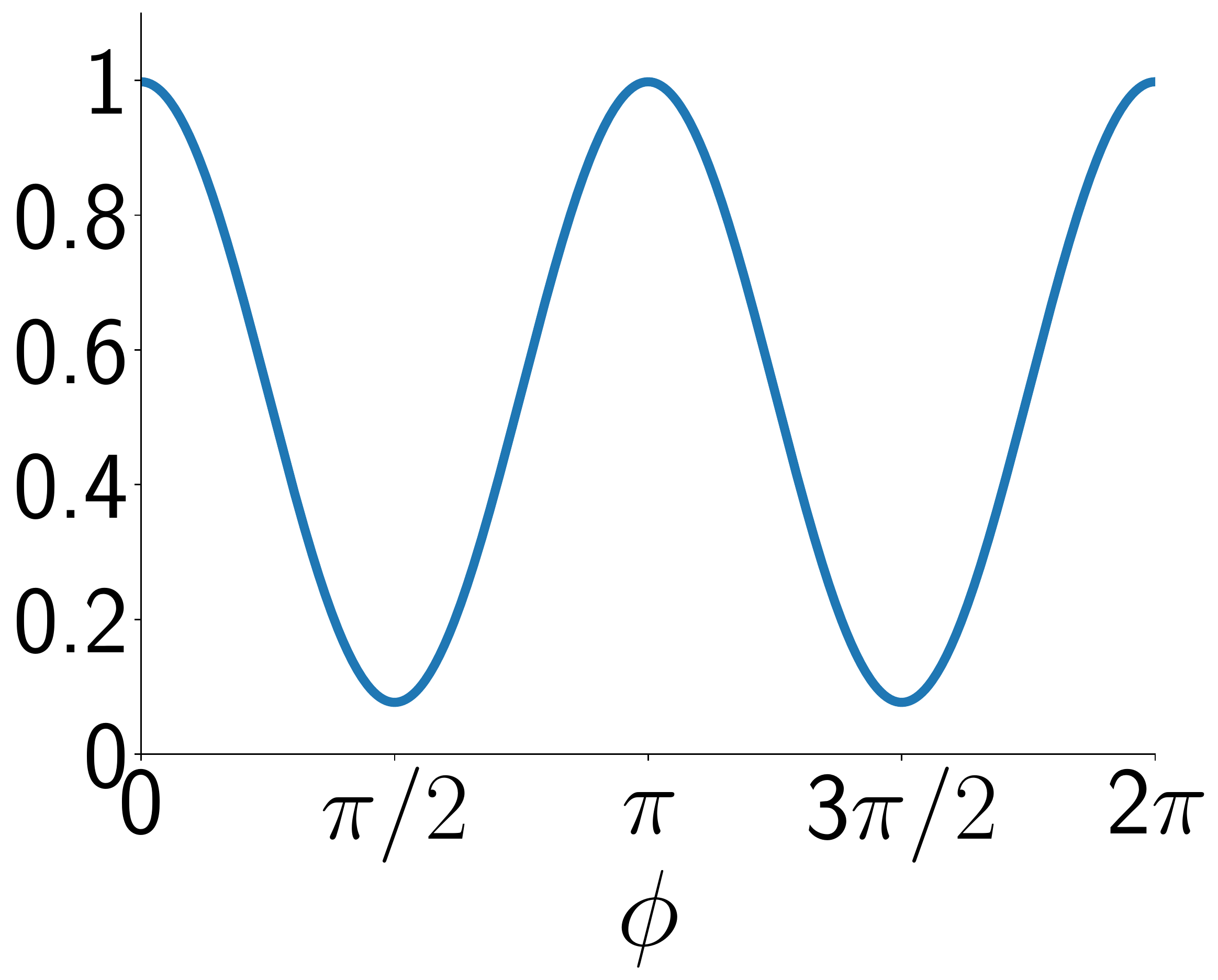}
		&\IncG[width=.23\textwidth,height=.12\textheight]{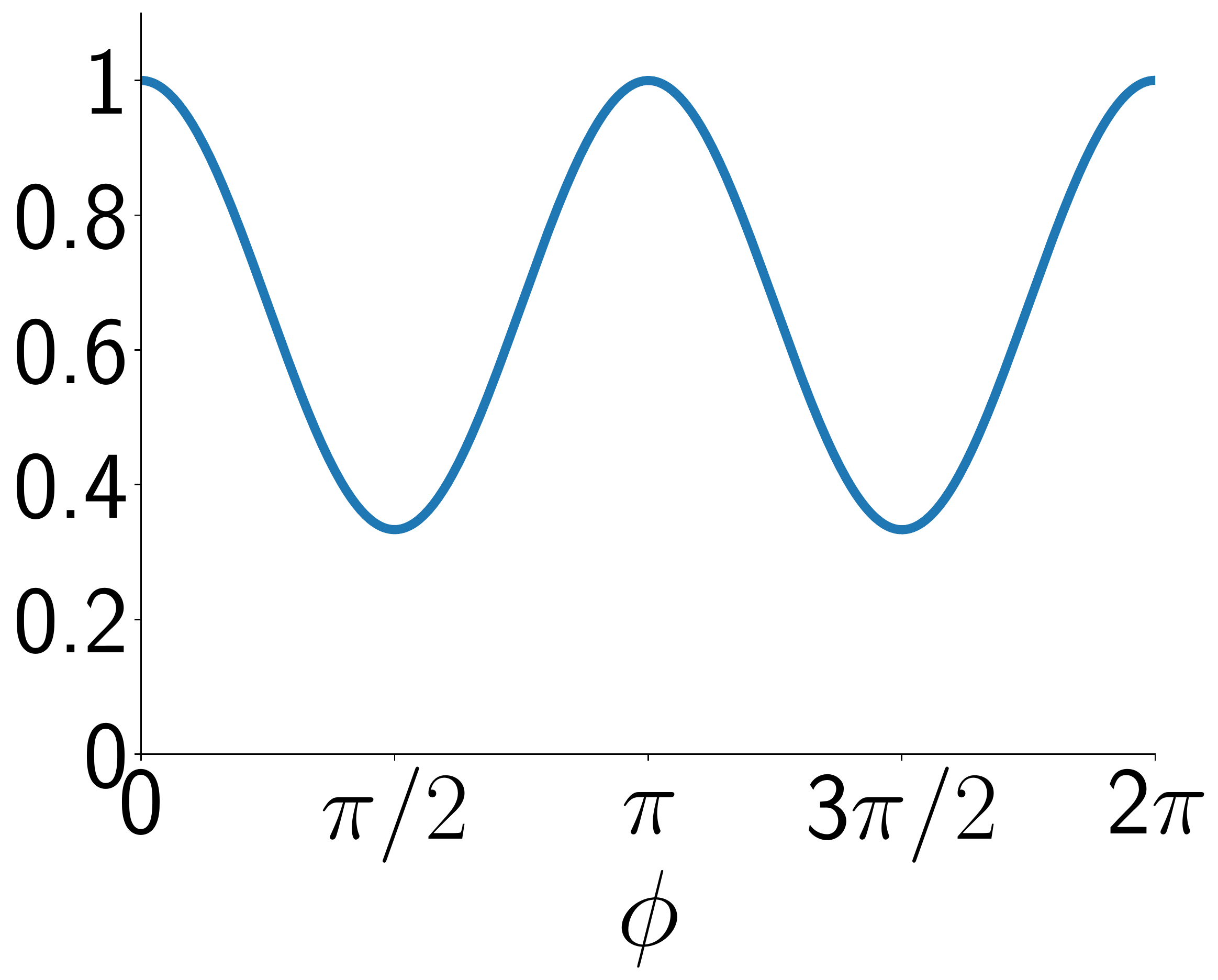}\\
		
		$D(1,0)$
		&\IncG[width=.23\textwidth,height=.12\textheight]{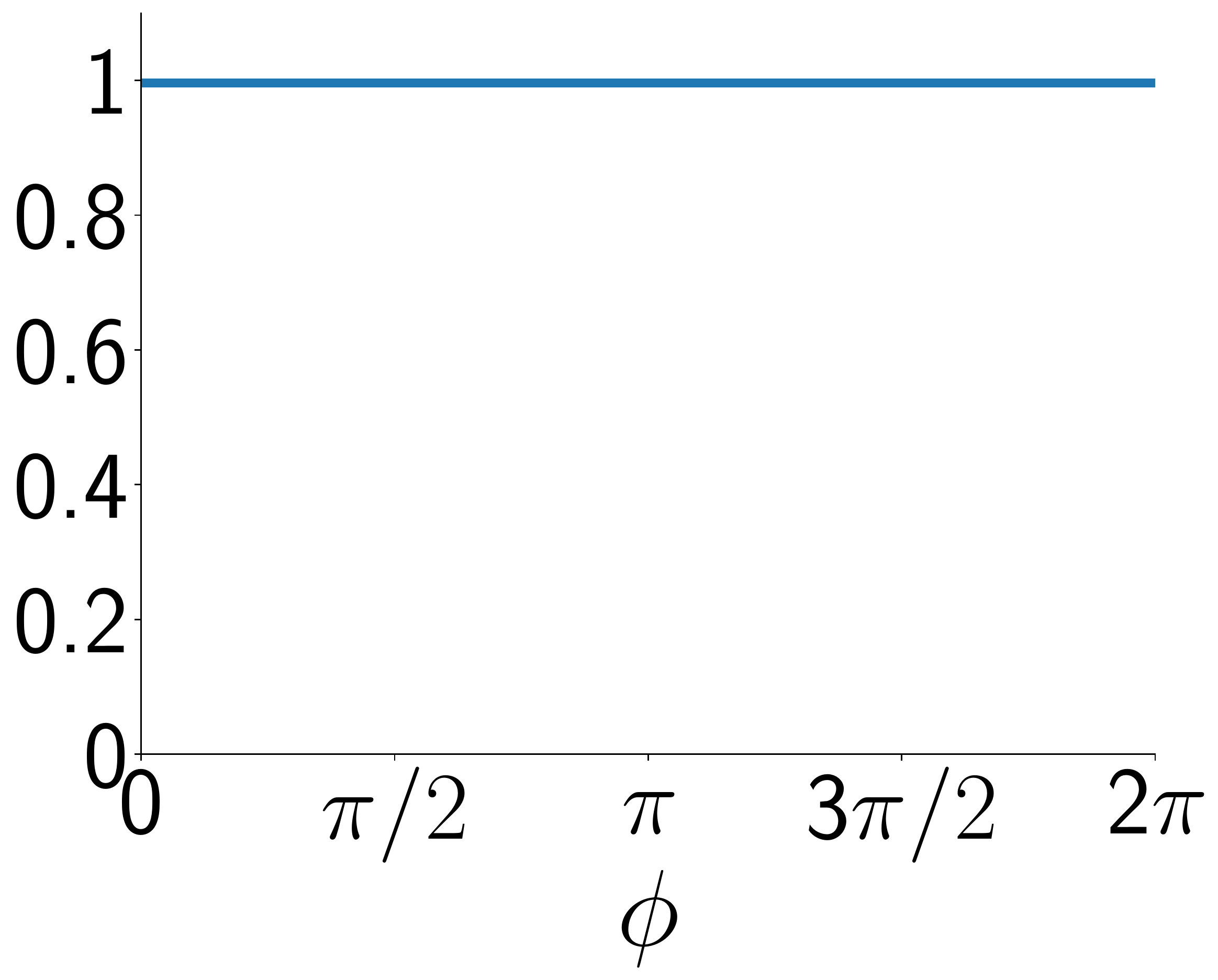}
		&\IncG[width=.23\textwidth,height=.12\textheight]{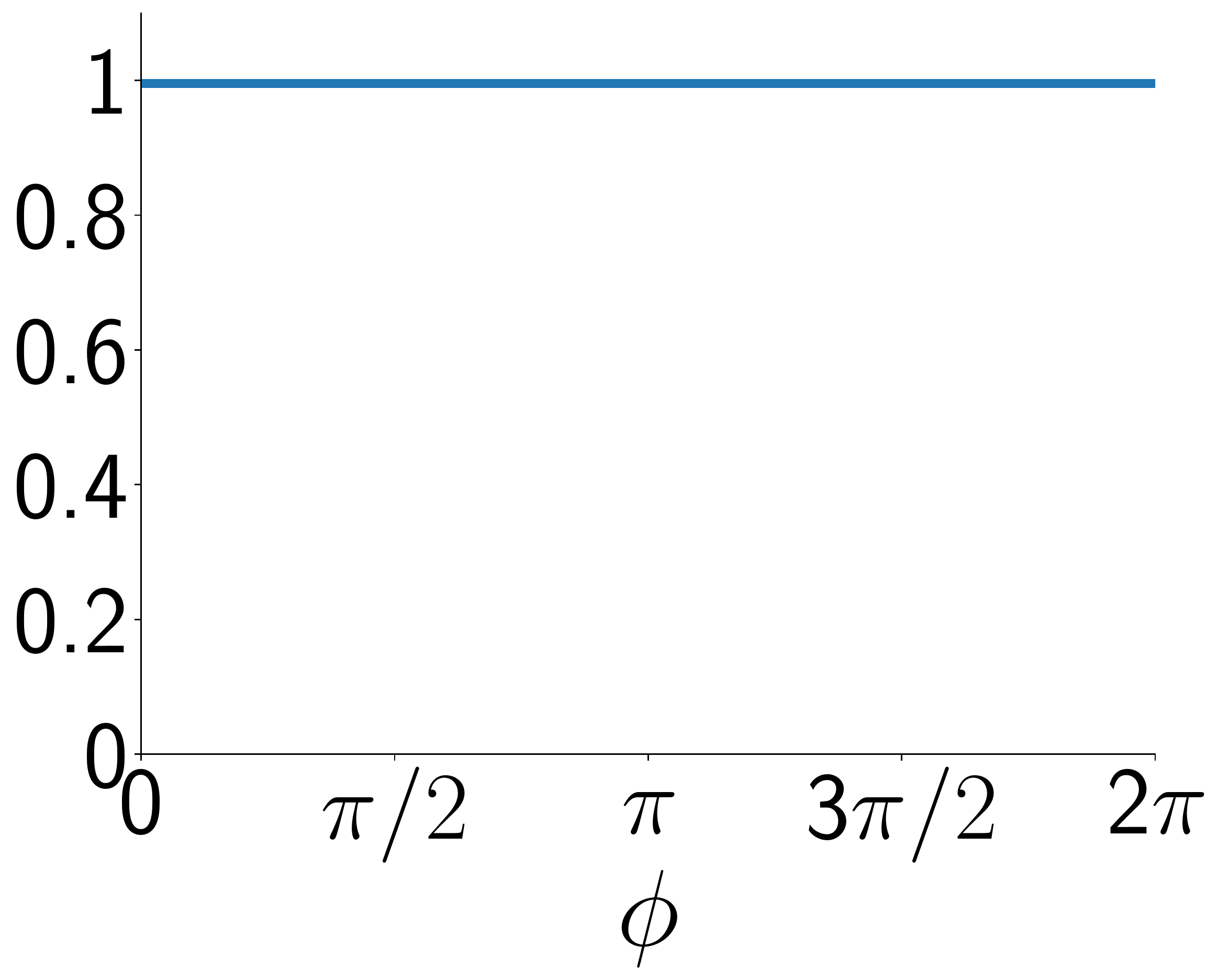}
		&\IncG[width=.23\textwidth,height=.12\textheight]{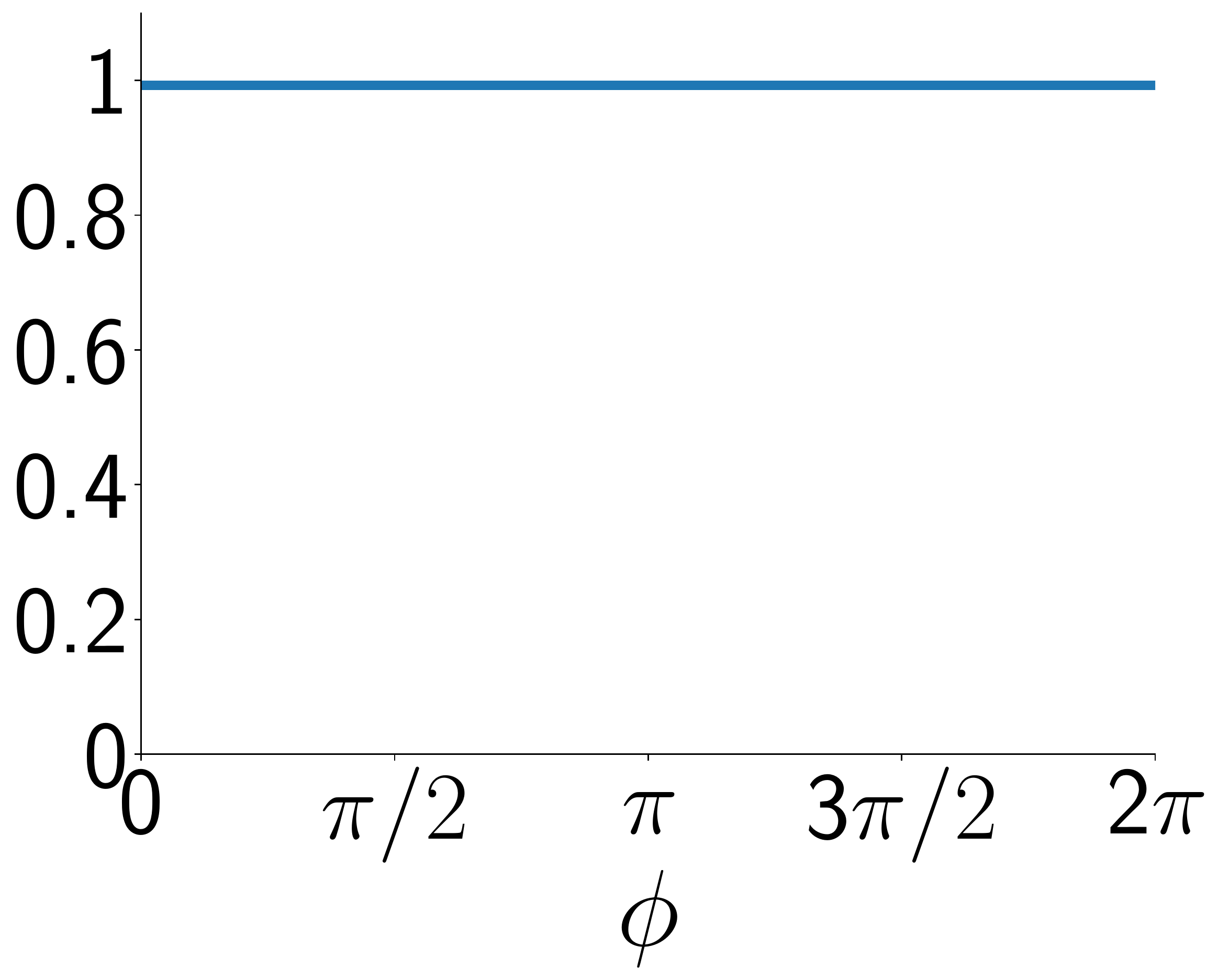}
		&\IncG[width=.23\textwidth,height=.12\textheight]{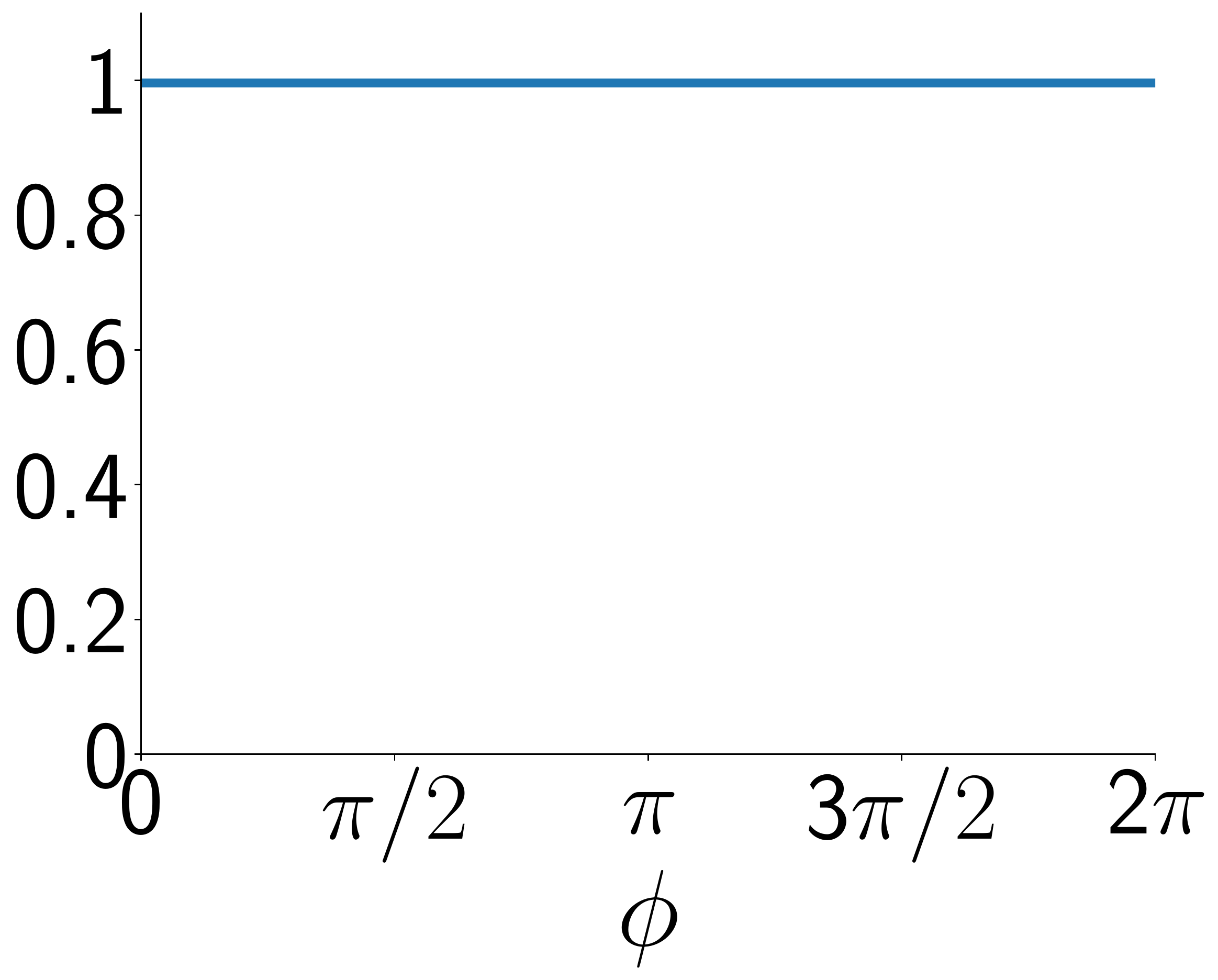}\\
		
	\end{tabular}

\end{table*}

\begin{figure}[h]
	\includegraphics[width=\columnwidth, height=8cm]{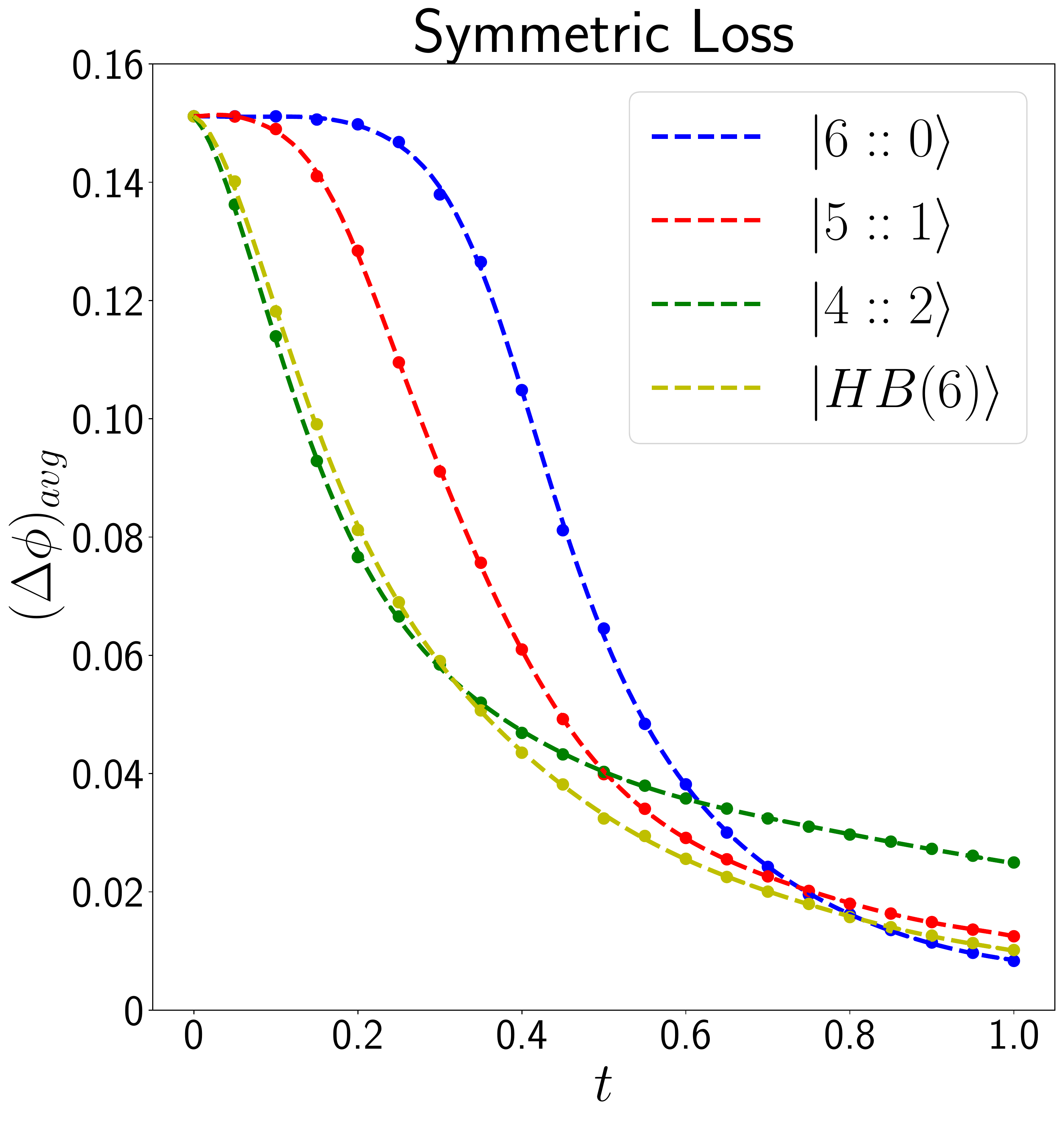}
	\caption{Plot of $(\Delta \phi)_{avg}$ as a function of $t$ for different input states for $N_r = 400$. Circle points are the values that are calculated by simulating the quantum circuits and the dashed lines are appropriate functions fitted to the data.}
	\label{dphi_sym}
\end{figure}
Table.~\ref{tab_interference_pattern} shows the unnormalized likelihood functions $(P_{n_0,n_1}(\phi))$ of the input states for $t=t_0=t_1=0.5$. 
In this symmetric case, $P_{n_0,n_1}(\phi)$ for other values of $t$ is similar; note that when $t=1$, always the number of photons detected is 6, therefore $P_{n_0,n_1}(\phi)=0$ for $n_0 +n_1 <6$. These probability distributions are calculated by simulating the corresponding quantum circuit at different phases in $[0,2\pi]$ for a large number of times and fitting an appropriate periodic function on them. 
They are plotted in a way that their maximum value is 1 (except $P_{3,3}(\phi)$ for $\ket{5::1}$ which is zero because this combination of photons never gets detected). They are not normalized, but this is not an issue because we normalize the final probability distribution at each step when we are estimating the unknown phase. 
Table.~\ref{tab_interference_pattern} is ordered in a way that the total number of detected photons $(n_0 + n_1)$ generally decreases as we go below the table. 
As we can see, whenever the number of detected photons is less than 6 (meaning that at least one photon is lost during the process), the probability distribution for $\ket{6::0}$ is a constant function, which means that we get no information about the phase shift in this case. 
Therefore, by increasing the photon loss probability, the precision of phase estimation with the $NOON$ states will decrease more rapidly than the other states. Compering $\ket{5::1}$ with $\ket{4::2}$ and $\ket{HB(6)}$, we can see that patterns of $\ket{5::1}$ gives us no information about the phase shift when the number of detected photons is less than 4, while patterns of $\ket{4::2}$ and $\ket{HB(6)}$ stop giving information only when the number of detected photons is less than 2. This means that at high loss rates, we expect $\ket{4::2}$ and $\ket{HB(6)}$ states to estimate phase better than $\ket{5::1}$.
Figure.~\ref{dphi_sym} shows the average standard deviations of the input states in estimating the phase shift, for $N_r = 400$ in the symmetric photon loss case.  Obviously, for $t>0.81$, $\ket{6::0}$ is the best input state and gives the least error in estimating the phase shift. However, when $ 0.32< t < 0.81$, $\ket{HB(6)}$ has the highest precision and when $t<0.32$, $\ket{4::2}$ becomes the best state to estimate the phase shift. When $t=0$, the photon loss rate is one, and this means that no photons get detected, and we gain no information about the phase shift, regardless of the input state. So, in this case, the information about the phase shift is the initial probability distribution for the phase shift, and at this point, $(\Delta \phi)_{avg}$ of all input states approach to a fixed value. 
\subsection{Asymmetric photon loss}
\label{result_asym}
\begin{figure}[h]
	\includegraphics[width=\columnwidth, height=8cm]{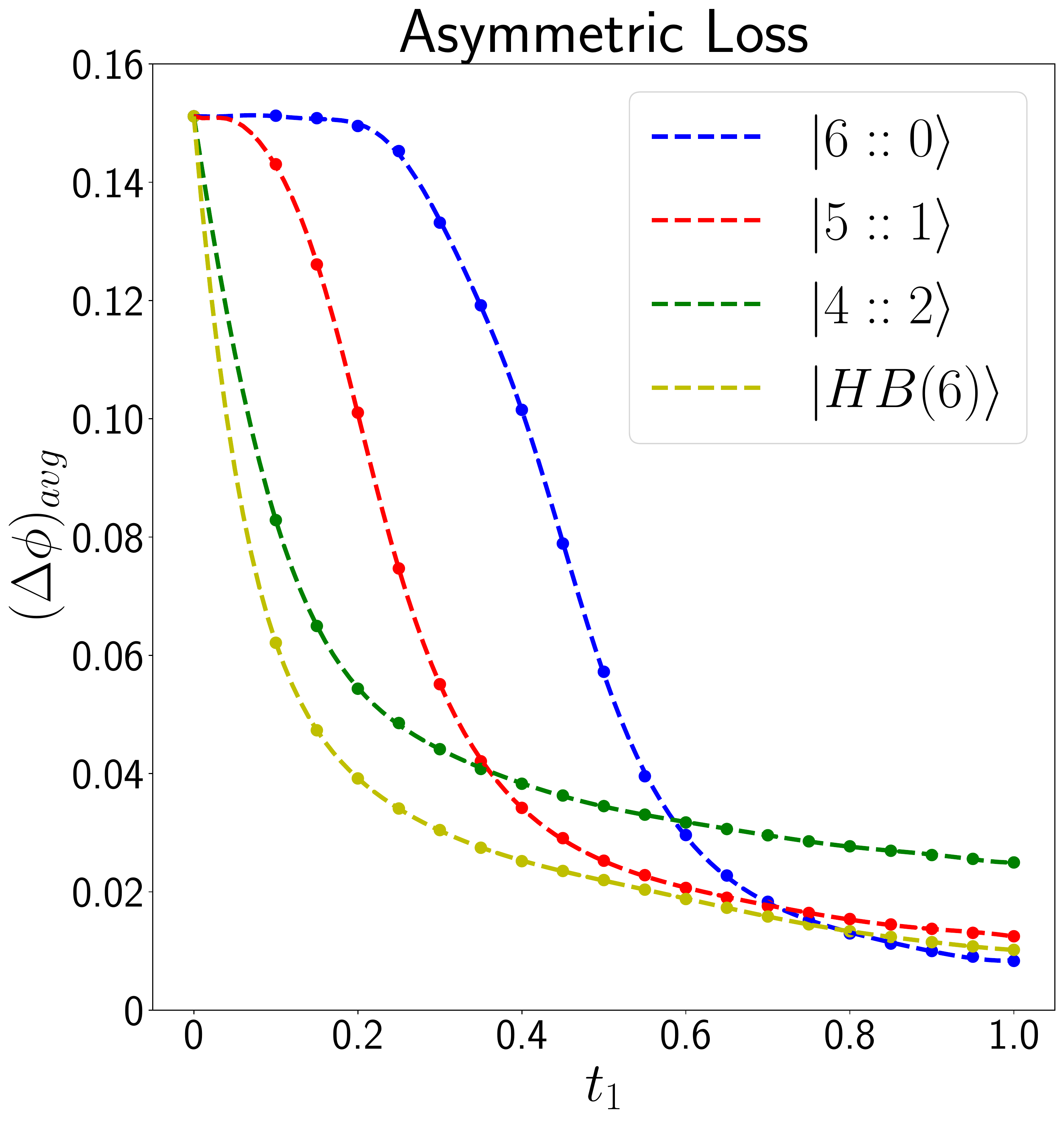}
	\caption{Plot of $(\Delta \phi)_{avg}$ as a function of $t_1$ for different input states for $N_r = 400$. Circle points are the values that are calculated by simulating the quantum circuits and the dashed lines are appropriate functions fitted to the data.}
	\label{dphi_asym}
\end{figure}
In asymmetric photon loss, our simulation results show that likelihood functions change by altering the value of $t_1$, and therefore we had to estimate them for different values of $t_1$ in $[0,1]$. Because of this, we have not shown them here. Fig.~\ref{dphi_asym} Shows the average standard deviation of the input states in estimating the phase shift for $N_r = 400$ in the asymmetric photon loss case. In this case when $t_1 > 0.79$, $\ket{6::0}$ gives the best precision while for other values of $t_1$, $\ket{HB(6)}$ has the best precision. Unlike the symmetric loss case, in asymmetric case, $\ket{4::2}$ never becomes the state with the lowest $(\Delta \phi)_{avg}$ in any loss rate. 
\begin{figure*}[t]	
	\subfloat{\includegraphics[width=0.32\textwidth, , height=0.25\textheight]{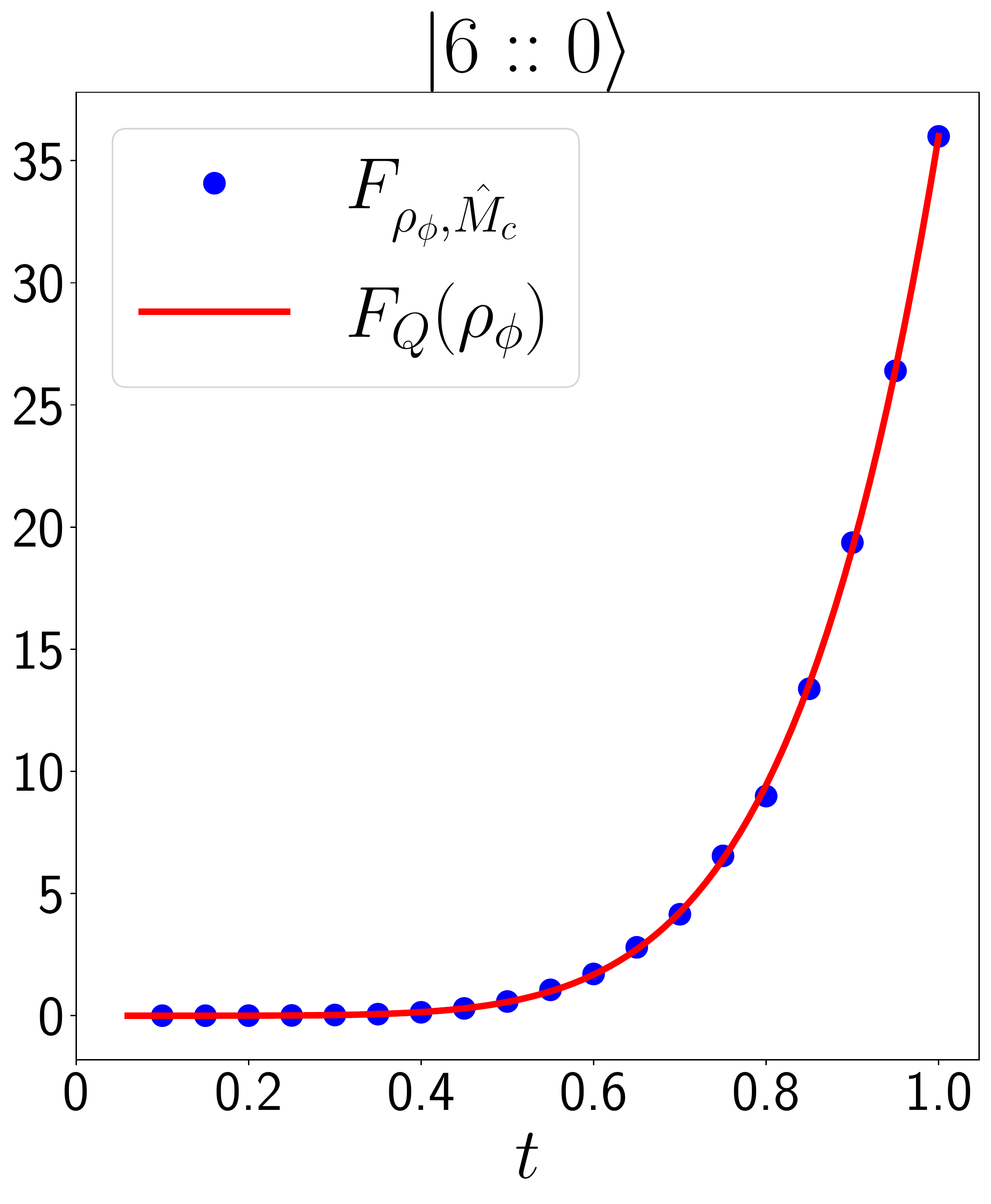}}         \quad
	\subfloat{\includegraphics[width=0.32\textwidth, , height=0.25\textheight]{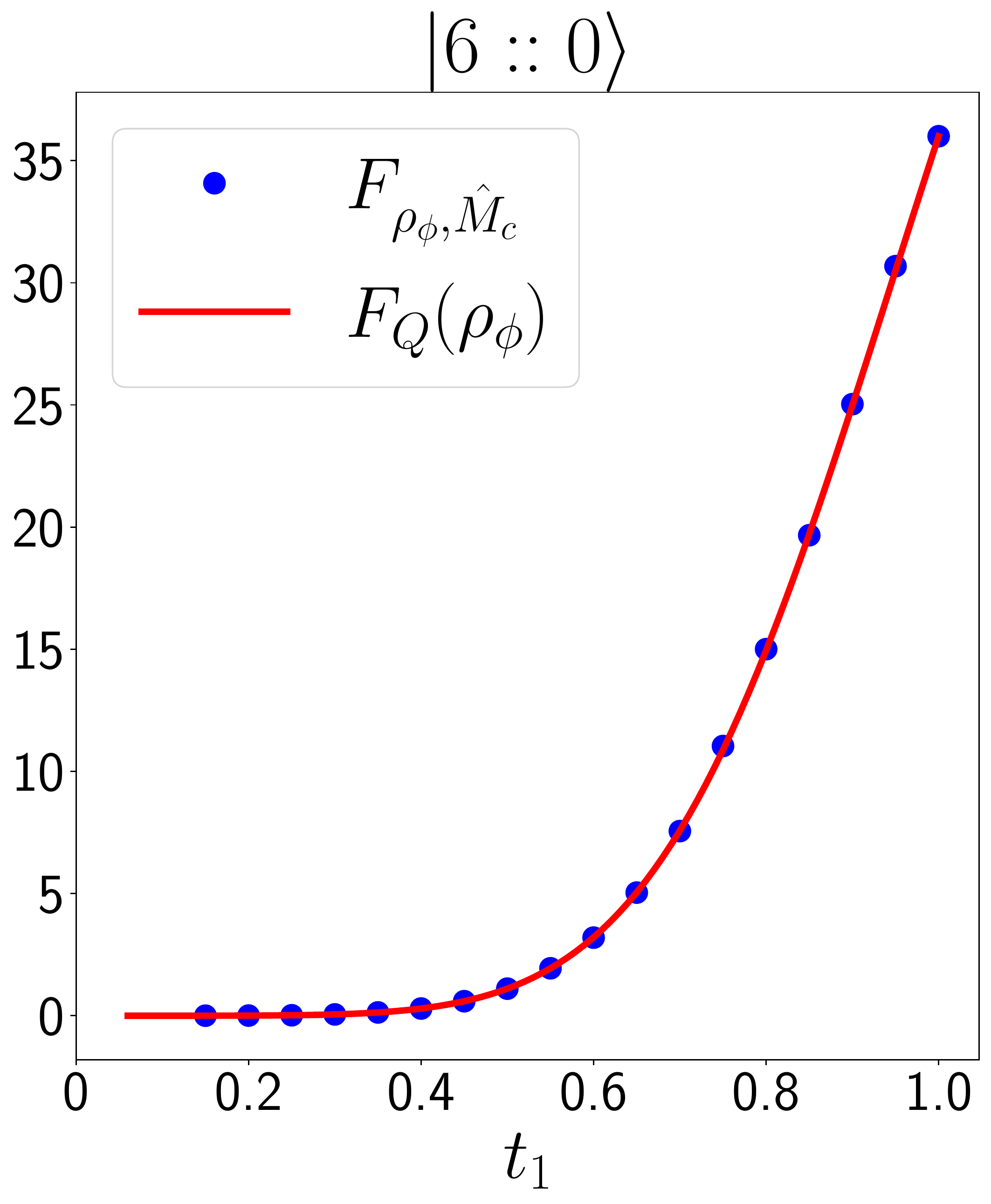}}	\quad 
	\subfloat{\includegraphics[width=0.32\textwidth, height=0.25\textheight]{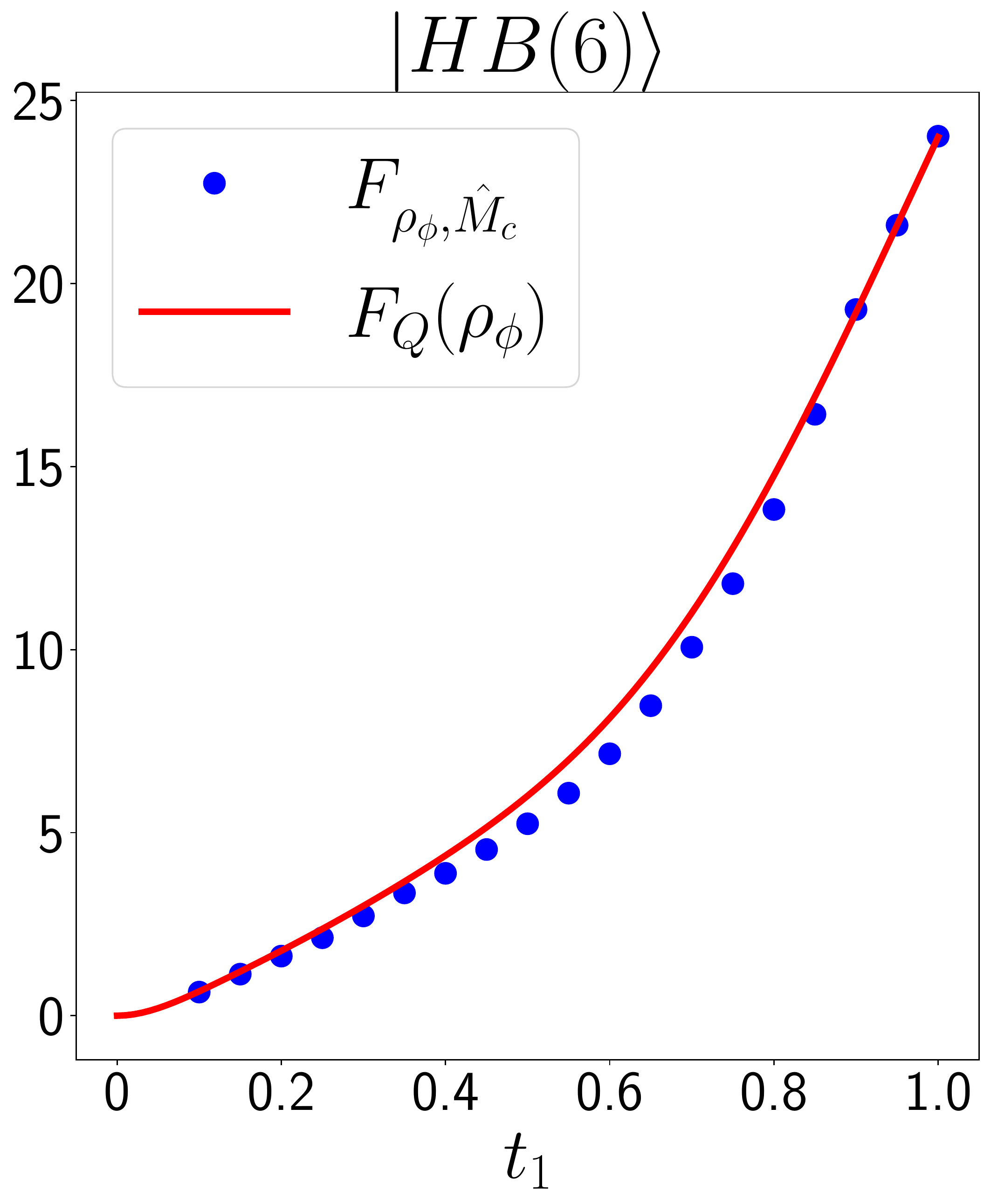}}					
	\caption[justification=justified,format=plain]{Comparing fisher information of our measurement scheme ($F_{\rho_{\phi}, \hat{M}_c}$) with QFI ($F_Q(\rho_{\phi})$) for $\ket{6::0}$ in symmetric (left plot) and asymmetric (middle plot) loss, and for $\ket{HB(6)}$ in asymmetric loss (right plot). It can be seen that for the $NOON$ state ($\ket{6::0}$) our measurement scheme ($\hat{M}_c$) is optimal. However, for $\ket{HB(6)}$, $F_{\rho_{\phi}, \hat{M}_c}$ at most points is a little lower than the QFI, which means that $\hat{M}_c$ is not the optimal measurement scheme at these points, but it is very close to the optimum value.}
	\label{fisher_sym}
\end{figure*}

\subsection{Estimating fisher information}
\label{estimating fisher}
\begin{figure*}[t]
	\subfloat{\includegraphics[width=0.43\textwidth, height = 0.3\textheight]{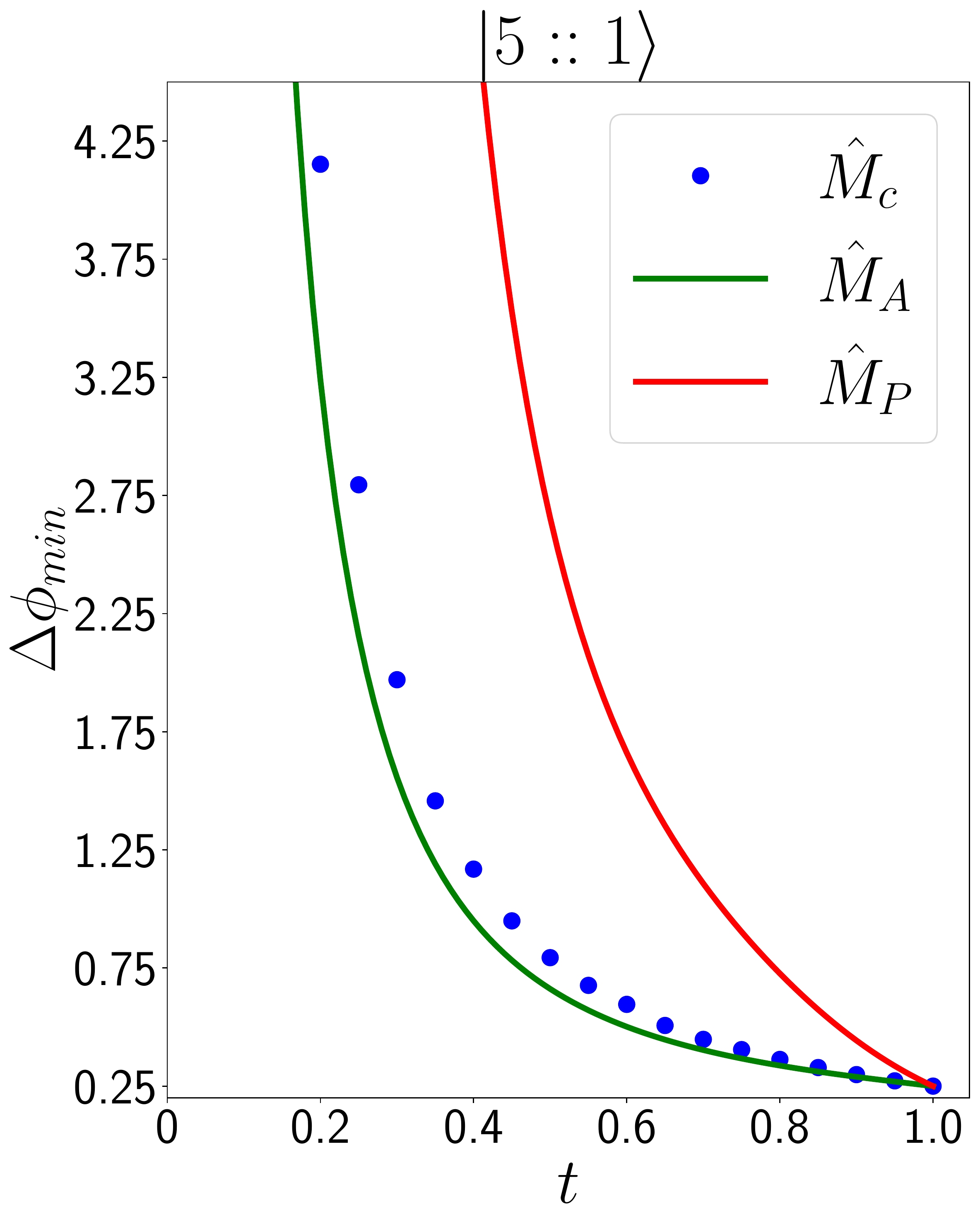}  }\quad
	\subfloat{\includegraphics[width=0.43\textwidth, height = 0.3\textheight]{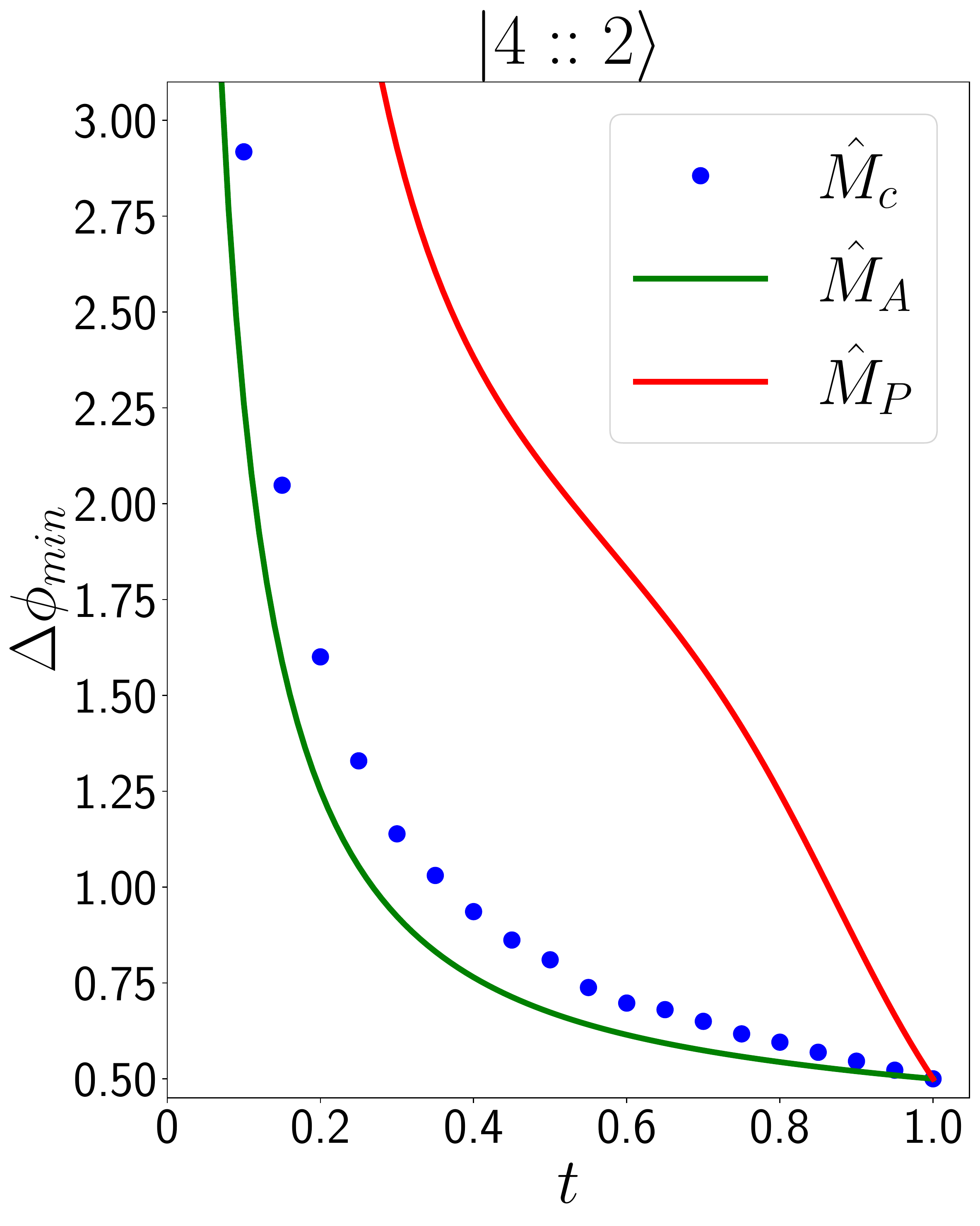}    }\hfil
	
	\subfloat{\includegraphics[width=0.43\textwidth, height = 0.3\textheight]{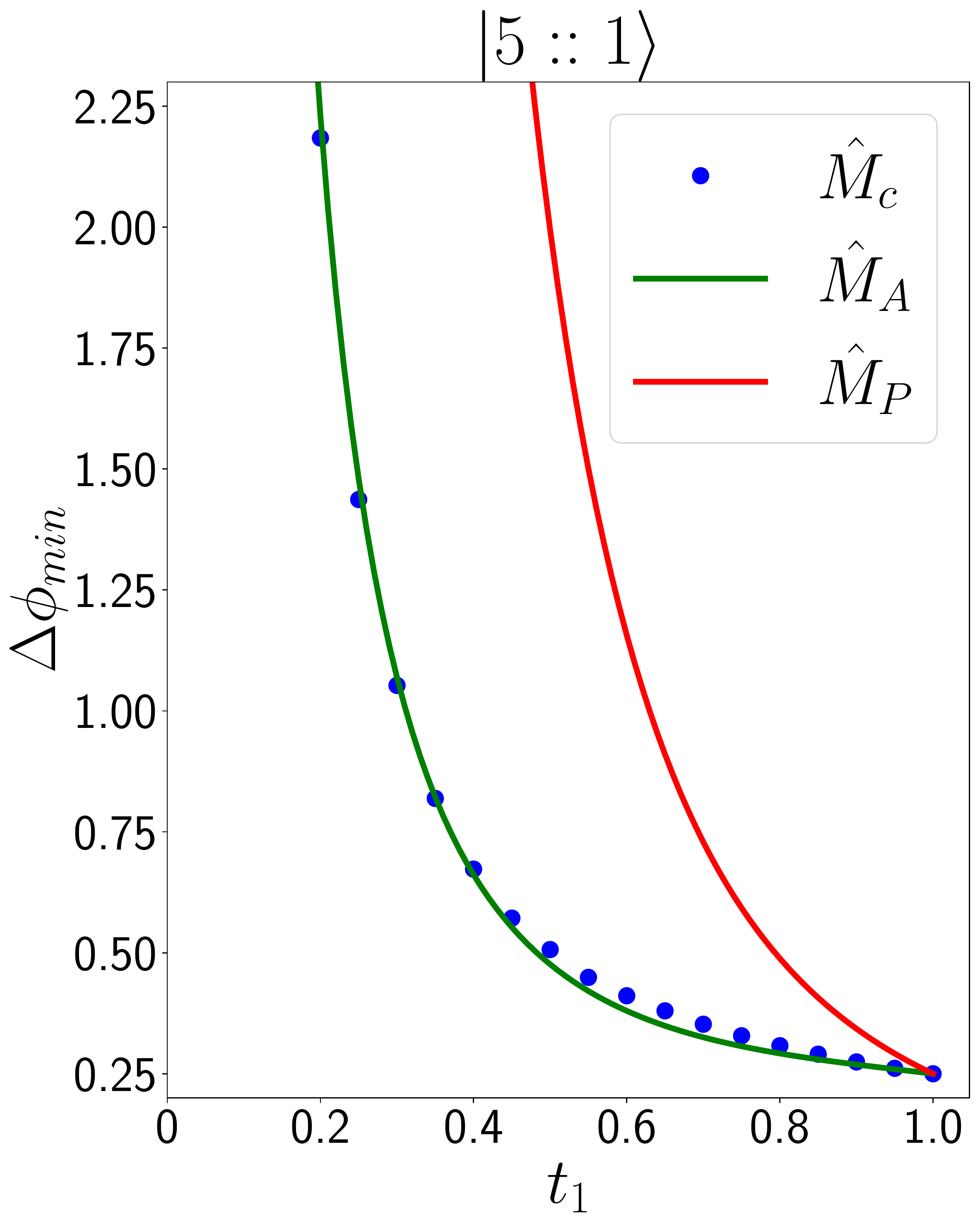}  }\quad
	\subfloat{\includegraphics[width=0.43\textwidth, height = 0.3\textheight]{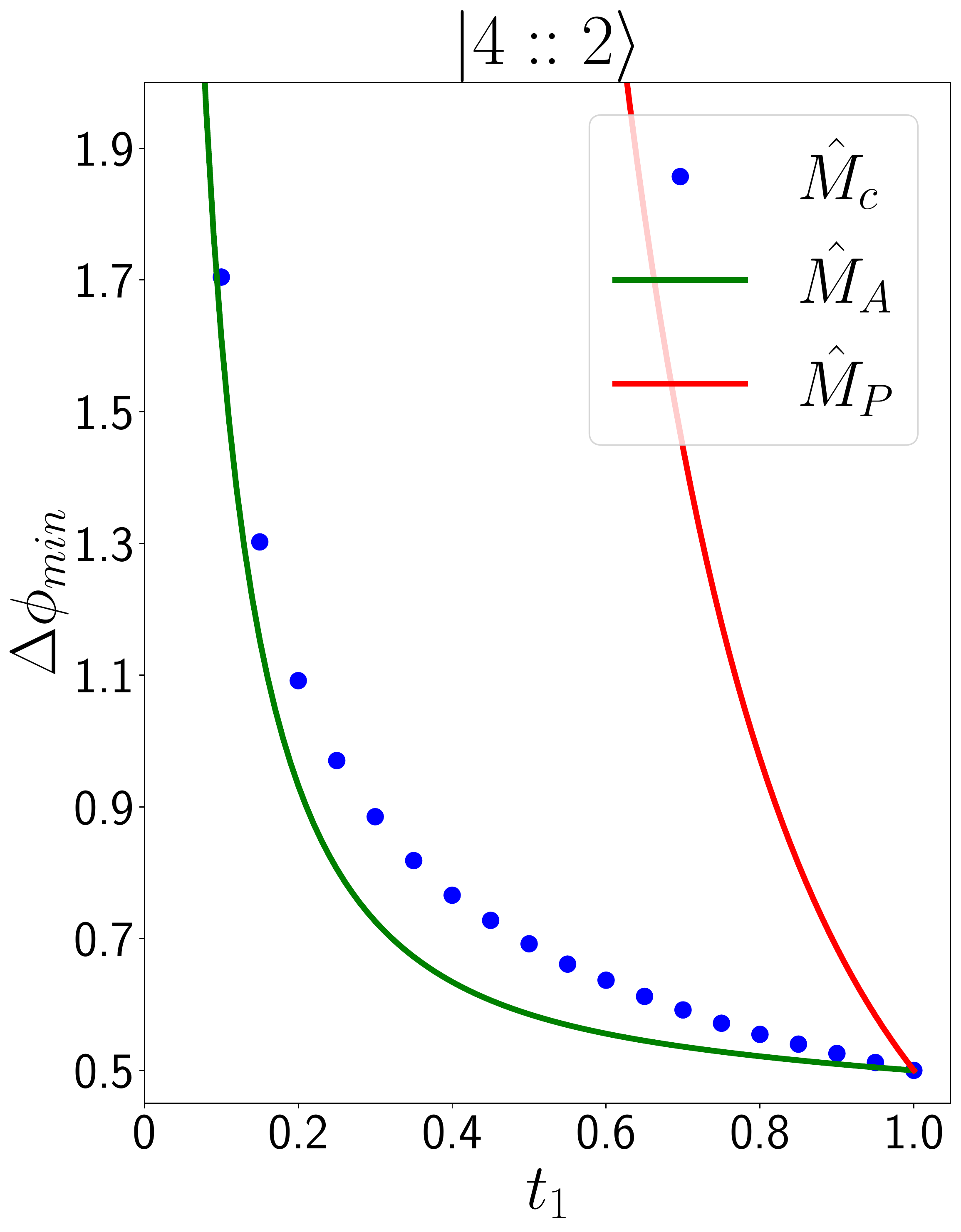}  }

	\caption{Comparing $\Delta \phi_{min}$ of the three measurement schemes ($\hat{M}_c$, $\hat{M}_A$ and $\hat{M}_p$) for $\ket{5::1}$ and $\ket{4::2}$ in symmetric (the two upper figures) and asymmetric (the two lower figures) loss. By increasing photon loss rate, $\Delta\phi_{min}$ of parity detection surges drastically compared to the other two schemes, which means that parity detection is not a good measurement scheme for these $mm^\prime$ states in the presence of loss. However, $\Delta\phi_{min}$ for our measurement scheme ($\hat{M}_c$), is close to the optimal scheme ($\hat{M}_A$), which means that $\hat{M}_c$ is a near-optimal measurement scheme for the two $mm^\prime$ states in the presence of loss.}
	\label{fig:mmprime_measurement_comparison}
\end{figure*}

Generally speaking, a phase estimation process can be divided into three steps: In the first step, one prepares a state ($\rho_{\phi}$), which contains information about the unknown phase that they want to estimate. In the second step, a measurement scheme ($\hat{M}$) is set up to extract the information of $\rho_{\phi}$, and in the last step, an appropriate estimator is chosen to estimate the unknown phase, based on the measurement results. The likelihood functions for the setup depends on $\rho_{\phi}$ and $\hat{M}$, so the fisher information of the phase estimation process depends on them can be calculated by the following formula~\cite{haase_2016_precision, rotondo_2017_quantum}: 
\begin{equation}
	F_{\rho_{\phi}, \hat{M}}(\phi) = \sum_{\vec{D}} \frac{[\partial P_{\rho_{\phi}, \hat{M}}(\vec{D}|\phi)/{\partial \phi}]^2}{P_{\rho_{\phi}, \hat{M}}(\vec{D}|\phi)},
	\label{eq:fisher_classic}
\end{equation} 
where the summation is over all possible outcomes of the measurement ($\vec{D}$).
In frequentist approach, the best possible precision that one can estimate (by using a locally unbiased estimator) the unknown phase is~\cite{li_2018_frequentist, polino2020photonic}
\begin{equation}
	\Delta \phi _{min} = \frac{1}{\sqrt{F_{\rho_{\phi}, \hat{M}}}}
\end{equation}
and the estimator giving this precision is an efficient estimator. By changing the measurement scheme ($\hat{M}$), one can change the fisher information of the setup, and as a result, the best possible precision in phase estimation. Maximizing the fisher information over all generic quantum measurement schemes yields the Quantum Fisher Information (QFI) of $\rho_{\phi}$~\cite{polino2020photonic}:
\begin{equation}
	F_{Q}(\rho_{\phi}) = \max_{\hat{M}} F_{\rho_{\phi}, \hat{M}}(\phi).
\end{equation} 
The measurement scheme that maximizes Fisher information is called the optimal measurement scheme for $\rho_{\phi}$. QFI depends only on $\rho_{\phi}$ and can ba calculated by~\cite{polino2020photonic}:
\begin{equation}
	F_{Q}(\rho_{\phi}) = (\Delta L_\phi)^2 = Tr[\rho_{\phi} L_\phi^2],
\end{equation} 
where $L_\phi$ is the symmetric logarithmic derivative operator and $(\Delta L_\phi)^2$ is its variance over $\rho_{\phi}$.

Regarding the interferometry setup that we are investigating (Fig.~\ref{Fig1}), $\rho_{\phi}$ is the quantum state of light (after tracing out environment modes) before the final beam splitter, and the final beam splitter together with the PNRDs is the measurement scheme, which we show it by $\hat{M}_c$. 
We estimate the fisher information for all the input states in both photon loss cases and discuss the measurement scheme's optimism. 
Instead of using Eq.~(\ref{eq:fisher_classic}), fisher information is estimated by using the Bayesian approach and simulating the quantum circuits a large number of times. 
\begin{figure*}[t]	
	\subfloat{\includegraphics[width=0.45\textwidth, height=.43\textheight]{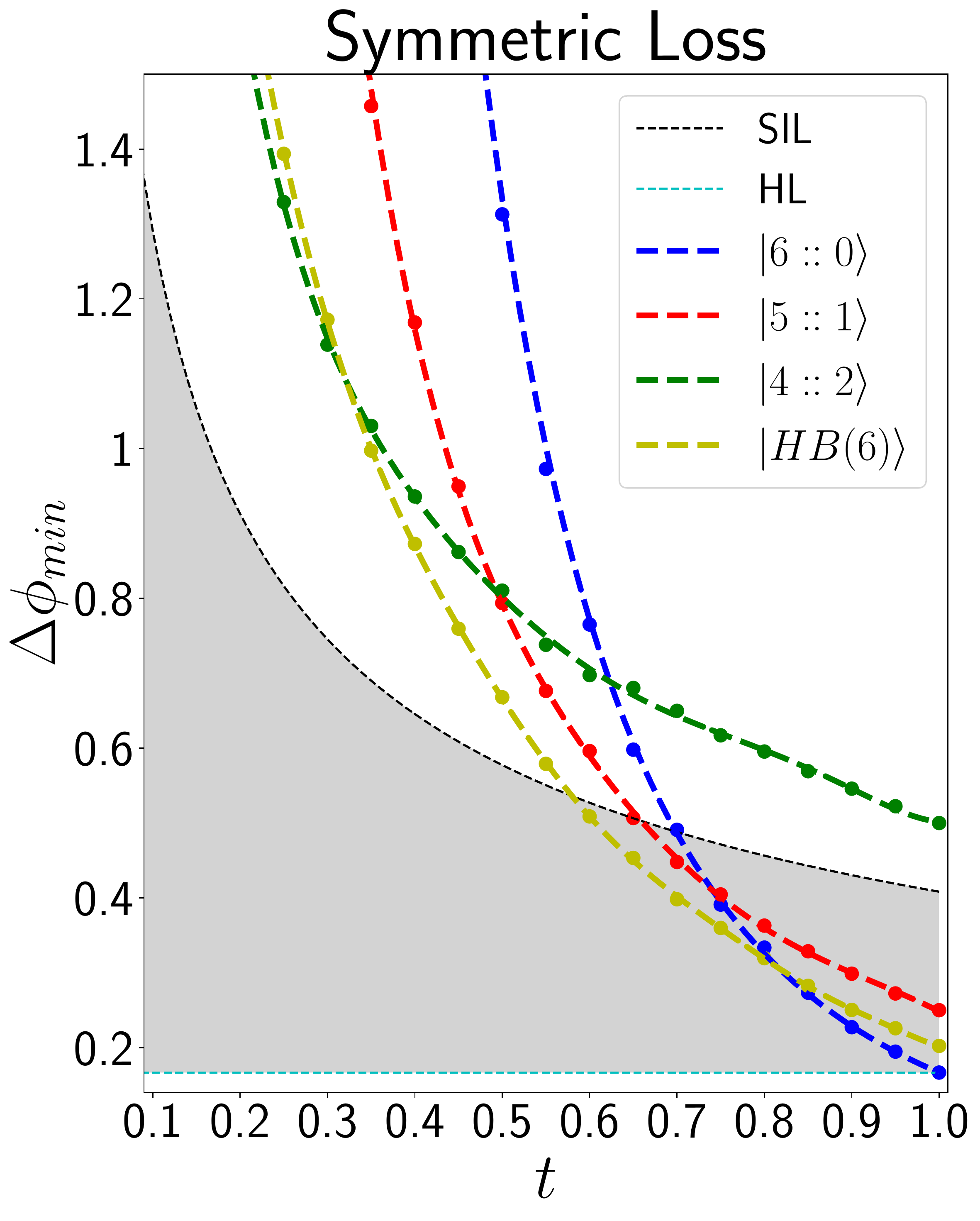}  }\quad
	\subfloat{\includegraphics[width=0.45\textwidth, height=.43\textheight]{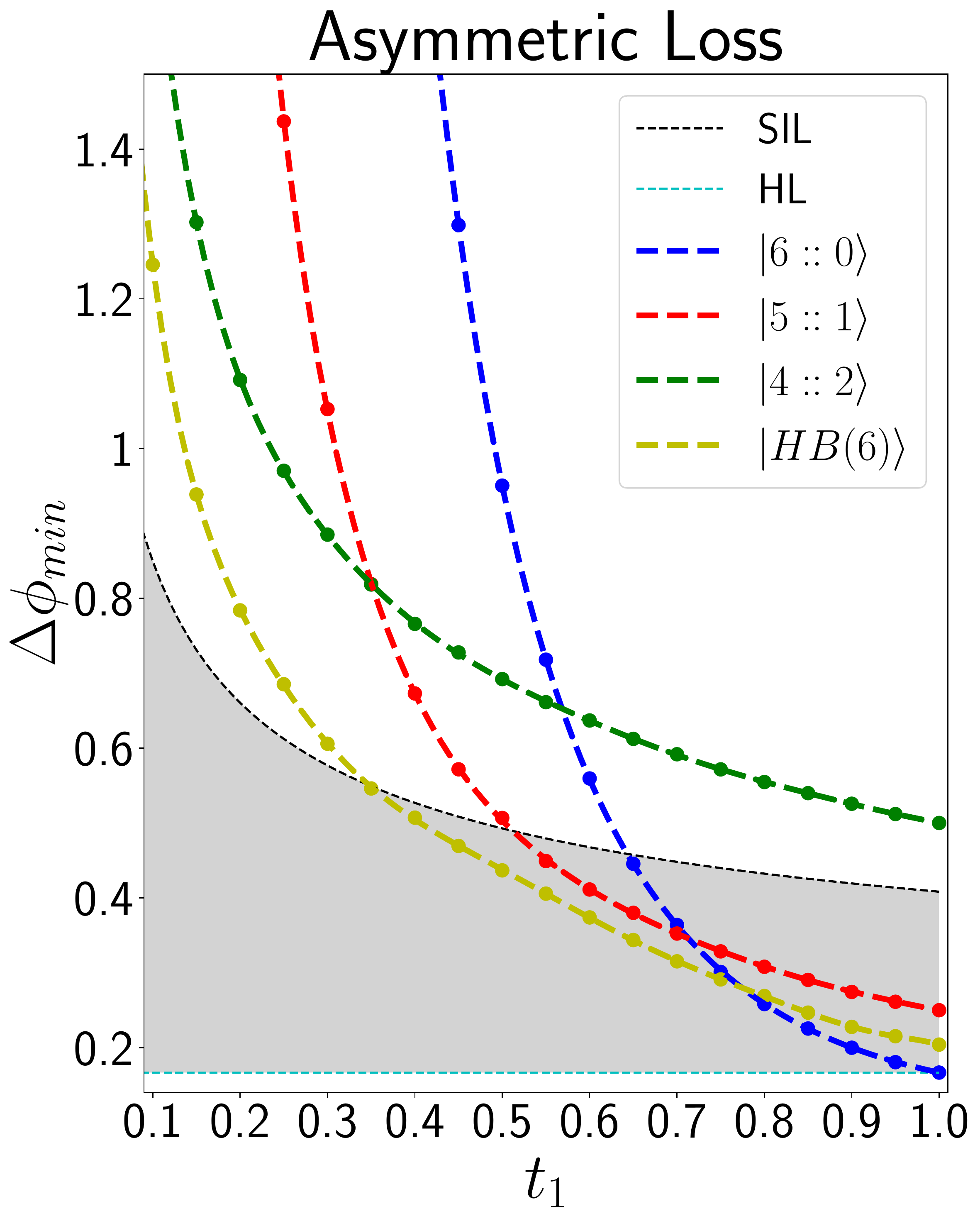}  }
	
	\caption{Comparing $\Delta \phi_{min}$ of our scheme with standard interferometry limit. Circle points are the values that are estimated by simulating the quantum circuits and the dashed lines are appropriate fits. The region between Heisenberg and standard classical limit is colored with light-gray. Whenever a line (corresponding to any input state) is in this region, it means that the corresponding input state has better precision than any classical interferometry setup, using the same number of resources. According to these figures, $\ket{6::0}, \ket{5::1}$ and $\ket{HB(6)}$ can beat the classical limit in some loss rates in both symmetric (the left figure) and asymmetric case (the right figure); however, $\ket{4::2}$ cannot beat the classical limit with our measurement scheme in any loss rates of symmetric and asymmetric losses.}
	\label{fig:classic_heisenberg_comparison}
\end{figure*}

	In the Bayesian estimation, when the number of measurements $N_{r}$ increases, the final probability distribution $P_{final}(\phi|\vec{D})$ approaches a Gaussian distribution~\cite{li_2018_frequentist}:
	\begin{equation}
		P_{final}(\phi|\vec{D}) \approx \sqrt{\frac{N_r F(\phi^*)}{2\pi}} \exp^{-\frac{N_rF(\phi^*)}{2} (\phi-\phi^*)^2}, \, (N_r >> 1).
	\end{equation} 
	This probability distribution is centered around the actual value of the unknown phase shift $\phi^*$, with the variance of $\frac{1}{N_r F(\phi^*)}$ in which $F(\phi^*)$ is the Fisher information of the setup.
	Using this result and simulating each quantum circuit a large number of times ($N_r>>1$), we have estimated the input states' Fisher information in symmetric and asymmetric photon loss conditions for the photon-counting measurement scheme. 
	QFI has been calculated For $NOON$ states in both symmetric and asymmetric photon loss~\cite{dorner_2009_optimal, zhang_2013_quantum-fisher-information-of-entangled-coherent-states, demkowicz_2009_quantum-phase-estimation-with-lossy-interferometer}, and for $HB$ states in asymmetric case~\cite{datta_2011_quantum}. Fig.~\ref{fisher_sym} shows comparison of these QFIs with the fisher information of our measurement scheme ($F_{\rho_{\phi}, \hat{M}_c}$). According to this figure, for $\ket{6::0}$ states in both loss cases, the photon-counting measurement scheme is the optimal measurement because $F_{\rho_{\phi}, \hat{M}_c}$ is equal to QFI. For $\ket{HB(6)}$ state in asymmetric loss, fisher information of our scheme is very close to QFI, and even at some points, it is equal to QFI; however, at most points, it is lower than the QFI which means that at these points our measurement scheme is not optimal.
	
	To the best of our knowledge, QFI has not been calculated for $mm^\prime$ states inside a lossy interferometer. However, the precision of phase estimation for this case has been calculated by two other measurement schemes. Bardhan \emph{et al.}~\cite{dowling_2013} by measuring the parity operator
	\begin{equation}
		\hat{P} = i^{(m+m^\prime)}\sum_{k=0}^{m} (-1)^k \ket{k, n-k} \bra{n-k, k},
	\end{equation}
	and using the error propagation formula have calculated the precision of phase estimation for parity detection. We show this measurement scheme by $\hat{M}_p$ here. Huver \emph{et al.} in~\cite{Huver_2008} have introduced another observable to extract phase information from $mm^\prime$ states:
	
	\begin{eqnarray}
		\label{operator}
		\hat{A}=\sum_{r,s=0}^{m'}|m'-r,m-s\rangle\langle m-r,m'-s| \nonumber\\+ |m-r,m'-s\rangle\langle m'-r,m-s|.
	\end{eqnarray}
	They have shown that their operator is theoretically optimal for $mm^\prime$ states, but they have not discussed the practical implementation of their optimal measurement scheme. We show this measurement scheme by $\hat{M}_A$. 
Figure.~\ref{fig:mmprime_measurement_comparison} shows comparison of $\Delta \phi_{min}$ of our measurement scheme with these two schemes, for $\ket{5::1}$ and $\ket{4::2}$ in symmetric and asymmetric losses. 
It reveals that, 1) when there is no photon loss, all three schemes have the same precision and $\Delta \phi_{min} = \frac{1}{m-m^\prime}$, which is the best possible precision for $mm^\prime$ states in the absence of loss~\cite{Huver_2008}. 
2) By increasing the photon loss rate (in both symmetric and asymmetric cases), parity detection's precision decreases with a much greater slope than the other two schemes. It means that parity detection is not a good measurement scheme to extract the phase information in the presence of loss. 3) The difference between the photon counting measurement scheme and the optimal scheme is tiny, especially for $\ket{5::1}$. Therefore the photon counting measurement scheme is near-optimal for these $mm^\prime$ states and can be implemented in the lab by using a 50-50 beam splitter and two PNRD detectors.

At the final part, in Fig.~{\ref{fig:classic_heisenberg_comparison}}, we have compared $\Delta \phi_{min} = \frac{1}{F_{\rho_{\phi}, \hat{M}_c}}$ of our phase estimation setups with the Heisenberg limit, $\Delta \phi_{HL} = \frac{1}{N}$, and standard interferometry limit~\cite{demkowicz_2009_quantum-phase-estimation-with-lossy-interferometer}:	
	\begin{equation}
		\Delta \phi_{min,SIL} = \frac{\sqrt{t_0} + \sqrt{t_1}}{2\sqrt{N t_0 t_1 }},
	\end{equation} 
	which is the best possible precision that a classical setup can reach.      
	The region between the standard classical limit and the Heisenberg limit is colored with light-gray. Whenever a line is in this colored region, there is an improvement over classical interferometry. According to these results, $\ket{4::2}$ cannot beat the classical limit in any loss rate. The other three input states can act better than the standard interferometry limit whenever the loss rate is lower than a specific value. 
	
	\section{Summary And Conclusion}
	\label{summary}
	We have simulated a lossy Mach-Zehnder interferometer by using quantum circuits, for four definite photon-number states $(\ket{6::0}$, $\ket{5::1}$, $\ket{4::2}$ and $\ket{HB(6)})$ as its input states. The measurement scheme that we have considered here is counting the number of photons detected at the interferometer's two different output ports.
	Because of the current quantum hardware's noise, the quantum circuits have been simulated by a classical computer. We have compared the phase estimation precision of the four input states when there is symmetric and asymmetric photon loss in the interferometer. According to our results, $\ket{HB(6)}$ state has the highest precision in most of the loss rates in both symmetric and asymmetric cases; in the symmetric loss whenever $0.68>L>0.19$, and in the asymmetric loss whenever $L_1>0.21$, $\ket{HB(6)}$ has the highest phase estimation precision. This means that $\ket{HB(6)}$ is better than the others for real-world quantum optical phase estimations.
	We have also checked the optimality of the photon-counting measurement scheme for these input states and found that it is optimum for $\ket{6::0}$ in both loss cases and is near-optimum for the other three states when there is loss.
	Our results also show the range of photon loss rates, in which one can beat the standard interferometry limit by using $\ket{6::0}$, $\ket{5::1}$ and $\ket{HB(6)}$ as input states in the Mach-Zehnder interferometer and photon-counting as measurement scheme. 
	
	By introducing appropriate subroutines into our quantum circuits, we can also implement phase fluctuations, input state preparation inefficiencies, and detector inefficiency. Furthermore, we can also implement any other measurement scheme by adding the appropriate subroutine in the last part of our quantum circuits. These features help us compare different definite photon-number input states' precision in different decoherence conditions with different measurement schemes.  	
	
	\appendix
	
	\section{Basis Gates}  
	\label{basis-gates}  
	We use the following basis gates, which are available at Qiskit Terra~\cite{Qiskit}:

	\begin{enumerate}
	\item One-qubit gates:
	\begin{flalign*}
		\begin{cases}
			X: &  X = \begin{pmatrix}
				0 & 1 \\
				1 & 0 
			\end{pmatrix}\\
			Z: &  Z = \begin{pmatrix}
				1 & 0 \\
				0 & -1 
			\end{pmatrix} \\
			H: &  H = \frac{1}{\sqrt{2}}\begin{pmatrix}
				1 & 1 \\
				1 & -1 
			\end{pmatrix} \\
			u3: &   u3(\theta,\phi,\lambda) = \begin{pmatrix}
				\cos(\frac{\theta}{2}) & -\exp^{i\lambda}\sin(\frac{\theta}{2}) \\
				\exp^{i\phi}\sin(\frac{\theta}{2}) & \exp^{i\phi + i\lambda}\cos(\frac{\theta}{2}) 
			\end{pmatrix} 
		\end{cases}	&&	
	\end{flalign*}	
		\item  Two-qubit gates 
	\begin{flalign*}
	\begin{cases}
	C_X(q\{c\},q\{t\}): &\text{$X$ is implemented on $q\{t\}$,}\\
	&\text{controlled by $q\{c\}$} \\  
	C_Z(q\{c\},q\{t\}): &\text{$Z$ is implemented on $q\{t\}$,}\\
	&\text{controlled by $q\{c\}$} \\  
	C_H(q\{c\},q\{t\}): &\text{$H$ is implemented on $q\{t\}$,}\\
	&\text{controlled by $q\{c\}$} \\  
	C_{u3}(q\{c\},q\{t\}): &\text{$u3$ is implemented on $q\{t\}$,}\\
	&\text{controlled by $q\{c\}$} \\  
	\end{cases}	&&	
	\end{flalign*}
	\item  Three-qubit gates 
	\begin{flalign*}
	\begin{cases}
	\begin{aligned}
	C_{CX}(q\{c1\},q\{c2\},q\{t\}): & \text{$X$ is implemented on $q\{t\}$,}\\
	&\text{controlled by $q\{c1\}$ and $q\{c2\}$}  
	\end{aligned}
	\end{cases}&&
	\end{flalign*}
	\end{enumerate} 

	We do not use the u3 gate unless other gates do not simply implement the operation.
	
	\begin{figure*}[t]
		\includegraphics[width=\linewidth]{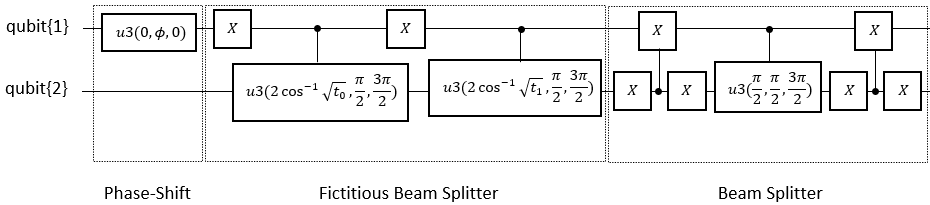}  
		\caption{Implementation of phase shift, fictitious beam splitter and the final beam splitter on the first photon.}
		\label{fig-one_photon_evolution}	
	\end{figure*}
	
	\section{Implementation of $U_{input}$ by Basis Gates}
	\label{U-input}
	Here we show how to implement $U_{input}$ by the basis gates. We write our input state in the following way:
	\begin{equation}
		\ket{\psi_{input}} = \sum_{i=1}^{n} c_i \ket{\psi_i},
	\end{equation}
	in which $c_i \neq 0$ and $\ket{\psi_i}$ is a product states of $2N$ qubits, where each qubit is either in $\ket{0}$ or $\ket{1}$ state and 
	\begin{equation}
		\braket{\psi_i|\psi_j} = 0 \,\, (for i\neq j).
	\end{equation}
	By the definition of $U_{input}$, 
	\begin{equation}
		U_{input}\ket{0}_1...\ket{0}_{2N} = \sum_{i=1}^{n} c_i \ket{\psi_i}.
	\end{equation}
	 The Following algorithm shows us how to implement $U_{input}$ by the basis gates.
	 First we transform $\ket{0}^{\otimes N}$ to $\ket{\psi_1}$ by implementing $X$ gate on qubits, which are not in $\ket{0}$ state in $\ket{\psi_1}$. Then by using the concept of Gray codes~\cite[pp.~191--193]{nielsen_chuang_2010} and introducing ancillary qubits to implement controlled operations with more than two control qubits~\cite[pp.~183--184]{nielsen_chuang_2010}, we implement the following unitary evolution on the subspace spanned by $\ket{\psi_1}$ and $\ket{\psi_2}$:
	\begin{equation}
		U_{(1,2)} = \begin{pmatrix}
			c_1 & -\sqrt{1-c_1 ^2} \\
			\sqrt{1-c_1 ^2} & c_1 
		\end{pmatrix}
	\end{equation}
	After this step, our state would be:
	\begin{equation}
		\ket{\psi} = c_1\ket{\psi_1} + c_2^{\prime}\ket{\psi_2}
	\end{equation}
	
	Here $c_2^{\prime} = \sqrt{1-c_1 ^2}$. It is clear that after this step our state has the right coefficient of projection onto $\ket{\psi_1}$ state and in a similar manner, by implementing following unitary evolution on the subspace spanned by $\ket{\psi_2}$ and $\ket{\psi_3}$ we can fix the coefficient of projection onto $\ket{\psi_2}$:
	\begin{equation}
		U_{(2,3)} = \begin{pmatrix}
			\frac{c_2}{c_2^{\prime}} & -\sqrt{1-\frac{c_2^2}{{c_2^{\prime} } ^2}} \\
			\sqrt{1-\frac{c_2^2}{{c_2^{\prime} } ^2}} & \frac{c_2}{c_2^{\prime}} 
		\end{pmatrix}
	\end{equation}
	We keep doing this until we fix the final projection coefficient ($c_n$), and afterward, our input state is ready.   
	
	\section{Implementation of phase shift, Fictitious Beam Splitter and the final Beam Splitter by Basis Gates}
	\label{ap-phase}	
	
	Figure \ref{fig-one_photon_evolution} shows how these three operations are implemented in our circuit for $qubit\{1\}$ and $qubit\{2\}$, which correspond to $photon\{1\}$. In implementing phase shift, we have used the fact that before the fictitious beam splitters, qubits corresponding to a photon can only be in state $\ket{0}\ket{0}$ or $\ket{1}\ket{1}$. These gates are implemented to other qubits (related to other photons) similarly.

	\bibliography{paper_ref}

\end{document}